 \documentclass[final,5p,times,twocolumn,authoryear]{elsarticle}

\usepackage{amssymb}
\usepackage{lipsum}
\usepackage{url}
\usepackage{threeparttable}
\usepackage{booktabs}
\usepackage{bm}
\usepackage{comment}
 \usepackage[bookmarks=true,         
    pdfnewwindow=true,      
   colorlinks=true,    
  linkcolor=xlinkcolor,     
 citecolor=xlinkcolor,     
filecolor=xlinkcolor,  
urlcolor=xlinkcolor,      
final=true,
 ]{hyperref}
 
\journal{Astronomy $\&$ Computing}

\newcommand{\soft}{\texttt}
\newcommand{\profund}{\soft{ProFound}}
\newcommand{\caesar}{\soft{CAESAR}}
\newcommand{\selavy}{\soft{Selavy}}
\newcommand{\pybdsf}{\soft{PyBDSF}}
\newcommand{\aegean}{\soft{Aegean}}
\newcommand{\claran}{\soft{CLARAN}}
\newcommand{\HeTu}{\soft{HeTu}}
\newcommand{\miriad}{\soft{MIRIAD}}
\newcommand{\stilts}{\soft{STILTS}}

\newcommand{\happy}{\soft{HAPPY}}

\begin{document}

\begin{frontmatter}



\title{Radio Sources Segmentation and Classification with Deep Learning}


\author[1]{B. Lao\corref{cor1}}
\ead{lbq19881213@gmail.com}
\cortext[cor1]{Corresponding Author}
\author[2]{S. Jaiswal}
\author[3]{Z. Zhao}
\author[4]{L. Lin}
\author[4]{J. Wang}
\author[1]{X. Sun}
\author[1]{S.-L. Qin}

\affiliation[1]{organization={School of Physics and Astronomy, 
Yunnan University},
            city={Kunming},
            postcode={650091}, 
            country={China}}

\affiliation[2]{organization={Shanghai Astronomical Observatory, Chinese Academy of Sciences},
            city={Shanghai},
            postcode={20030}, 
            country={China}}

\affiliation[3]{organization={School of Electrical and Information Engineering, University of Sydney},
            city={Sydney},
            postcode={2006}, 
            country={Australia}}

\affiliation[4]{organization={School of Information and Communication, Guilin University Of Electronic Technology},
            city={Guilin},
            postcode={541004}, 
            country={China}}

\begin{abstract}
Modern large radio continuum surveys have high sensitivity and resolution, and can resolve previously undetected extended and diffuse emissions, which brings great challenges for the detection and morphological classification of  
extended sources. We present \HeTu-v2, a deep learning-based source detector that uses the combined networks of Mask Region-based Convolutional Neural Networks (Mask R-CNN) and a Transformer block to achieve high-quality radio sources segmentation and classification. The sources are classified into 5 categories: Compact or point-like sources (CS), Fanaroff-Riley Type I (FRI), Fanaroff-Riley Type II (FRII), Head-Tail (HT), and Core-Jet (CJ) sources. \HeTu-v2 has been trained and validated with the data from the Faint Images of the Radio Sky at Twenty-one centimeters (FIRST). We found that \HeTu-v2 has a high accuracy with a mean average precision ($AP_{\rm @50:5:95}$) of 77.8\%, which is 15.6 points and 11.3 points higher than that of \HeTu-v1 and the original Mask R-CNN respectively. We produced a 
FIRST morphological catalog (FIRST-\HeTu) using \HeTu-v2, which contains 835,435 sources and achieves 98.6\% of completeness and up to 98.5\% of accuracy compared to the latest 2014 
data release of the FIRST survey. \HeTu-v2 could also be employed for other astronomical tasks like building sky models, associating radio components, and classifying radio galaxies. 
\end{abstract}



\begin{keyword}
Radio continuum survey \sep Radio sources \sep Image segmentation \sep Morphological classification \sep Deep learning



\end{keyword}

\end{frontmatter}



\section{Introduction}\label{sec:intro}
The modern large radio telescopes, $e.g.$ the Square Kilometre Array (SKA) pathfinders like Low Frequency Array \citep[LOFAR;][]{2013A&A...556A...2V}, Murchison Widefield Array \citep[MWA;][]{2013PASA...30....7T}, and Australian SKA Pathfinder \citep[ASKAP;][]{2021PASA...38....9H}, are designed with the novel technologies, such as phased array, digital beamforming, or Phased Array Feed (PAF), and thus have wide field of view, high resolution, high sensitivity, and broad frequency coverage. The radio continuum surveys carried out by these telescopes have an effective combination of depth and area, and can detect a lot of previously unseen or unresolved emissions \citep{2020A&A...635A...5D,2021MNRAS.501.3833D,2021Galax...9...99A}.
The new detection presents 
various complicated morphologies beyond simple point-like sources.
Examples are the LOFAR Two-metre Sky Survey \citep[LoTSS;][]{2017A&A...598A.104S}, GaLactic and Extragalactic All-sky MWA eXtended (GLEAM-X) survey \citep{2022PASA...39...35H}, MeerKAT International GHz Tiered Extragalactic Exploration (MIGHTEE) survey \citep{2016mks..confE...6J}, ASKAP Evolutionary Map of the Universe (EMU) survey \citep{2011PASA...28..215N}, Apertif imaging survey \citep{2022A&A...667A..38A}, and Karl G. Jansky Very Large Array Sky Survey \citep[VLASS;][]{2020PASP..132c5001L}. The EMU survey, as an instance, is expected to produce a catalog containing about 70 million radio sources 
\citep{2011PASA...28..215N}. With this, considerable efforts in recent years have focused on developing algorithms for imaging and source-finding in a rapid, reliable, and automated way \citep{lao2019parallel,2020ASPC..527..591W,2018PASA...35...11H,riggi2016automated}.

In radio astronomy, a source-finding software, also called source finder, is typically used to identify and extract radio sources from images. The 
source finders widely used in existing surveys are \aegean\ \citep{2012MNRAS.422.1812H,2018PASA...35...11H}, Python Blob Detection and Source Finder \citep[\pybdsf;][]{2015ascl.soft02007M}, and \selavy\ \citep{2012PASA...29..371W}; and they all form catalogs by fitting Gaussian components to sources. In the fitting procedure, compact or point-like sources are fitted with a single Gaussian component, and extended sources are fitted with a combination of Gaussian components. This approach works well for compact or point-like sources but is not necessarily true for extended sources because there can be over-fitting of components where Gaussians are not as appropriate for the emission shape \citep{hale2019radio}.
To address this problem, several new source finders have been proposed, such as Compact And Extended Source Automated Recognition \citep[\caesar;][]{riggi2016automated,riggi2019caesar} and \profund\ \citep{2018MNRAS.476.3137R,hale2019radio}. Both source finders are developed based on the pixel segmentation technique and specifically designed for the extraction of extended sources, but their performances are still not as good as that achieved for compact sources. In particular, they cannot work for the extended source that has two-lobed jets with a large separation \citep{hale2019radio}.

With the artificial intelligence awakening and developments of 
Convolutional Neural Networks (CNN) in recent times, deep learning algorithms have been widely used as an automated approach in many specific areas of radio astronomy \citep{2020MNRAS.496.1517M}, 
including astronomical source detection \citep{2019Galax...8....3L,2019MNRAS.484.2793V,2019MNRAS.482.1211W,burke2019deblending,arslan2020radio,farias2020mask,2021SciBu..66.2145L,2023A&C....4200682R}. The 
source-finding methods based on deep learning, called source detectors, allow automatic localization and classification of sources, and their detection speeds are significantly faster than the standard source finders by utilizing stronger computing power, viz. Graphics Processing Units (GPUs) or Neural Processor Units (NPUs). Moreover, 
a source detector performs classification of sources based on their morphologies, and the separate components of larger extended sources are automatically linked and classified. The existing source detectors, however, have some limitations on the grounds of accuracy and performance for extended sources. For example, the accuracy and performance of \claran\ \citep{2019MNRAS.482.1211W} are limited by the lack of layers of a backbone network it uses.       

\citet{2021SciBu..66.2145L} developed a source detector \HeTu~ (to be termed as \HeTu-v1) using Faster Region-based CNN \citep[Faster R-CNN;][]{ren2015faster}. The \HeTu-v1 is able to locate and classify four classes of radio sources, 
but its accuracy is limited 
because of the approximation method used in the network to get the bounding boxes of the sources, and 
fewer samples of the extended sources used in the model training. 
Recently, it has been found that the methods \citep[e.g.][]{burke2019deblending,arslan2020radio,farias2020mask,2023A&C....4200682R} using Mask R-CNN \citep{he2017mask} have excellent performance in localization of the bounding boxes, as well as provide segmentation masks for each source that is implemented similar to the pixel segmentation-based source finders. However, the segmentation masks obtained by the original Mask R-CNN based software are still coarse and cannot satisfy scientific research. The quality of the segmentation can profoundly impact the accuracy of the flux density estimation of radio sources, especially for extended sources. Thus, high-quality segmentation is critical for source-finding in modern radio surveys.

This paper focuses on the upgrade of \HeTu-v1 into
\HeTu-v2 using the combination network of Mask R-CNN and state-of-the-art object detection algorithm $i.e.$ Transformer block \citep{carion2020end} to achieve the high-quality segmentation and morphological classification of radio sources. 
\HeTu-v2 
can perform multiple astronomical tasks and has distinct advantages over 
other source detectors: 
\begin{description}
\item \textbf{Building morphological catalog.} The main function of \HeTu-v2 is to build a catalog from radio image. Each source in the catalog 
is described by the detected information (class name, mask, bounding box, and score) and standard properties (flux density and location) of the source. The morphological classification covers five categories (see Section \ref{sec:scheme}). 
The cataloged properties include Gaussian-fit information for the compact sources and deep-learning estimated information for the extended sources. Such a catalog allows its direct use in statistical analysis and astronomical scientific research of specific classes. 
\item \textbf{Building accurate sky models.} \HeTu-v2 can produce high-quality segmentation, which can be used to build a sky model that is typically used in the radio data processing pipeline 
like \profund\ and \caesar, and avoids the Gaussian over-fit when modeling the extended emission. More importantly, \HeTu-v2 can automatically associate the separated radio lobes to build a complete model, which is more accurate than \profund\ and \caesar.
\item \textbf{Automatic associate radio components.} \HeTu-v2 can be extended to find the 
radio components associated with a radio source in an automated way from the components-based catalog. The association method is based on the predicted segmentation mask of the radio source that is generated by \HeTu-v2, and therefore the result is more accurate than the methods based on the bounding box. This greatly benefits 
in finding the corresponding host galaxies of radio sources and inferring their physical parameters.  
\item \textbf{Classifying radio galaxies.} If there are a number of input radio images and each input image is considered to have exactly one interesting source, it is troublesome to morphologically classify the source in each image manually through visual inspection when using traditional source finders. 
\HeTu-v2 can, however, be used as a classifier to automatically classify compact and extended radio galaxies into the five classes mentioned above.  Different classes of radio galaxies can be used as tracers of the cosmic environment to address key cosmological questions. It is an important step towards achieving the scientific goals of modern radio surveys.
\end{description}

The rest of this paper is organized as follows: Section \ref{sec:re-work} provides a brief overview of the 
deep learning based 
source detectors. Section \ref{sec:method_data} introduces the framework and network of \HeTu-v2, the dataset and its related methods including data generation, the data augmentation, the new morphological classification scheme, and the \HeTu-v2 model training. 
The results of the evaluation and inference of the \HeTu-v2 model are given in Section \ref{sec:result-disc}. Section \ref{sec:result-disc} also describes the application prospects of \HeTu-v2. Finally, Section \ref{sec:conlu} concludes the paper based on the current state of \HeTu-v2. 

\section{Source detectors based on deep learning methods} \label{sec:re-work}
Due to the advantages of deep learning methods 
in terms of their running speed and accuracy, many automated source detectors based on deep learning models have been developed in recent times mainly along two directions: either using a particular CNN architecture \citep{2018MNRAS.480.3749G,2019Galax...8....3L,2019MNRAS.484.2793V} or a object detection algorithm \citep{2019MNRAS.482.1211W,2021SciBu..66.2145L,arslan2020radio,burke2019deblending,farias2020mask,2022PASP..134f4503Z,2023A&C....4200682R}.

In CNN architecture based methods, COSMODEEP \citep{2018MNRAS.480.3749G} is a classifier
that first divide the input simulated radio image into a number of tiles, performs the source detection by classifying each tile
as containing or not a radio source, and then mosaicing all tiles with the classification results to form the final detection. Such a method only detects the parts of radio sources, and lacks classification capability as it only performs binary classification. ConvoSource \citep{2019Galax...8....3L} first reconstruct a solution map of the input real map ($e.g.$ simulated SKA image), and then determine the locations of predicted sources by post-processing using reconstruction threshold. DeepSource \citep{2019MNRAS.484.2793V} first generates a high signal-to-noise ratio image of the input simulated 
radio image, then finds the best threshold, detects blobs, and assigns one point source to each blob using Thresholded Blob Detection (TBD) technique. Both ConvoSource and DeepSource are only used in image processing, not to directly perform source detection. Furthermore, ConvoSource and DeepSource are not able to extract the high-level features which can be a problem in the case of more complicated or faint sources as they are implemented in the basics of CNNs. In summary, the described CNN architecture based methods only utilize ideal simulated images and lack of practical application to real complex observational data. 

To improve the CNN architectures, 
people have adopted the object detection algorithms that make use of Region Proposal Network \citep[RPN;][]{ren2015faster}. \claran\ \citep{2019MNRAS.482.1211W} is based on the Faster R-CNN \citep{ren2015faster} framework and uses 16 layers of the Visual Geometry Group Network (VGGNet-16) as the backbone, which can automatically locate and classify individually separate and extended components of radio sources on the 
input radio image and its counterpart infrared image. Generally, the deeper the network, the more accurate the predictions, but the VGGNet only supports up to 19 layers on account of vanishing gradient and exploding gradient during the increase in the number of layers. Moreover, \claran\ only uses the highest feature map, and therefore it loses local details of objects from the lower feature maps that causes lower accuracy for multi-sources detection. 

\HeTu-v1 \citep{2021SciBu..66.2145L} is based on the improved Faster R-CNN algorithm using the backbone network architecture of the Residual Network \citep[ResNet;][]{he2016deep} with Feature Pyramid Network \citep[FPN;][]{lin2017feature}.
ResNet adopts the back-propagation algorithm, which avoids the problems of vanishing gradient and exploding gradient, and the network can be deeper than VGGNet; 101 layers of ResNet (ResNet101) have been used in \HeTu-v1. The combination of ResNet and FPN through the bottom-up pathway, top-down pathway, and lateral connection design allows extraction of the multi-scale feature maps that not only ensures the advantages of small object detection but also improves the detection accuracy of complex objects. However, the methods based on Faster R-CNN have limited positioning accuracy due to the use of Region of Intersect (RoI) Pooling for double rounding. For this, high positioning accuracy methods have been developed by using Mask R-CNN \citep{he2017mask}, which improves the positioning accuracy by using the RoI Align method instead of using the RoI Polling method and adding a branch to generate pixel masks for objects \citep{he2017mask}.

\citet{arslan2020radio} used Mask R-CNN to implement a source detector on the same dataset defined in \claran, and showed that it can classify radio source morphologies and generate masks at the same time. Astro R-CNN \citep{burke2019deblending} applies Mask R-CNN to perform star and galaxy detection, classification, and deblending on simulated images. 
Mask Galaxy \citep{farias2020mask} also uses Mask R-CNN to perform detection, segmentation, and morphological classification for only one class of (Spiral or Elliptical) galaxy. However, Astro R-CNN and Mask Galaxy methods focus on processing optical images. 

Caesar-mrcnn \citep{2023A&C....4200682R} is a Mask R-CNN based source detector which is able to detect and classify compact, extended, spurious, and poorly imaged sources in radio continuum images. However, Caesar-mrcnn performs poorly on the extended sources due to the ambiguity of the segmentation masks obtained by the Mask R-CNN, and the morphological classification scheme of Caesar-mrcnn lacks astrophysical relevance for most extended sources. Furthermore, the current version of Caesar-mrcnn does not support building catalogs like the standard source finders, such as compact source Gaussian fitting and extended source properties estimation which are the essential features of a source detector.      


As shown above, the implementations of Mask R-CNN have been found to provide powerful detectors that are well-suited for executing source-finding and classification in radio surveys. However, it should be noted that they provide less accurate predicted masks in the boundary regions of objects \citep{tang2021look}, which is a crucial factor for the segmentation of high-resolution extended sources. 
Hence, the main contribution of this work is to explore a high-quality segmentation approach based on the improved Mask R-CNN that provides an appropriate baseline for future works.

\section{Data and method}\label{sec:method_data}
\subsection{Classification scheme and data} \label{sec:cla_data}
\subsubsection{Morphological classification scheme of \HeTu-v2} \label{sec:scheme}
The morphological classification of radio sources is a method of grouping sources based on their visual appearance. Studying the morphological classification of radio sources is able to 
reveal the physical properties of the sources, allowing us to achieve many critical scientific goals, such as the cosmological evolution of radio galaxy counts and luminosities \citep{van2012radio}, formation and evolution of radio galaxies and peculiar morphological radio sources ($e.g.$ Super Massive Black Holes (SMBH) and Active Galactic Nuclei (AGN) feedback \citep{fabian2012observational,padovani2017active}), and identifying low brightness diffuse radio galaxies ($e.g.$ nearby Star-Forming Galaxies \citep[SFG;][]{hopkins2004evolution}). 

The radio sky usually includes a variety of compact point-like and extended sources. A compact source (CS) is an unresolved source with a single non-diffuse emission component or is a point-like source, and an extended source is a resolved source that has at least one diffuse component. Traditionally, the extended sources have been classified as Fanaroff-Riley type I (FRI) and Fanaroff-Riley type II (FRII) sources using the Fanaroff-Riley scheme \citep{fanaroff1974morphology}. In this scheme,
the ratio of the distance between the regions of the highest brightness on opposite sides (radio jets) of the central core, to the total extent of the source up to the lowest brightness, is used to distinguish these two classes. Sources with ratios less than 0.5 are defined as FRI, and those with ratios greater than 0.5 are defined as FRII. 

Besides the typical FR sources, there is a morphological subclass called bent-tailed sources or Head-Tail (HT) sources that present more complex bent morphologies of the two radio jets caused by environmental factors or intrinsic properties \citep{2011ApJS..194...31P}. The bending of the two jets forms an opening angle of less than $180^\circ$ at the center of the source, giving the overall shape of the radio source like a ``L", ``C" or ``V" shape. Depending on the size of the opening angle, HT sources can be further classified into two sub-classes. 
If the opening angle of the sources is greater or equal to $90^\circ$, the source is classified 
as Wide-Angle Tailed (WAT); if the opening angle is less than $90^\circ$, the source is classified 
as Narrow-Angle Tailed (NAT) source. HT sources are generally found in galaxy clusters in the local Universe and can be used in observations as possible tracers of galaxy clusters at high redshifts ($z\geq 1$) \citep{blanton2003discovery,dehghan2014bent}. 

\begin{figure}[!ht]
\centering
\includegraphics[scale=0.35]{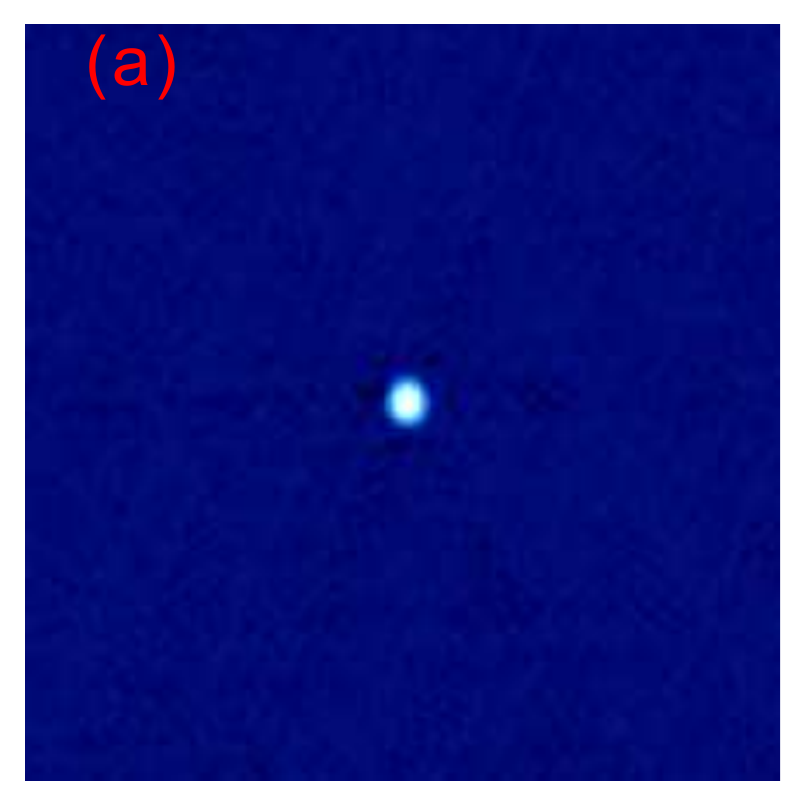}
\includegraphics[scale=0.35]{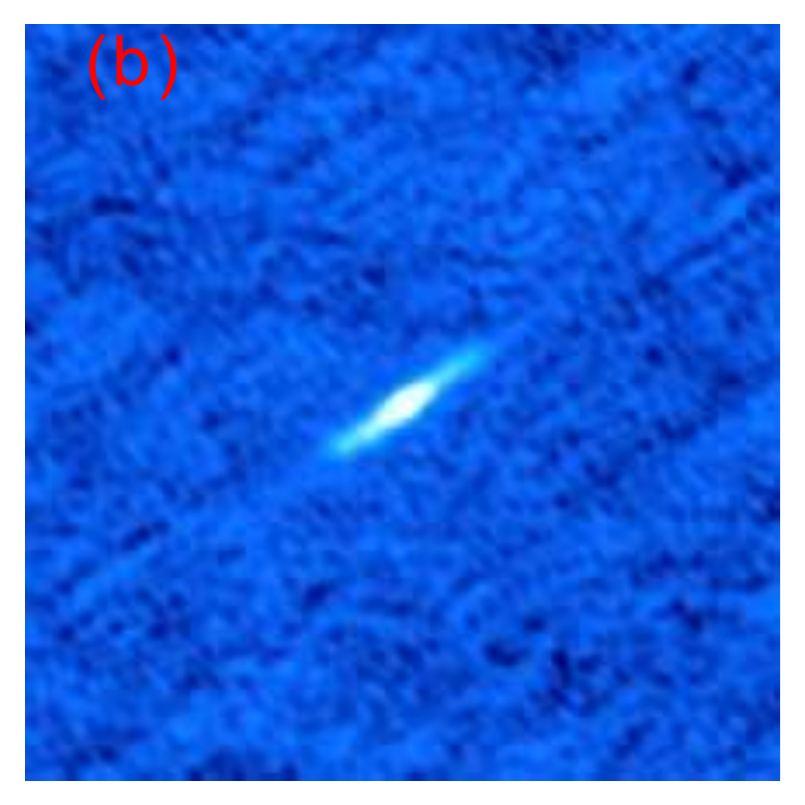}
\includegraphics[scale=0.35]{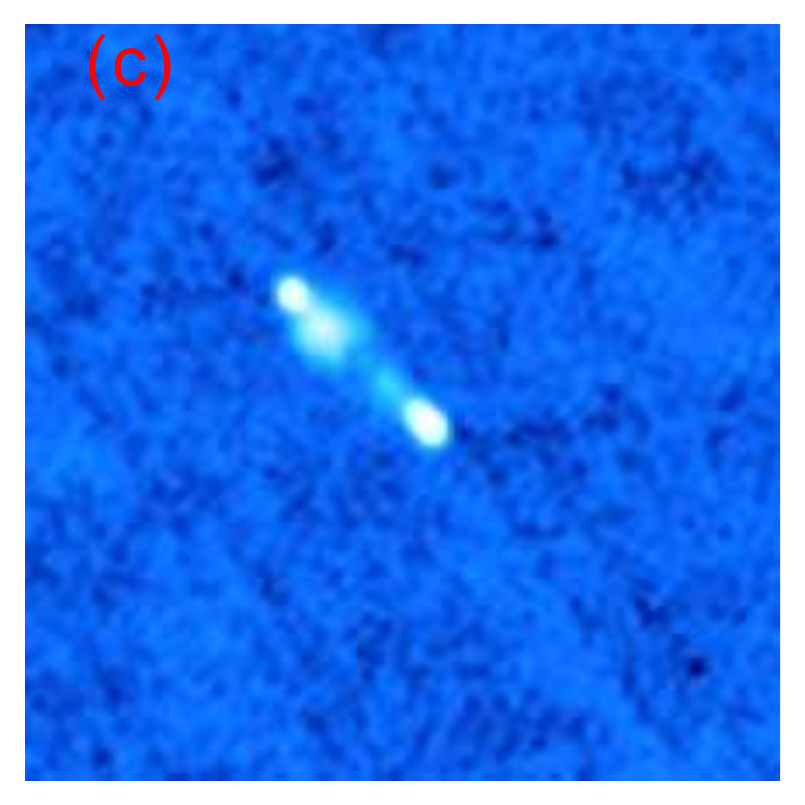}
\hspace{20mm}
\includegraphics[scale=0.35]{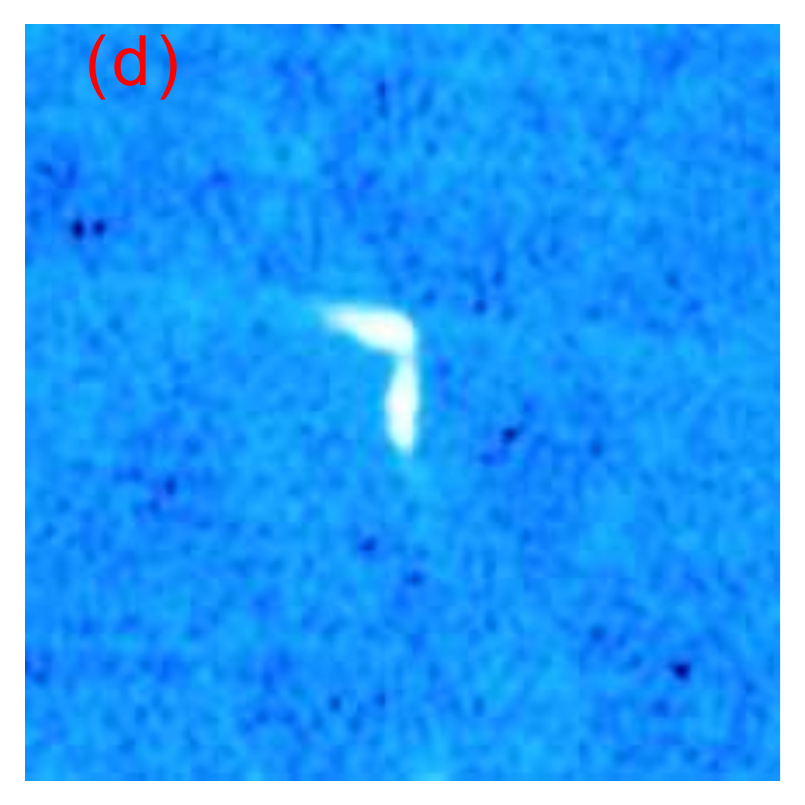}
\includegraphics[scale=0.35]{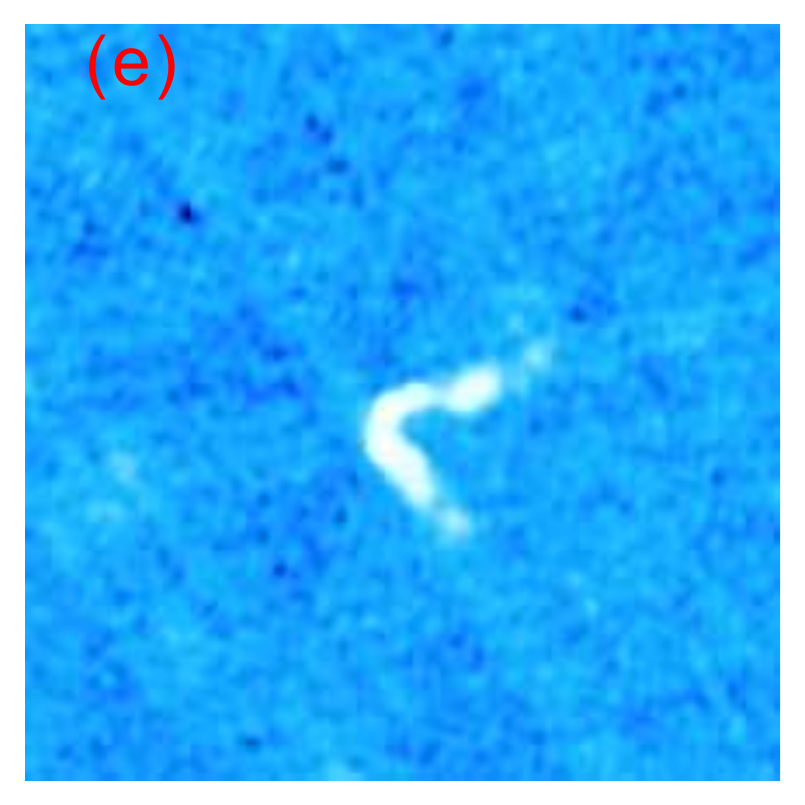}
\includegraphics[scale=0.35]{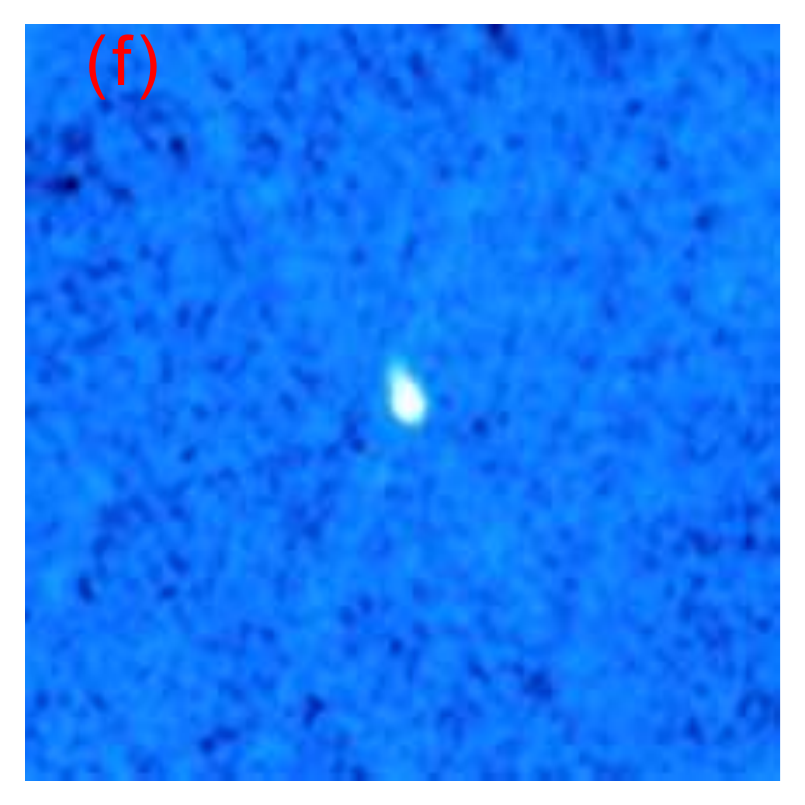}
\caption{Examples of morphological classes. From panels (a) to (f) are CS, FRI, FRII, WAT, NAT, and CJ sources. Panels (d) and (e) are combined as HT sources in this work.}
\label{fig:class-scheme}
\end{figure}

In addition to the above four classes, there is another category of sources with a prominent compact component and one-sided weak extended feature. We called it as Core-Jet (CJ) sources. CJ sources 
form a considerable group and their radio emission characteristics are obviously different from FR sources, so this category also needs to be taken into 
account. In summary, we refine the morphological classification scheme of \HeTu-v1 ($i.e.$ CS, FRI, FRII, and CJ) and consider five morphological classes in this work: CS, FRI, FRII, HT, and CJ. Example images of the five classes are shown in Fig. \ref{fig:class-scheme}.

\subsubsection{Data} \label{sec:data}
We used the images from the VLA FIRST radio survey \citep{becker1995first}, and the catalog 
from the latest data release version 14Dec17 (FIRST-14dec17) that contains 946,432 sources at a detection limit of 1 mJy~\citep{helfand2015last}. FIRST-14dec17 catalog was produced by the Astronomical Image Processing System \citep[AIPS;][]{1990apaa.conf..125G} based source-finding program \happy\ \citep{1997ApJ...475..479W} whose
key algorithm is similar to \aegean\ and \pybdsf. The FIRST images were 
downloaded and 
pre-processed, and then 
input to the trained model of \HeTu-v1 for preliminary source finding and classification. Next, the training and validation datasets of \HeTu-v2 were
generated based on preliminary results using the new morphological classification scheme (see Section \ref{sec:scheme}).
Finally, all processed images were
used for \HeTu-v2 inference experiments (see Section \ref{sec:inference}). This section describes data download, data pre-processing, conducting preliminary experiments, and building training and validation datasets. 

To speed up the image downloads, we have implemented a Message Passing Interface (MPI)-based program to batch download the FITS images in parallel using multiple processes. In this program, the FIRST-14dec17 catalog has been equally divided into multiple sub-catalogs by the total number of MPI processes used. Then, FITS images of these sub-catalogs are downloaded in parallel. In each download operation, the coordinates of each source in the sub-catalog have been extracted as the image center of the FITS image and downloaded this image from the FIRST cutout webpage \citep{haridas2005fits} using the Python {\bf{request.post}}
method. The size of each FITS image is $3.96\times3.96$ arcmins ($i.e.$ $132\times132$ pixels) in equinox of J2000. 
There are a total of 946,366 FITS images available for download, all are selected to flow preliminary experiments.

The image pre-processing is performed to generate the PNG images from the original FITS images as the PNG image is one of the essential formats for the input of the \HeTu\ model. In the pre-processing, all PNG images are generated using the log-max-min color scale and `cool' colormap. Following \cite{2003ASPC..295..489J} the log-max-min color scale is given in Eq. (\ref{eq:log}): 
\begin{equation}\label{eq:log}
{I_{\rm out}} = \frac{1}{3}\log_{10}\left( {\frac{{{I_{\rm in}} - {I_{\rm min }}}}{{{I_{\rm width}}}}1000 + 1} \right),
\end{equation}
where $I_{\rm width}=I_{\rm max}-I_{\rm min}$ represents the width pixel values in an image, $I_{\rm max}$ and $I_{\rm min}$ refer to the maximum and minimum pixel value in the image, respectively. $I_{\rm out}$ and $I_{\rm in}$ represent the final output and original input pixel values in the image, respectively.

The pre-processed PNG images were fed into the radio source detector \HeTu-v1 
for source finding and classification. 
The trained model of \HeTu-v1 is 
the same 
as that used in the final testing experiments from our previous work~\citep{2021SciBu..66.2145L}. This trained model can automatically identify and locate sources with the bounding box, label name, probability score, and other physical information from input images. These parameters are 
presented in the final catalog in CSV format. Finally, 
a total of 955,208 sources 
are detected by \HeTu-v1 model, including 852,271 CS sources, 11,134 FRI sources, 35,904 FRII sources, and 55,719 CJ sources. 
Based on this result, new training and validation datasets can be quickly built through visual inspection and automated ways.

Based on the new classification scheme in Section \ref{sec:scheme}, we utilized visual inspection to select images for \HeTu-v2's training and validation datasets from the above preliminary results generated by \HeTu-v1. This involved carefully selecting images and fixing any annotation errors, such as reclassifying false positives and adjusting predicted bounding box positions. Finally, a total of 5,118 corrected and annotated images have been selected through this process.
Among these, 3,172 images contained 3,601 sources for training, while 1,946 images contained 2,164 sources for validation. 
Details of the two datasets are shown in Table \ref{tab:train-val}.

\begin{table}[!ht]
\footnotesize
\centering
\begin{threeparttable}\caption{Number of sources in training and validation datasets}\label{tab:train-val}
\doublerulesep 0.1pt \tabcolsep 4.3pt 
\begin{tabular}{cccc}
\toprule
  Class & Training & Rotation-augmented training & Validation   \\\hline
  CS     & 1,673& 82,105 & 894\\
  FRI    & 326& 16,313 & 217\\
  FRII   & 738& 36,577 & 479\\
  HT     &406& 20,466 & 270\\
  CJ     &458& 23,132  & 304\\ \hline
  Total & 3,601 & 178,593& 2,164\\
\bottomrule
\end{tabular}
\end{threeparttable}
\end{table}

\subsection{Method}
Radio sources segmentation in deep learning is called instance segmentation which can be roughly divided into two-stage \citep[e.g. Mask R-CNN;][]{he2017mask}, one-stage~\citep[e.g. You Only Look At CoefficienTs, YOLCAT;][]{bolya2019yolact}, and query-based~\citep[e.g. Segmenting Objects by Learning Queries, SQLQ;][]{dong2021solq} methods. However, the quality of segmentation generated by using any of these methods alone is still unsatisfactory.

Recently, inspired by the powerful capability of Transformers \citep{carion2020end} in object detections, 
a hybrid method, Mask Transfine \citep{ke2022mask}, was proposed for high-quality instance segmentation. On top of the Mask R-CNN, the image regions have been decomposed to build a hierarchical quadtree, which is subsequently fed into a Transformer block to predict the final mask labels. Benefiting from the Transformer-based refining, 
significant performance improvements have been achieved on large-scale natural image datasets such as the Common Objects in COntext \citep[COCO;][]{lin2014microsoft}, Cityscapes \footnote{\url{https://www.cityscapes-dataset.com/}}, and BDD100K \footnote{\url{https://www.bdd100k.com/}}.

In 
the images obtained from radio observations, the boundary of the 
sources is more smooth and not as sharp as in the natural images. Thus, correctly segmenting radio sources is more challenging, especially 
in the presence of noise.
Inspired by Mask Transfiner, we 
embark on 
the ambiguity areas and exploit the long-distance dependencies of the Transformer block to refine the segmentation boundary obtained from the Mask R-CNN. As can be seen from below, significant improvements are achieved.

\subsubsection{Framework of \HeTu-v2}\label{sec:hetu2}
\HeTu-v2 
implements Mask Transfine \citep{ke2022mask} which is built on the Pytorch \citep{paszke2019pytorch} based object detection library \textsc{Detectron2} \footnote{\url{https://github.com/facebookresearch/detectron2}}. In the training phase \HeTu-v2 processes 
images in the PNG format only. In order to better adapt to astronomical tasks, in the inference phase \HeTu-v2 
accommodates multiple formats of astronomical data, 
including the PNG, \miriad\ files, and FITS image format
(see Section \ref{sec:inference} for details). \HeTu-v2 is well suited for solving the source finding, classification, and segmentation problems in radio continuum surveys, and the framework is able to form a single software package to perform all tasks.

 \begin{figure*}[!ht]
 \centering
 \includegraphics[scale=0.3]{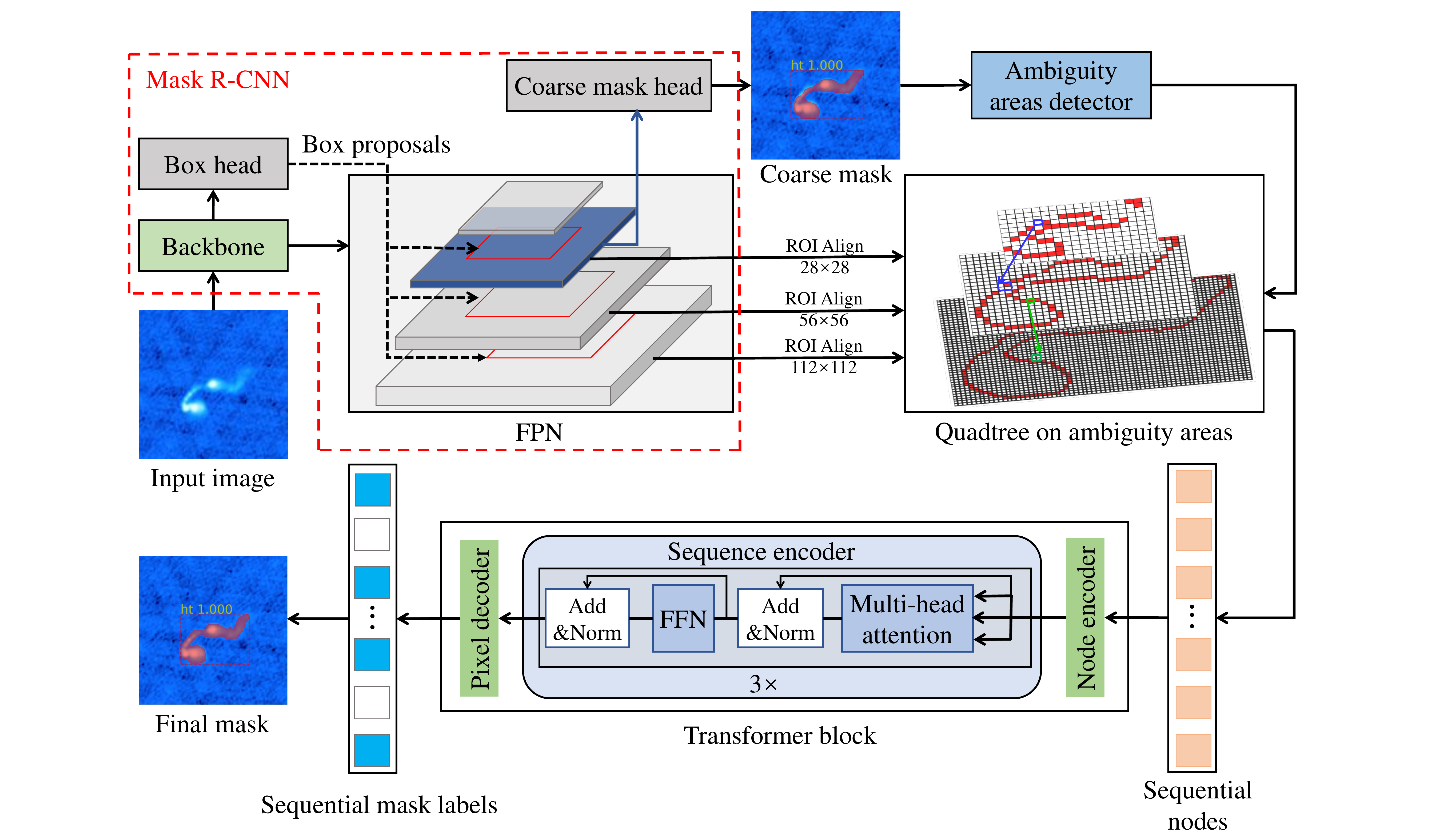}
 \caption{The framework of \textsc{HeTu}-v2 for radio sources segmentation and classification. Source: 
 Modified from \cite{ke2022mask}.}
 \label{fig:framework}
 \end{figure*}

\begin{figure}[!ht]
 \centering
 \includegraphics[scale=0.15]{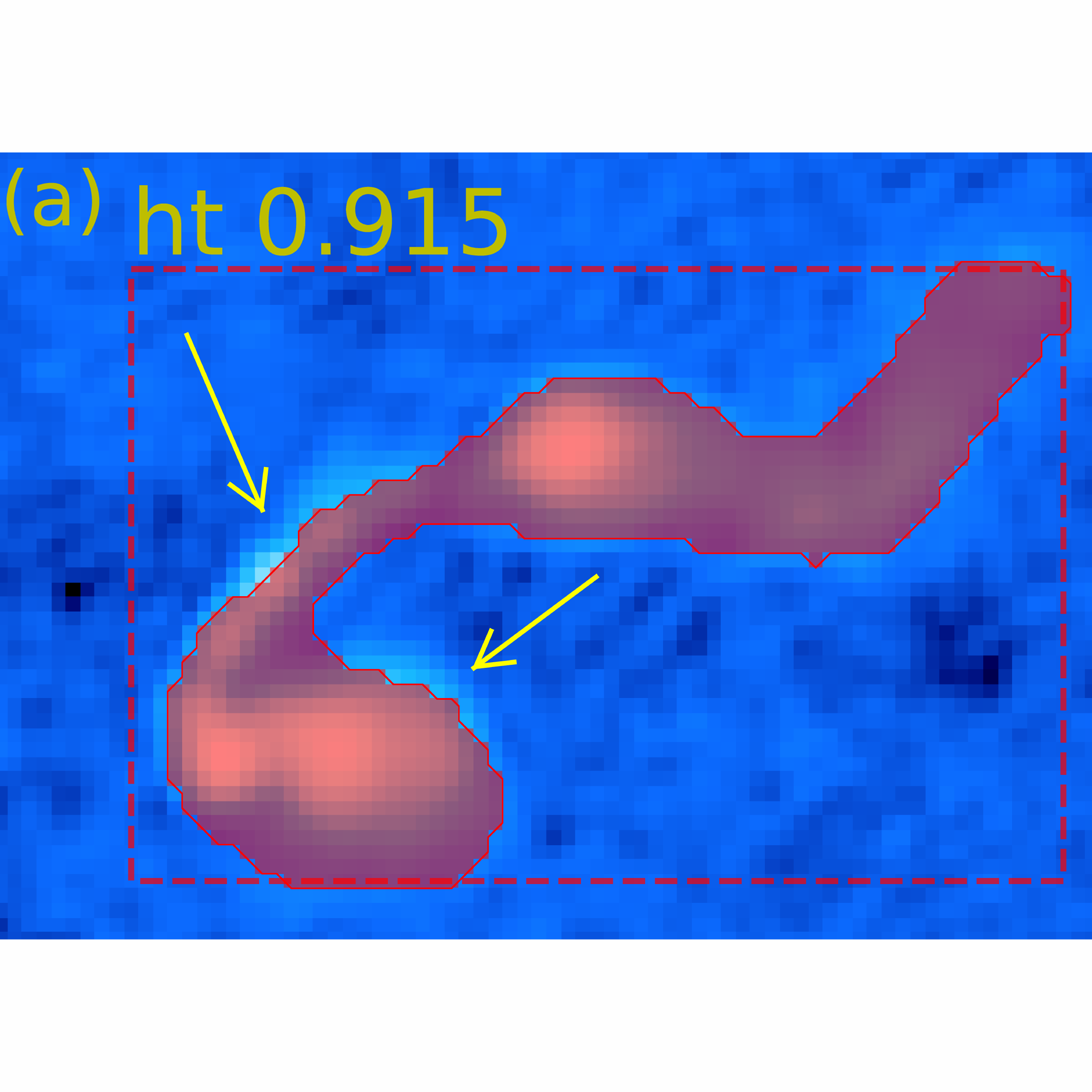}
 \hspace{0.1mm}
 \includegraphics[scale=0.15]{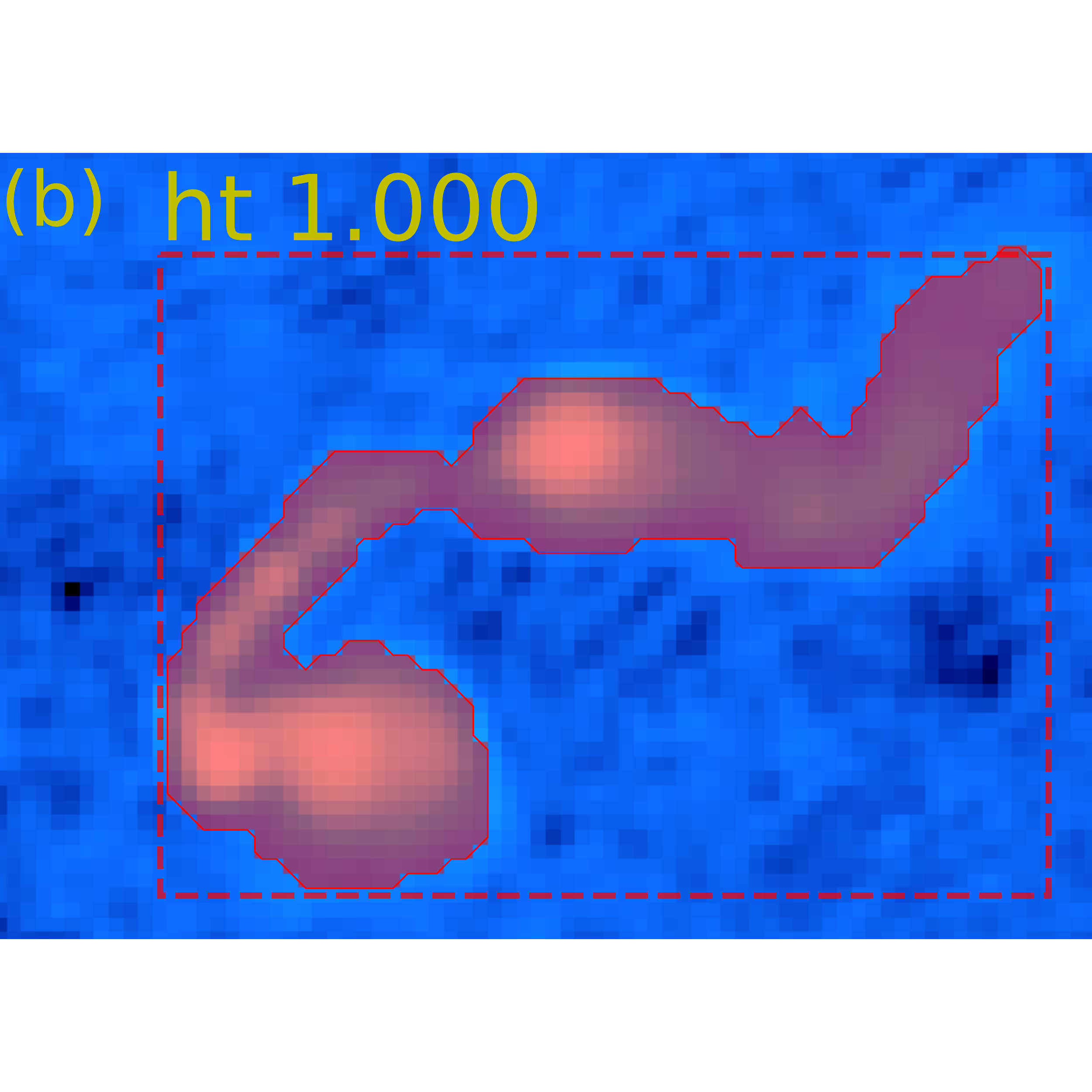}
 \vspace{0.1mm}
  \includegraphics[scale=0.15]{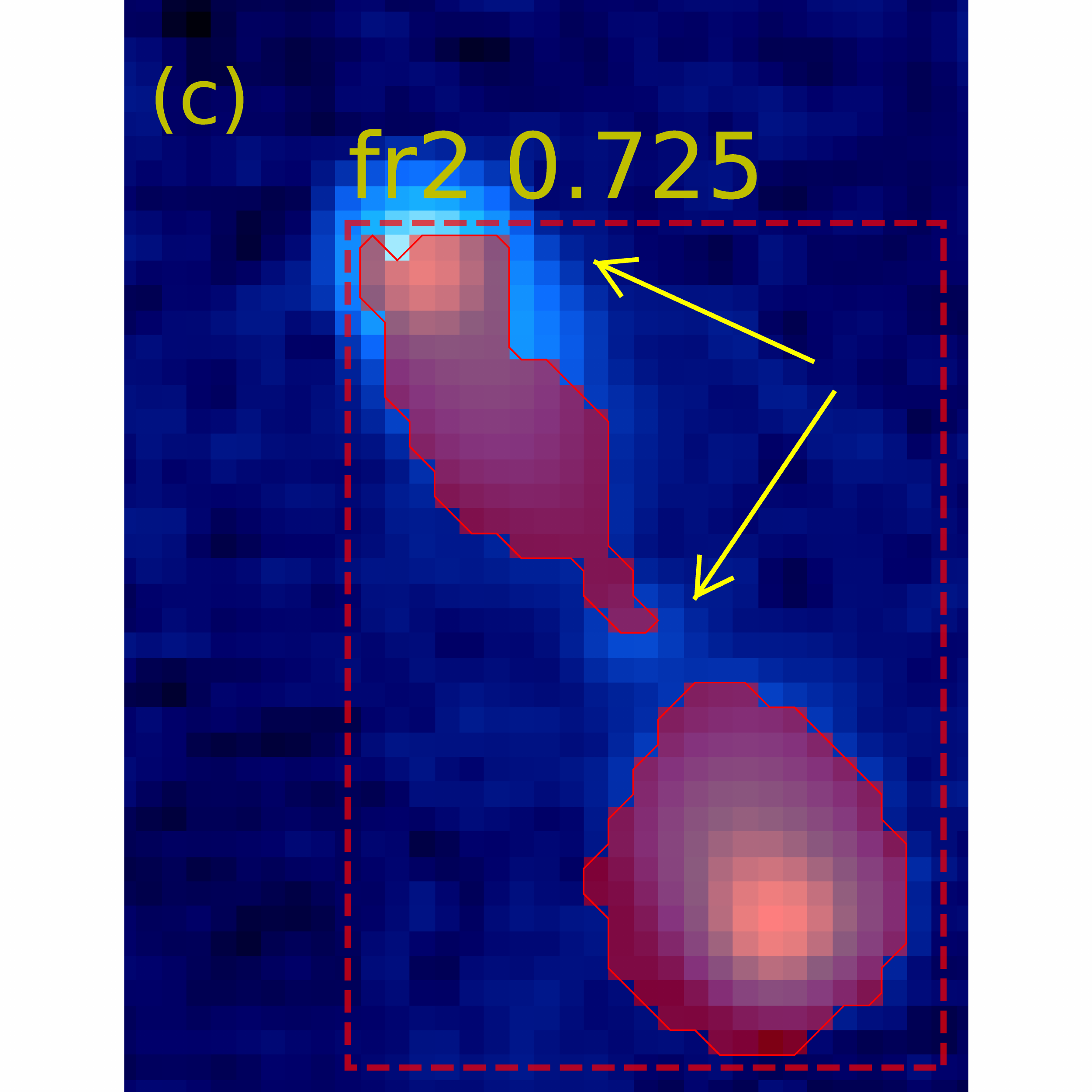}
 \hspace{0.1mm}
 \includegraphics[scale=0.15]{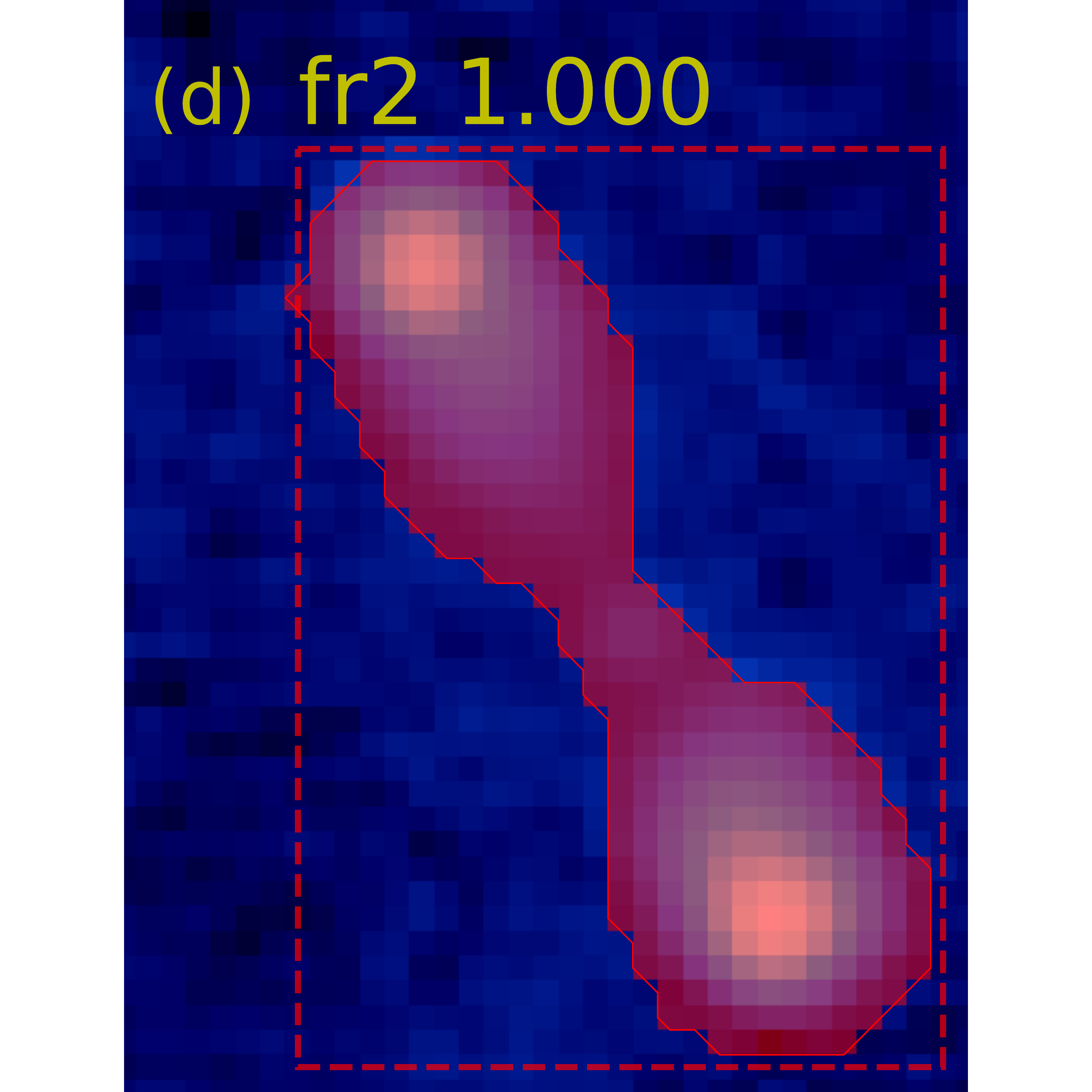}
 \caption{Examples of coarse and refined mask prediction. Panels (a) and (c) are the coarse mask predictions by Mask R-CNN, and panels (b) and (d) are the final refined mask predictions by \textsc{HeTu}-v2. Coarse masks appear at the boundaries of the source where the yellow arrows point in panels (a) and (c). The rectangular boxes and masks represent the identified sources. Each source is labeled with a class name (ht: HT, fr2: FRII) and a score between 0 and 1 on the top left of the bounding box.}
 \label{fig:coarse-fine-mask}
 \end{figure}

As shown in Fig. \ref{fig:framework}, the framework of \textsc{HeTu}-v2 mainly includes a two-stage detector Mask R-CNN, an ambiguity areas detector, and a Transformer block. Firstly, Mask R-CNN is used to generate an initial coarse mask and three levels RoI feature pyramid for a given input radio image. Examples of coarse mask predictions are seen in panels (a) and (c) in Fig. \ref{fig:coarse-fine-mask}, where pixel misclassification occurs at the edge of the bottom left lobe of the HT source and at the edge of the top left lobe of the FRII source respectively. Secondly, the coarse mask and RoI feature pyramid are fed to the ambiguity areas detector to detect the ambiguity areas and output a quadtree \citep{finkel1974quad} responding to the feature level. Finally, the sequence of quadtree nodes is fed into the Transformer block to reclassify the ambiguity pixels and output a fine mask, seen in panels (b) and (d) in Fig. \ref{fig:coarse-fine-mask}, showing that the ambiguity regions of the HT and FRII sources are successfully detected, and the final mask results generated by the Transformer block are corrected and refined compared to panels (a) and (c) respectively. Details of the designed framework's network are described below.

Mask R-CNN uses 101 layers deep Residual Network (ResNet101) \citep{he2016deep} combined with Feature Pyramid Network \citep[FPN;][]{lin2017feature} as backbone network (ResNet101-FPN) to extract multi-scale level feature maps from a given input image. These feature maps are represented as [$P2$, $P3$, $P4$, $P5$, $P6$] with pixels of size from large to small. They are fed into the RPN \citep{ren2015faster} with the input image to generate region proposals, and the RoI is extracted from the feature maps 
based on the region proposals using a pooling operation RoI Align \citep{he2017mask}, where [$P2$, $P3$, $P4$, $P5$] of the feature maps have been utilized. High-level (low-resolution) RoI features contain more contextual and semantic information, and low-level (high-resolution) RoI features resolve more local details. RoI has been used to generate the bounding box, segmentation, or mask and label by the object detection head ($i.e.$ box head) \citep{lin2017feature} and coarse mask head network, respectively. The coarse mask head network consists of a Fully Convolutional Networks (FCN) with four $3\times3$ convolutional layers. Due to the adoption of an advanced backbone network, Mask R-CNN achieves excellent performance. 
RoI Align is used in the later steps of the network, which greatly reduces the problem of feature mapping and averaging errors, so the quality of the bounding box produced by Mask R-CNN is higher. However, the segmentation masks are still very coarse due to the downsampling operation used in the feature pyramid and RoI Align steps \citep{zhang2021refinemask}. For example, in Fig. \ref{fig:framework}, the high-level RoI feature ($i.e.$ $28\times28$) has been selected to perform mask prediction, which 
reduces spatial resolution, so the generated segmented mask is called a coarse mask.

High-quality instance segmentation needs to consider the high-level semantics of the image as well as high-resolution deep features, but the large size of high-resolution features requires much more computational and memory 
resources \citep{PMID:32248092}. To refine the segmented mask with low resource usage, the loss information of multi-levels caused by downsampling operation 
are defined as ambiguity regions 
which are detected 
by an ambiguity areas detector. As shown in Fig. \ref{fig:framework}, the ambiguity areas detector follows the cascaded design 
to detect information loss nodes at three levels on the RoI feature pyramid: $28\times28$, $56\times56$, and $128\times128$. Taking the information loss point detected from the lowest level ($28\times28$) as the root node, the four sub-quadrant points are recursively expanded from top to bottom, and a multi-level quadtree is constructed. In the ambiguity areas detector, the lowest-level RoI feature and the predicted coarse mask are first fed to a lightweight FCN ($i.e.$ four $3\times3$ convolutions) to 
detect the root node of the quadtree. Each root node is decomposed into 4 sub-nodes in the previous layer, combined with the corresponding RoI feature, uses a single $1\times1$ convolutional layer to predict finer information loss nodes to keep the detection module lightweight until the mask of the highest layer is generated.

The Transformer block is composed of a node encoder, a sequence encoder, and a pixel decoder. The node encoder first enriches the feature representation of each point, such as the position encoding of the supplementary point and the local detail information of the neighboring points. Then, the node sequence as a Query into the sequence encoder to perform feature fusion and update between points on the input sequence. Finally, the pixel decoder predicts the instance label corresponding to each point. The node encoder encodes each quadtree node using four different pieces of information: fine-grained deep features, coarse mask, relative position encoding, and surrounding point information. Sequence encoder is a Transformer based module, each layer mainly consists of a multi-head self-attention module and a Fully Connected Feed-forward Network (FFN). Pixel decoder is a simple two-layer Multi-Layer Perceptron \citep[MLP;][]{kirillov2020pointrend} without a multi-head attention module. 

\begin{figure*}[!ht]
\centering
\includegraphics[scale=0.5]{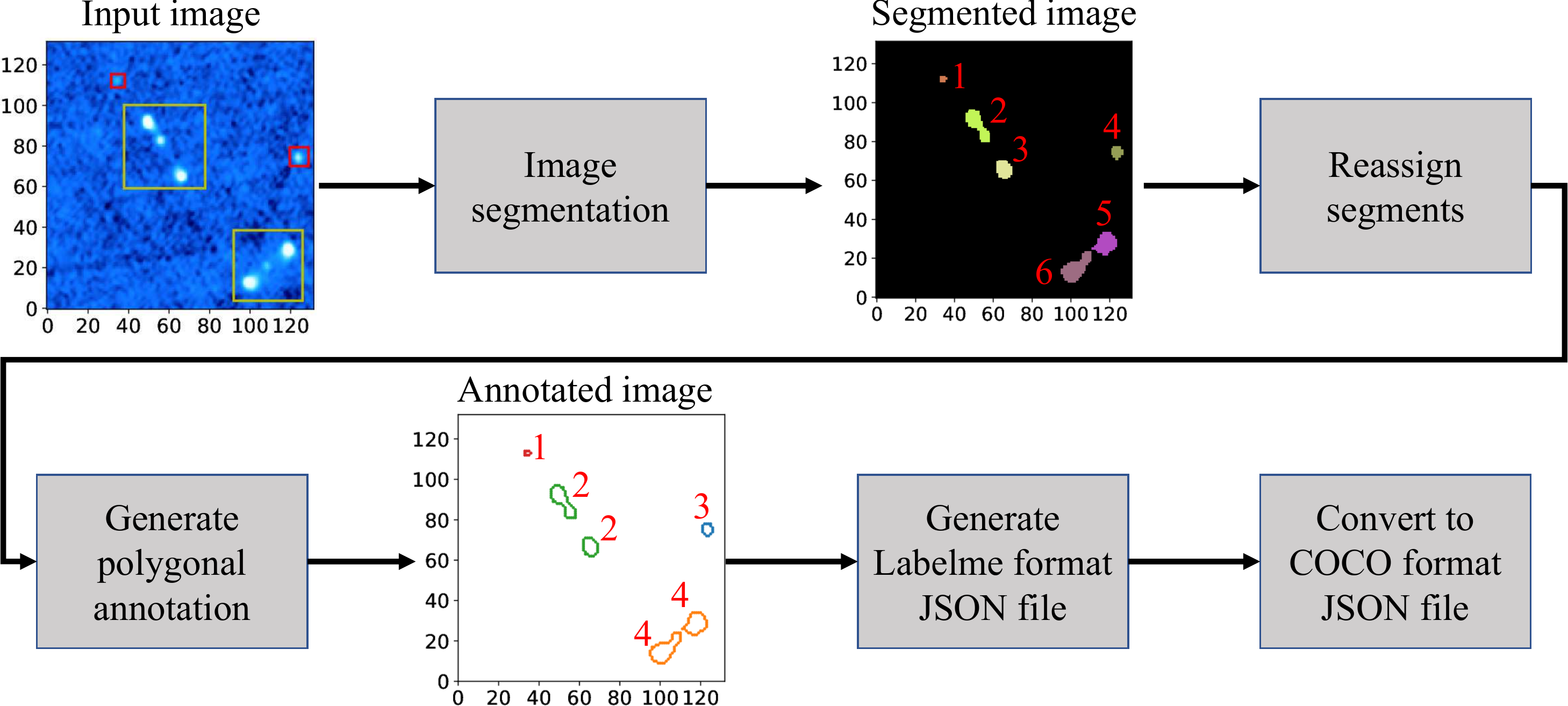}
\caption{The flow chart of the auto-generate polygonal annotation.}
\label{fig:auto-poly}
\end{figure*}

\subsubsection{Auto-generate polygonal annotation}\label{sec:auto-poly}
The supervised learning method has been adopted in the training procedure of \HeTu-v2, in which machines are trained and validated using well-labeled datasets. In these datasets, the ground truth of each object is annotated. In \HeTu-v1, the ground truth annotation information includes bounding boxes and labels, while in \HeTu-v2 
the polygonal annotation of object segmentation is added. Currently, the widely used annotation tools are \textsc{Labelme} \citep{russell2008labelme} and VGG Image Annotator (\textsc{VIA}) tools \citep{dutta2016vgg}. Both \textsc{Labelme} and \textsc{VIA} support image polygonal annotation through manual mode. The polygonal annotation generated in this mode is greatly affected by the visual. Due to the complex morphology of the radio source, the annotation task requires a lot of 
human effort and time. We have, therefore, developed a script to automatically generate polygonal annotation using the results from \textsc{HeTu}-v1 model. The input data of this script includes the above FIRST FITS images in Table \ref{tab:train-val}, and the bounding boxes and the labels name of the detected results of these images. 
The flow chart of this automated method is shown in Fig.~\ref{fig:auto-poly}, where
2 FRII sources and 2 CS sources are detected by \textsc{HeTu}-v1 and indicated with 
the yellow and red bounding boxes in the input image respectively. The flow chart is described as follows.

Firstly, image segmentation is performed to the input image 
using Python module {\bf{photutils.segmentation}}\footnote{\url{https://photutils.readthedocs.io/en/stable/segmentation.html}}. Image segmentation is the process of assigning a label to each pixel in an image, 
and pixels with the same label belong to the same source. Detected sources must have a minimum number of connected pixels, each greater than a specified threshold for the image. The threshold level is usually defined as a multiple of the background noise ($\sigma$ level) above the background. Here, we use threshold level of 4 $\sigma$ and find sources that have at least 5 connected pixels. In the segmented image in Fig.~\ref{fig:auto-poly}, for example, there are 6 radio sources that have been detected and labeled with numbers 1-6, and those segments are filled-in with 6 different colors. 
However, some extended sources, detected by {\bf{photutils.segmentation}} and presented as different segments in the segmented image, could belong to one true extended source. For example, there are two FRII sources given by (2,3) and (5,6) labels in Fig.~\ref{fig:auto-poly}; each FRII source in the segmented image are detected as 2 segments or sources in two different colors.

Secondly, the segments within the bounding box of the true source are automatically reassigned with a new and same label number. Then, the coordinates of the boundary points of the newly labeled segments are calculated and used to generate the polygonal annotation. For example, in Fig.~\ref{fig:auto-poly} the segments within the yellow color bounding box in the input image are labeled to the same number, and the polygonal points of the newly labeled sources are plotted with the same edge colors, see the annotated image. After steps of reassigning segments and generating polygonal annotation, the separate segments of each of the two FRII sources have been combined into one segment, namely 2 and 4 with their polygon points plotted using edge colors of green and orange, respectively.

Thirdly, the polygonal annotation of each image and its related information, including image name, image width and height, image data, label name, and polygonal point coordinates of each radio source, are written to a \textsc{Labelme} format JSON file. These files can be opened with \textsc{Labelme} with visualization to double-check the correctness of the labeled information.

Finally, as \textsc{HeTu}-v2 uses the COCO format \citep{lin2014microsoft} dataset in training and validation experiments, all of \textsc{Labelme} format JSON files for the training and validation sets are converted to 
COCO format JSON files respectively.

\subsubsection{Image data augmentation}\label{sec:data-aug}
As the motive power of deep learning, data are crucial for model training. Sufficient training data can not only alleviate the over-fitting problem of the model during training but also further expand the parameter search space and help the model to further optimize towards the global optimal solution. For training of large models, the original amount and style of samples in Table \ref{tab:train-val} are inadequate. Data augmentation is an effective technique to increase training samples and makes samples as diverse as possible, which has become a necessary part of the successful application of deep learning models on image data \citep{2022arXiv220408610Y}. 

To reduce the over-fitting of \textsc{HeTu}-v2 network and to improve the generalization ability and robustness of the network, an augmentation method combining offline and online is adopted in training. For online mode, we directly use the image resize and flip methods 
in \textsc{Detetron2}. The input images are first resized to the shorter edge range from 640 to 800 pixels with step size of 32 pixels, 
and then each image performed a horizontal flip. 

\begin{figure}[!ht]
\centering
\includegraphics[scale=0.15]{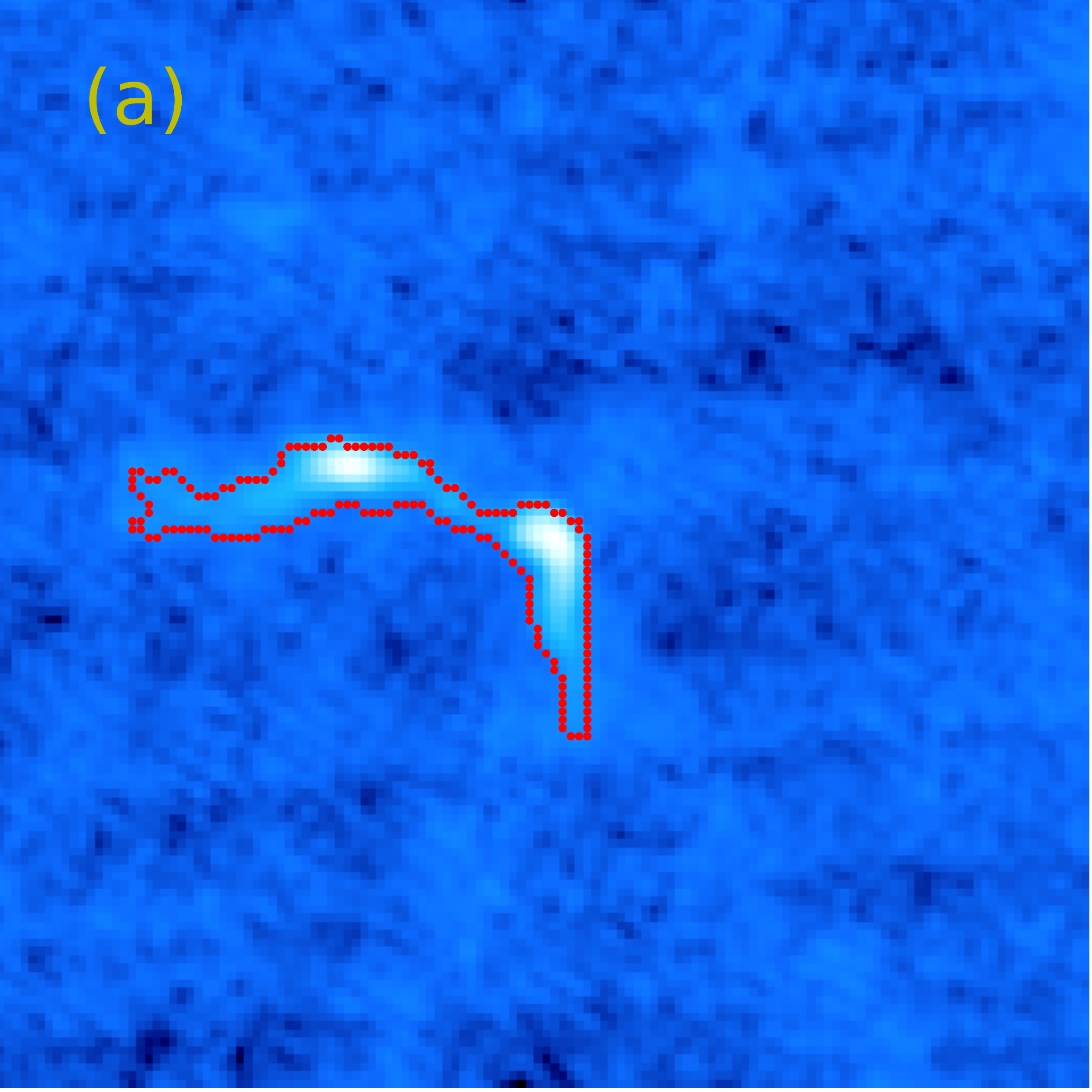}
\hspace{0.1mm}
\vspace{0.1mm}
\includegraphics[scale=0.15]{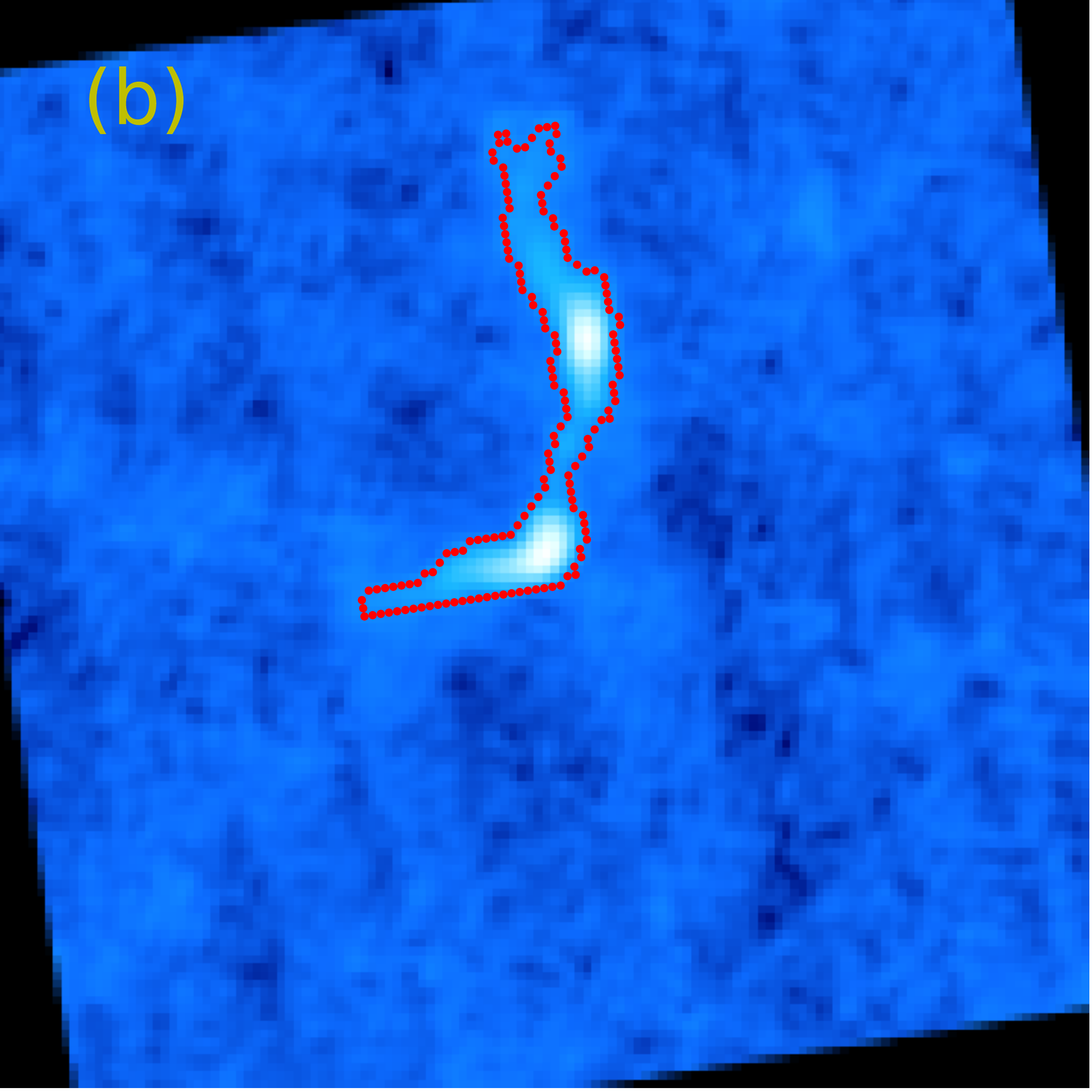}
\hspace{0.1mm}
\vspace{0.1mm}
\includegraphics[scale=0.15]{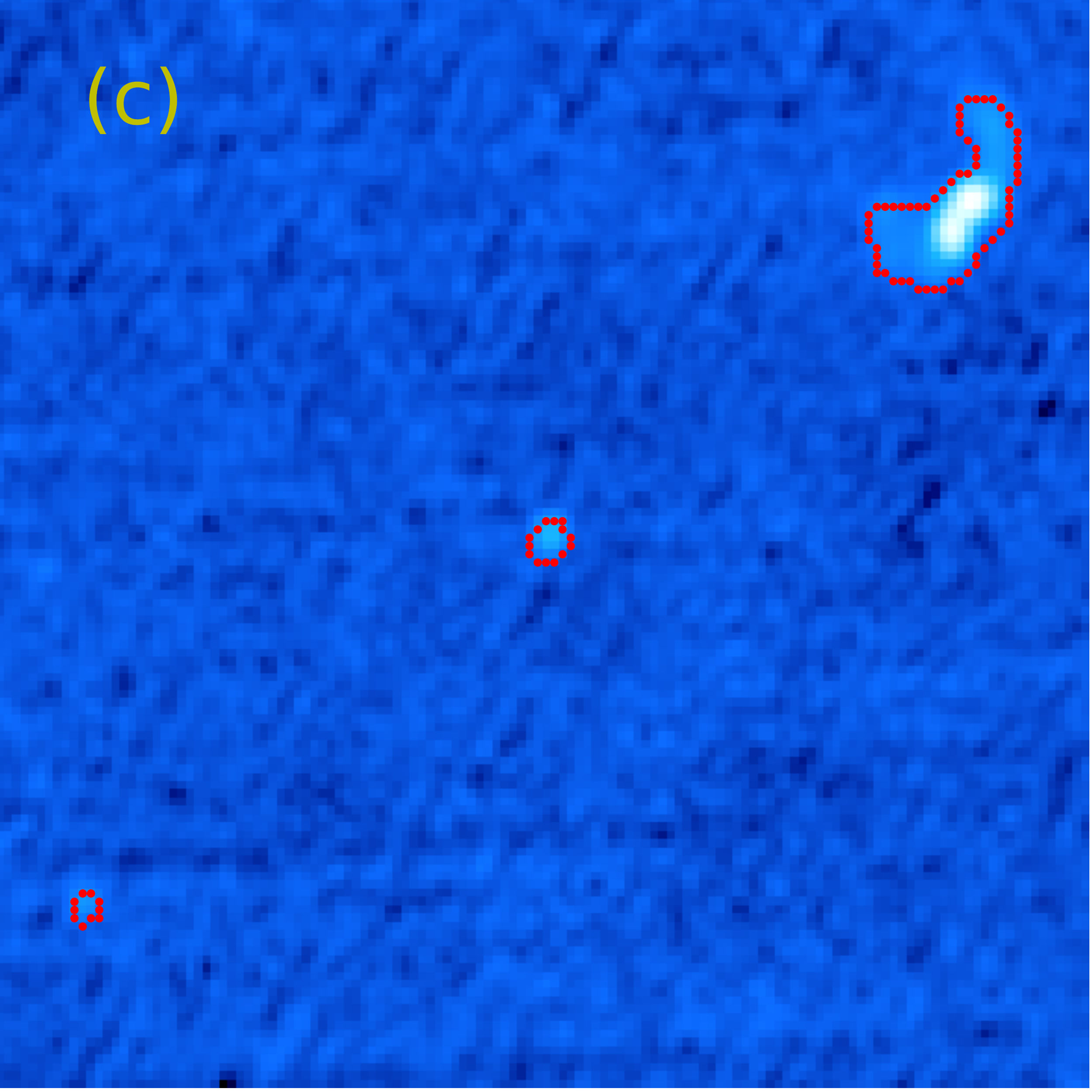}
\hspace{0.1mm}
\includegraphics[scale=0.15]{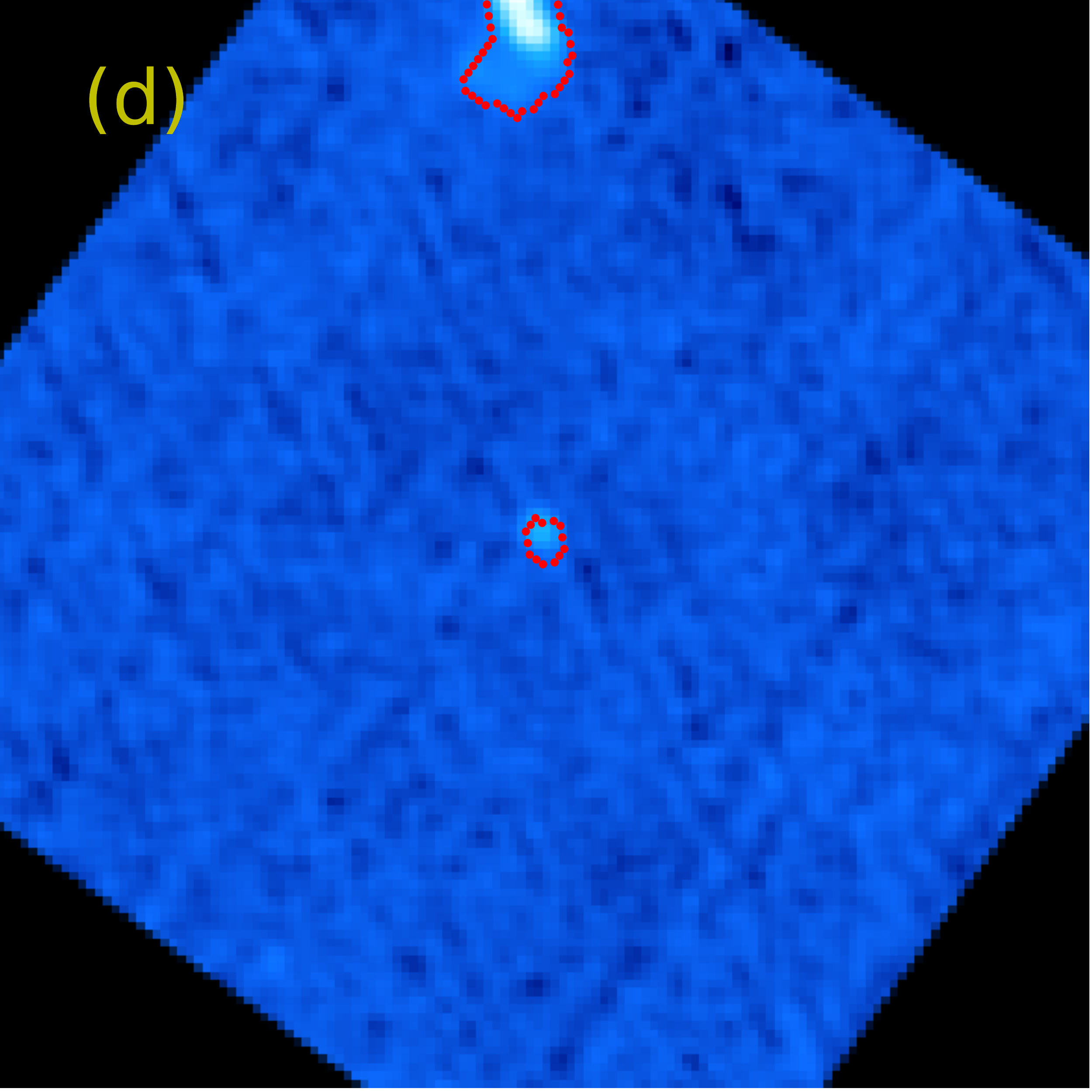}
\caption{Examples of random augmentation in the training set, the red polygon points in these examples are the polygon annotation of radio source instances. Panels (a) and (c) are the original images, and panels (b) and (d) are their augmented images, respectively. In panel (b), after the image and instance are rotated, the entire instance is still within the pixel size of the image, and such augmented images can be used for model training. In 
panel (d), two instances are missing or incomplete after rotation, and such augmented images must be removed.}
\label{fig:aug-example}
\end{figure}

The number of training sets after online augmentation is still unable to reach the requirements of \textsc{HeTu}-v2 model training, so we consider adding image random-rotation methods 
before online augmentation. When the image pixel size is kept unchanged, the radio source instance in the image after rotation may overflow beyond the image pixel size, see Fig. \ref{fig:aug-example}. It should be noted that the blank regions in the rotated images, such as those in panels (b) and (d) of Fig. \ref{fig:aug-example}, do not impact the model training. These overflowing images need to be removed manually, so the rotation method is more suitable for offline mode. Our offline augmentation method is implemented 
using the Python package \textsc{imgaug}\footnote{\url{https://github.com/aleju/imgaug}}, 
in which each image is randomly selected 
for rotation from $-$180 to 180 degrees,
and 50 selections are conducted. The rotation-augmented dataset has 159,303 images and 178,593 sources. More details can be seen in Table \ref{tab:train-val}.      

\subsection{Model training}\label{sec:model-train}
All experiments including training, evaluation, and inference are conducted in an Intel Xeon Silver 4314 CPU server running at 2.4 GHz (64 cores, 256 GB RAM memory) equipped with three NVIDIA GPUs. 
Two of the GPUs are NVIDIA RTX A4000, 
and each has 3,072 CUDA cores and 16~GB global memory; the remaining one is NVIDIA A40, and has 5,376 CUDA cores and 48~GB global memory. 
We used all the three GPUs for training.

The goal of model training is to find the model parameters through the sample data to minimize the loss function. The loss function of the \HeTu-v2 model consists of 
multiple losses, and the total loss is calculated by \citep{ke2022mask}: 
\begin{equation}\label{eq:loss}
L_{\rm total} = {\lambda _1}{L_{\rm detection}} + {\lambda _2}{L_{\rm coarse}} + {\lambda _3}{L_{\rm refine}} + {\lambda _4}{L_{\rm ambiguity}},
\end{equation}
Where $L_{\rm detect}$ is the detection loss including the bounding box and label losses from the Mask R-CNN, $L_{\rm coarse}$ denotes the loss of the initial coarse mask predicted by the Mask R-CNN, $L_{\rm refine}$ is the refinement with mean absolute error ($L1$) loss of labels for ambiguity nodes from the Transformer block, $L_{\rm ambiguity}$ is a Binary Cross Entropy loss of ambiguity regions detection by an ambiguity areas detector, and [$\lambda_{1}$, $\lambda_{2}$, $\lambda_{3}$, $\lambda_{4}$] are hyperparameter weights [1.0, 1.0, 1.0, 0.5].

\HeTu-v2 model is trained using the $3\times$ training schedule \citep{he2019rethinking}. With 9 images per batch on 159,303 radio images, the learning rate decreases by a factor of 0.1 at 210,000 and 250,000 iterations, and finally terminates at 270,000 iterations, where an iteration refers to the process of training on a batch of images. To optimize training and speed up model convergence, Stochastic Gradient Descent (SGD) with a momentum of 0.9 and 1,000 iteration linear warm-up with an initial learning rate of 0.01 and weight decay of 0.0001 is used. In addition, distributed data-parallel strategy has been employed to reduce training time. The model is replicated on each GPU device, and the total batch data are equally divided into 3 different GPUs for calculation, each GPU worker loads its own data from disk. During training, each GPU process communicates with other processes through the ``All-Reduce'' method and exchanges their gradients to obtain the average gradient of all processes. Each GPU process uses the averaged gradient to update its own parameters, making each GPU parameter synchronize.

During training, the training and validation metrics ($e.g.$ time, multi-task losses, mask, and class accuracy) have been recorded into a JSON file every 20 iterations, and the trained model file is saved per 10,000 iterations. The training has taken $\sim67.5$ hours to finish. The trained \HeTu-v2 model can be performed on radio images with size of $132\times132$ pixels in less than $\sim200$ milliseconds per sample using one GPU. 

\begin{figure}[!ht]
\centering
\includegraphics[scale=0.33]{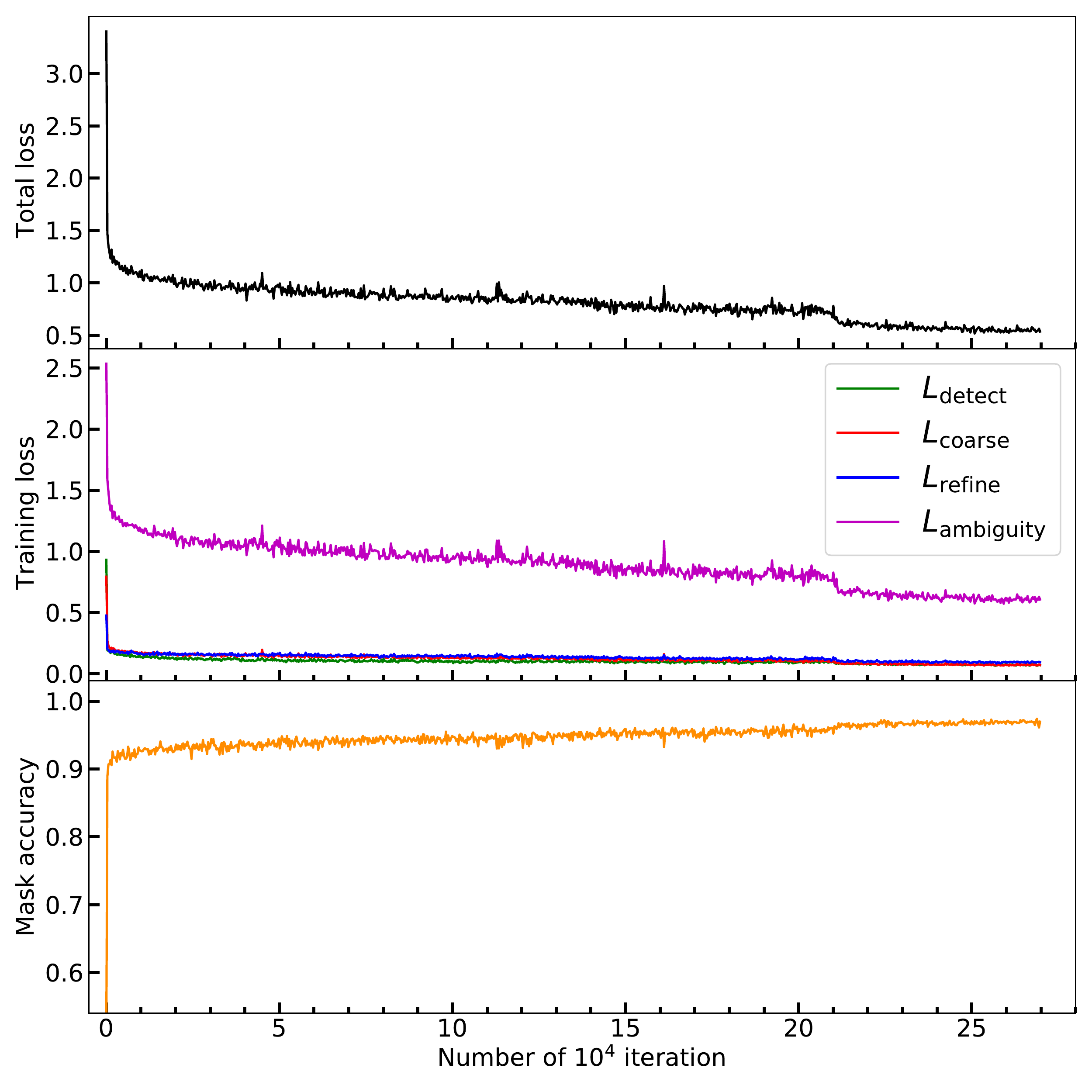}
\caption{Training loss and mask accuracy curves with the number of iterations. The top panel shows the total loss, the middle panel shows the loss of the tasks, and the bottom panel shows the mask accuracy.}
\label{fig:train-loss}
\end{figure}

Fig. \ref{fig:train-loss} shows the training loss and mask accuracy curves. As the number of iterations increases, the loss of the model on the training dataset continues to decrease. The first 10,000 iterations show a significant drop in the total loss due to a large difference between the parameters of the model and the optimal parameters. By continuously optimizing the model parameters on the validation set, the model will eventually approach the optimal value, resulting in improved performance in locating and classifying radio sources. The total loss steadily decreases from 10,000 to 209,999 iterations as the model undergoes local optimization. 
At the 210,000th iteration, there appears to be a sudden drop in the total loss, indicating that the model had deviated from its previous locally optimized state as the learning rate decreased tenfold from 0.01 to 0.001 at this iteration. Finally, the total loss function begins to converge at the 229,999th iteration, and the mask accuracy is the highest for the first time 
at this iteration. After this iteration, the total loss flattens out and stabilizes at around $L_{\rm total}=0.53$, with $L_{\rm detection}=0.07$, $L_{\rm coarse}=0.07$, $L_{\rm refine}=0.09$, and $L_{\rm ambiguity}=$0.6. Therefore, the output model of the 229,999th iteration is used as the final trained model of \HeTu-v2.

\section{Results and discussion}\label{sec:result-disc}

\subsection{Model performance and evaluation} \label{sec:model-performance}
To quantitatively evaluate the \HeTu-v2 network performance, we have applied the trained \HeTu-v2 model to the validation dataset (see Table \ref{tab:train-val}) and calculated the evaluation metrics. Since the validation dataset is in COCO format, the Microsoft COCO Evaluation Metrics \citep{lin2014microsoft} are used to quantitatively evaluate the performance of \HeTu-v2 network, including Average Precisions ($AP$s) and Average Recalls ($AR$s), but $AR$s are not commonly used. 
Compared with the Pascal VOC metrics \citep{everingham2010pascal}, Microsoft COCO Evaluation Metrics are more comprehensive.

\begin{table*}[!ht]
\footnotesize
\centering
\begin{threeparttable}\caption{Summary of $AP$ results on validation dataset using different methods. $AP_{\rm @50:5:95}$ represents the $AP$ at $IoU$ threshold values ranging from 0.5 to 0.95. Meanwhile, $AP_{\rm 50}$ and $AP_{75}$ refer to the $AP$ at $IoU$ thresholds of 0.5 and 0.75, respectively. Additionally, $AP_{\rm small}$, $AP_{\rm medium}$, and $AP_{\rm large}$ denote the $AP$ values for sources with small (area $< 16^2$ pixels), medium ($16^2$ pixels $<$ area $<32^2$ pixels), and large (area $> 32^2$ pixels) sizes, respectively, at $IoU$ threshold values ranging from 0.5 to 0.95. $mAP$ is the mean $AP$ across all classes.} \label{tab:APs}
\doublerulesep 0.2pt \tabcolsep 10pt 
\begin{tabular}{c|ccccccc}
\toprule
 Methods & Class & $AP_{\rm @50:5:95}$ & $AP_{\rm 50}$ & $AP_{75}$ & $AP_{\rm small}$& $AP_{\rm medium}$  & $AP_{\rm large}$ \\\hline
 \HeTu-v2 & CS     & 75.8\%& 97.3\%& 91.1\% & 75.8\% & -& -\\
  &FRI    & 75.4\%& 97.4\%  & 91.3\% & 75.6\% & 79.0\% & 79.2\%\\
  &FRII   & 82.0\%& 99.9\% & 95.9\%& 81.9\% &84.1\% & 78.2\%\\
  &HT     &73.4\%& 98.4\% & 88.1\% & 68.2\% & 77.5\% & 72.7\%\\
  & CJ     &82.5\%&  98.9\%  & 95.3\% &  82.7\% & 81.4\% & -\\\cline{2-8}
  & $\bm{mAP}$ & \bf{77.8\%} & \bf{98.4\%} & \bf{92.3\%} &\bf{76.8\%} & \bf{80.5\%} & \bf{76.7 \%}\\\hline
   \HeTu-v1   & CS     & 42.2\%& 89.1\%& 33.3\%& 42.2\%  & -& -\\
  &FRI    &58.7\%& 94.7\%& 72.4\%& 60.4\%& 59.8\%& 45.4\%\\
 &FRII&79.4\%& 99.9\%& 96.6\%& 73.4\%& 81.9\% & 85.4\%\\
  &HT     &68.6\%& 99.2\%& 85.4\%& 62.3\%& 70.2\%& 70.2\%\\
  & CJ     & 62.1\%& 98.9\%& 73.7\%& 62.0\%& 67.9\%& -\\\cline{2-8}
  & $\bm{mAP}$ & \bf{62.2\%}& \bf{96.4\%}& \bf{72.3\%}& \bf{60.1\%}& \bf{69.9\%}& \bf{67.0\%}\\\hline
  Mask R-CNN   & CS     &57.6\%& 93.7\%& 65.1\%& 57.6\% & -& -\\
  &FRI    &67.1\%&98.2\%& 82.0\%& 69.8\%&65.0\%& 45.8\%\\
  &FRII&68.2\%& 99.7\%&86.0\%& 74.4\%& 68.3\% & 54.9\%\\
  &HT     &65.6\%& 98.5\%& 82.8\%& 67.2\%& 68.0\%& 58.0\%\\
  & CJ     & 73.1\%& 98.8\%& 93.1\%& 73.3\% & 73.5\%& - \\\cline{2-8}
  & $\bm{mAP}$ & \bf{66.3\%}& \bf{97.8\%} & \bf{81.8\%}& \bf{68.5\%}& \bf{68.7\%}& \bf{52.9\%}\\
\bottomrule
\end{tabular}
\end{threeparttable}
\end{table*}

To understand $AP$, we first introduce Precision ($P$) and Recall ($R$), which represent the reliability (accuracy) and completeness of the detection results generated by the \HeTu-v2 model on the validation set. $P$ is the True Positive ($TP$) prediction rate of all positive predictions for a class: 
\begin{equation}\label{eq:Precision}
P = \frac{TP}{TP+FP}=\frac{TP}{\#predictions},
\end{equation}
where $TP$ denotes the number of positive class sources that are correctly predicted to be in the positive class and whose Intersection over Union ($IoU$) value exceeds the specified threshold ($IoU_{\rm threshold}\in 0.0\;...\;1.0$), and $FP$ represents the number of negative class sources that are incorrectly predicted to be in the positive class or whose $IoU$ value is below the $IoU_{\rm threshold}$. The $IoU$ is given by the ratio of the area of intersection ($\cap$) between the predicted mask ($M_{p}$) and ground truth mask ($M_{gt}$) to the area of their union ($\cup$): 
\begin{equation}\label{eq:IOU}
IoU = \frac{{area({M_p}\; \cap \;{M_{gt}})}}{{area({M_p}\; \cup \;{M_{gt}})}}.
\end{equation}

$R$ is the $TP$ prediction rate of all ground truth sources in a class: 
\begin{equation}\label{eq:Recall}
R = \frac{TP}{TP+FN}=\frac{TP}{\#ground\,\,truths},
\end{equation}
where False Negative ($FN$) is the number of positive class sources but are incorrectly predicted to be in the negative class.

The area under the $P-R$ curve of a given class is called Average Precision ($AP$). A higher $AP$ indicates better detection performance for the network. In practice, $AP$ represents the precision averaged across all recall values that are equally spaced between 0.0 and 1.0. These equally spaced recall values are typically achieved by adjusting the $IoU$ threshold level to obtain the desired recall. Here, the $AP$ is calculated as the mean of precision values at a set of 101 equally spaced recall levels $R \in [0,0.01,...,1.0]$:
\begin{equation}\label{eq:AP}
AP = \frac{1}{{101}}\sum\limits_{R \in [0,0.01,...,1.0]} {P(R)},
\end{equation}
where $P(R)$ is the maximum precision in interval $\Delta R$.

In the Microsoft COCO Evaluation Metrics, the performance of the network in distinguishing objects is refined by varying the $IoU_{\rm threshold}$ from 0.5 to 0.95 with a step size of 0.05, resulting in 10 sets of $P-R$ curves. The 10 $AP$s calculated from the 10 sets of $P-R$ curves are averaged and expressed as $AP_{\rm @50:5:95}$, which is the primary challenge metric.
When the $IoU_{\rm threshold}$ is fixed, such as $IoU_{\rm threshold}=0.5$ and $IoU_{\rm threshold}=0.75$, the metric is written as $AP_{\rm 50}$ ($i.e.$ Pascal VOC metric) and $AP_{\rm 75}$, respectively. In addition, the $AP$ of sources with different pixel areas is also calculated to evaluate the network's detection performance on objects of varying sizes. Regarding radio source detection, sources with an area less than $16^2$ pixels are represented as $AP_{\rm small}$, those with an area larger than $16^2$ pixels but less than $32^2$ pixels are represented as $AP_{\rm medium}$, and those with an area larger than $32^2$ pixels are represented as $AP_{\rm large}$. These metrics are computed using the same $IoU$ threshold range as that of $AP_{\rm @50:5:95}$. The definitions of $AP_{\rm small}$, $AP_{\rm medium}$, and $AP_{\rm large}$ used for radio source detection differ from those defined in the original COCO Evaluation Metrics because the pixel area of radio sources is typically much smaller. In object detection, the mean $AP$ ($mAP$) is typically used as the final evaluation metric. This metric is defined as the average of the $AP$ metric across all classes in a model, and is given by:

\begin{equation}\label{eq:mAP}
mAP = \frac{{\sum\limits_{i = 1}^{{N_{\rm c}}} {A{P_i}} }}{{{N_{\rm{c}}}}},
\end{equation}
where ${N_{\rm{c}}}$ is the number of classes.
 
In Table \ref{tab:APs}, various $AP$ metrics are to evaluate the performance of \HeTu-v2, \HeTu-v1, and Mask R-CNN networks. These metrics show the model's accuracy across different $IoU$ thresholds and object scales. The three methods use the same training and validation datasets. \HeTu-v2 obtains the highest mean $AP_{\rm @50:5:95}$ of 77.8\% and outperforms \HeTu-v1 and Mask R-CNN by 15.6 points and 11.3 points, respectively. 
Other $mAP$s of \HeTu-v2 also have varying degrees of improvement compared to \HeTu-v1 and Mask R-CNN.
The significant advancements reveal the effectiveness of the framework upgrade from \HeTu-v1 to \HeTu-v2.
This also indicates that \HeTu-v2 network can effectively solve the loss of spatial information problem in Mask R-CNN, especially for large-scale extended sources that the mean $AP_{\rm large}$ increased by 23.8 points. Furthermore, the $AP_{\rm 50}$s of all classes are no less than 95\%, particularly the $AP_{\rm 50}$ of FRII is 99.9\% $i.e.$ close to 100\%. \HeTu-v2 model has the highest accuracy on a medium scale of sources with mean $AP_{\rm medium}$ of 80.5\% and improved by 10.6 points and 11.8 points than \HeTu-v1 and Mask R-CNN, respectively. This suggests that \HeTu-v2 is the most appropriate method for modern radio surveys as they can resolve more complex sources.

\begin{figure*}[!ht]
\centering
\includegraphics[scale=0.375]{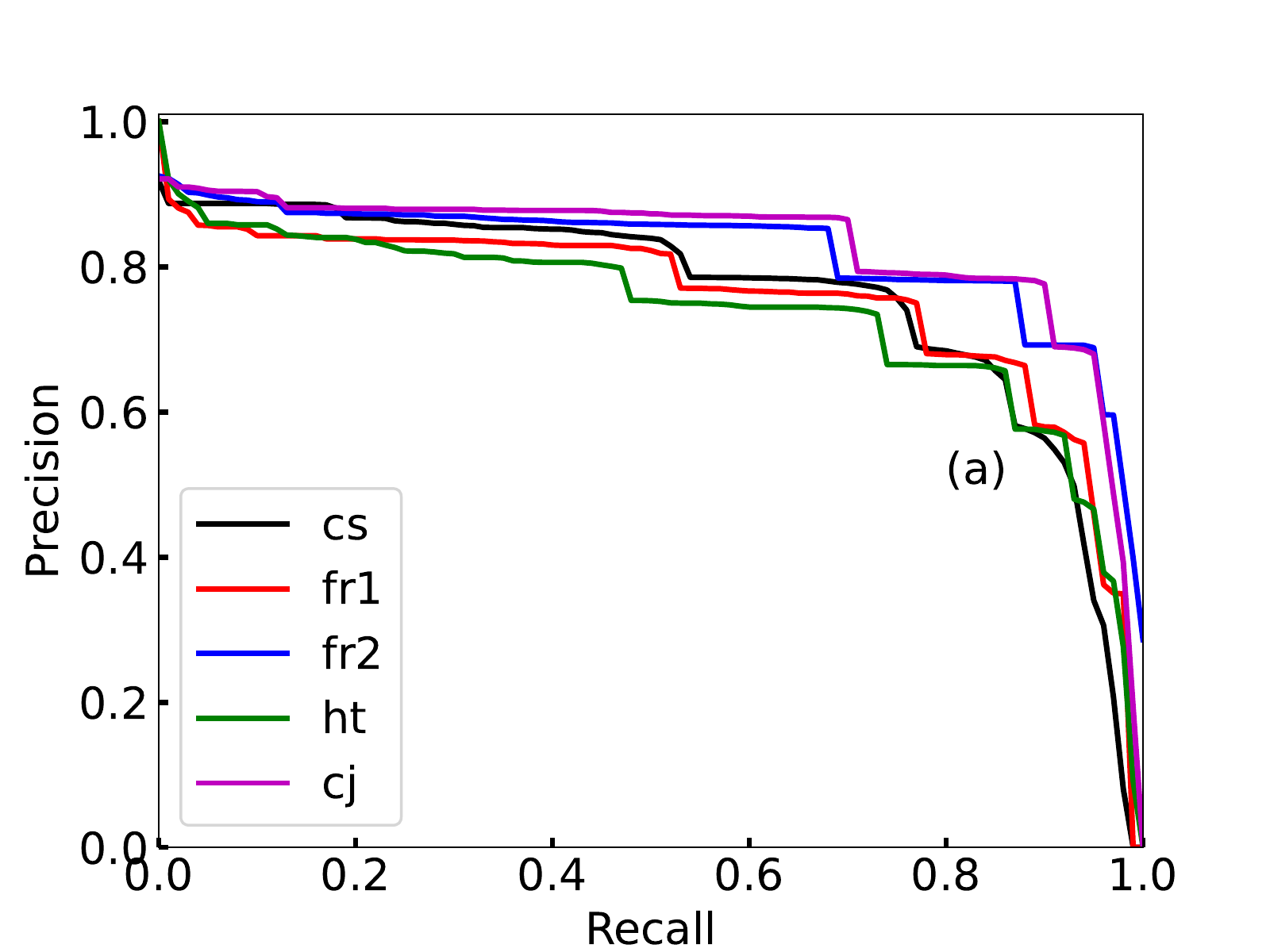}
\includegraphics[scale=0.375]{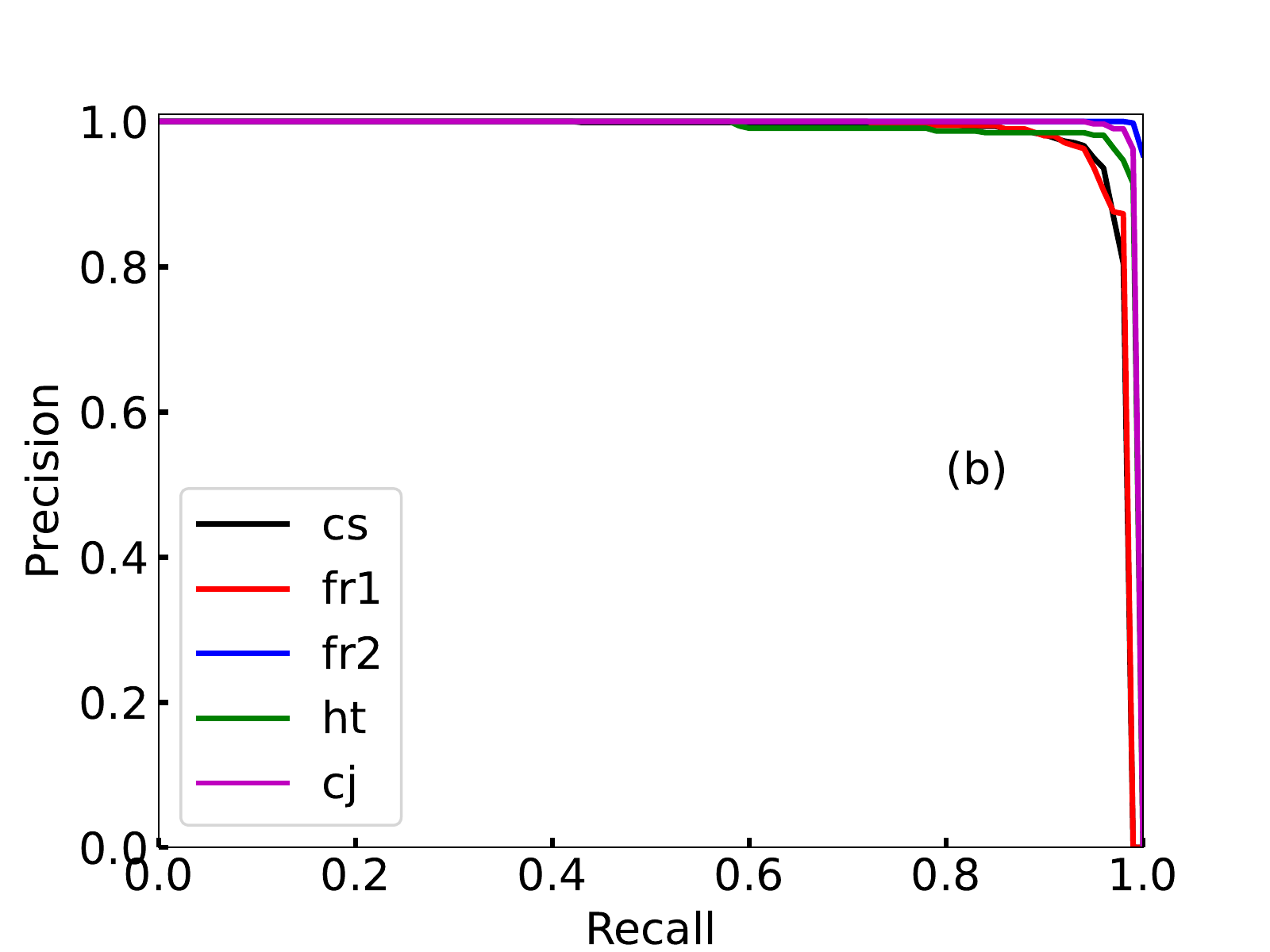}
\includegraphics[scale=0.375]{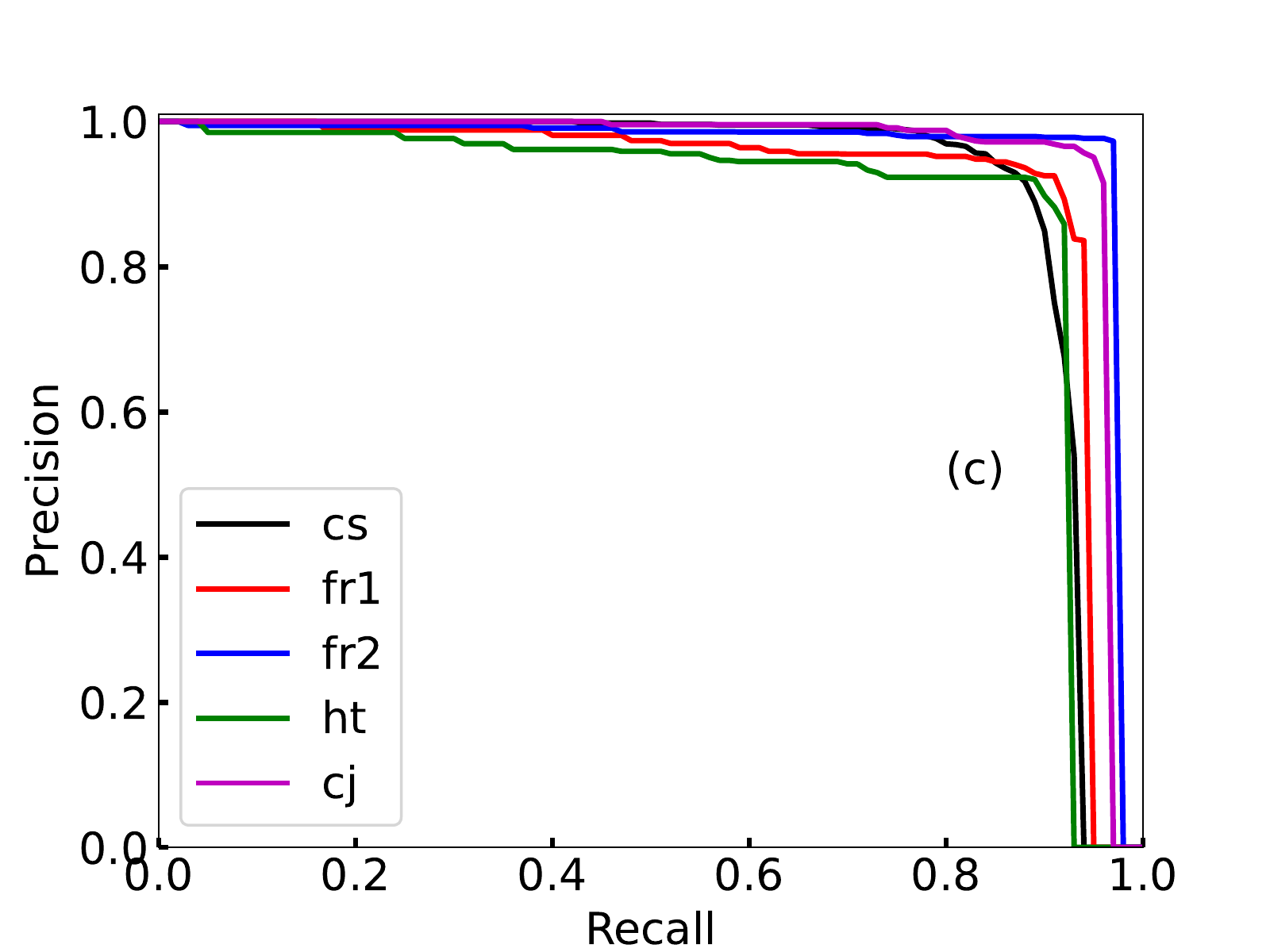}
\caption{Precision-recall ($P-R$) curves of \HeTu-v2 calculated at different $IoU$ thresholds averaged over the validation data samples for five classes. Panel (a) is calculated at $IoU$ threshold ranging from 0.5 to 0.95. Panel (b) is calculated at an $IoU$ threshold of 0.5, and Panel (c) is calculated at an $IoU$ threshold of 0.75.}
\label{fig:P-R}
\end{figure*}

The results shown in Fig. \ref{fig:P-R} summarize the performance of \HeTu-v2 network against ground truth images in the validation data set. In the validation data, each sample is assigned a prediction score by the \HeTu-v2 model, which represents the probability that the sample is positive. The samples for each class are then sorted based on their prediction scores. Samples with prediction scores larger than the threshold are classified as $TP$, while those smaller than the threshold are $FP$. As the number of samples for evaluation increases, the \HeTu-v2 model identifies more $TP$ or $FP$ detections according to the division rule, resulting in an increase in $R$ but a decrease in $P$. This pattern may oscillate depending on the characteristics of the data being evaluated. Taking panel (a) of Fig.~\ref{fig:P-R} as an example, all $P-R$ curves show a decrease in $P$ as $R$ increases, suggesting the relationship of a trade-off between $P$ and $R$. In general, a higher $P$ at a certain $R$ indicates better model detection ability. This is reflected in the larger area under the $P-R$ curve, i.e., the higher the $AP$, the better the detection ability of the model. Panels (a)-(c) in Fig. \ref{fig:P-R} are used to calculate the $AP_{\rm @50:5:95}$, $AP_{\rm 50}$, and $AP_{\rm 75}$ through Eq. (\ref{eq:AP}), indicating that the area under $P-R$ curves consistent with the $AP$s in the Table \ref{tab:APs}. The $P-R$ curves of the object scales are similar to Fig. \ref{fig:P-R} and therefore are not shown in this paper.

Moreover, the final precision and recall of \HeTu-v2 on all validation data were calculated at thresholds of 0.5 and 0.75,
and the results are shown in Table \ref{tab:final-precision-recall}. Table \ref{tab:final-precision-recall} shows that the final precision for the five classes is 93.4\% and 91.2\% at thresholds of 0.5 and 0.7, respectively. Additionally, the final recall for the five classes is 95.1\% and 87.1\% at thresholds of 0.5 and 0.7, respectively. These findings highlight the high accuracy and completeness of the \HeTu-v2 model in locating and classifying radio sources within validation images. As the threshold increases, the number of true or false positives decreases. This leads to a decrease in the overall recall and a fluctuation in the precision. The change in final recall and precision for FRII sources is minimal, with only a 0.2\% decrease in recall and a 1.3\% fluctuation in precision. This indicates that the \HeTu-v2 model has a high level of confidence in detecting FRII sources. 
 
\begin{table}[!ht]
\footnotesize
\centering
\begin{threeparttable}\caption{The final precision and recall of \HeTu-v2 on all validation data using thresholds of 0.5 and 0.75 respectively}\label{tab:final-precision-recall}
\doublerulesep 0.2pt \tabcolsep 10pt 
\begin{tabular}{c|ccc}
\toprule
 $IoU$ threshold & Class & Precision  & Recall  \\\hline
 0.5 & CS   & 92.5\% & 92.9\% \\
     & FRI  & 93.0\% & 91.2\% \\
     & FRII & 96.7\% & 99.8\% \\
     & HT   & 88.1\% & 95.9\%\\
     &CJ    & 96.1\% & 96.4\%\\\cline{2-4}
     &ALL   & 93.4\% & 95.1\%\\\hline
 0.75& CS   & 87.2\% & 80.0\% \\
     &FRI   & 93.8\% & 82.9\%  \\
     &FRII  & 96.9\% & 98.5\%  \\
     &HT    & 87.3\% & 91.9\%   \\
     &CJ    & 95.4\% & 89.1\%  \\\cline{2-4}
     &ALL   & 91.2\% & 87.1\%\\
\bottomrule
\end{tabular}
\end{threeparttable}
\end{table}

\begin{figure*}[!ht]
\centering
\includegraphics[scale=0.15]{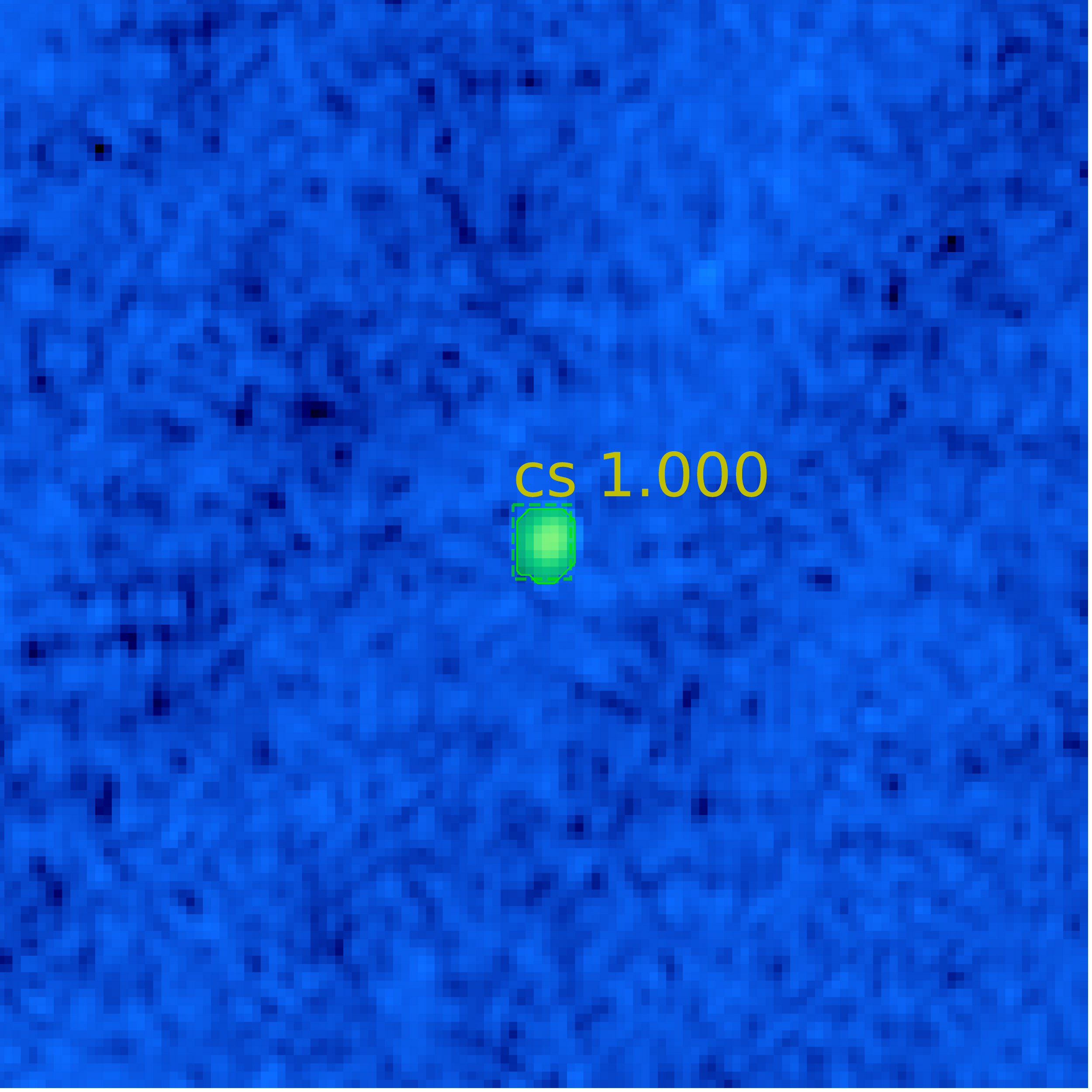}
\hspace{0.1mm}
\vspace{0.1mm}
\includegraphics[scale=0.15]{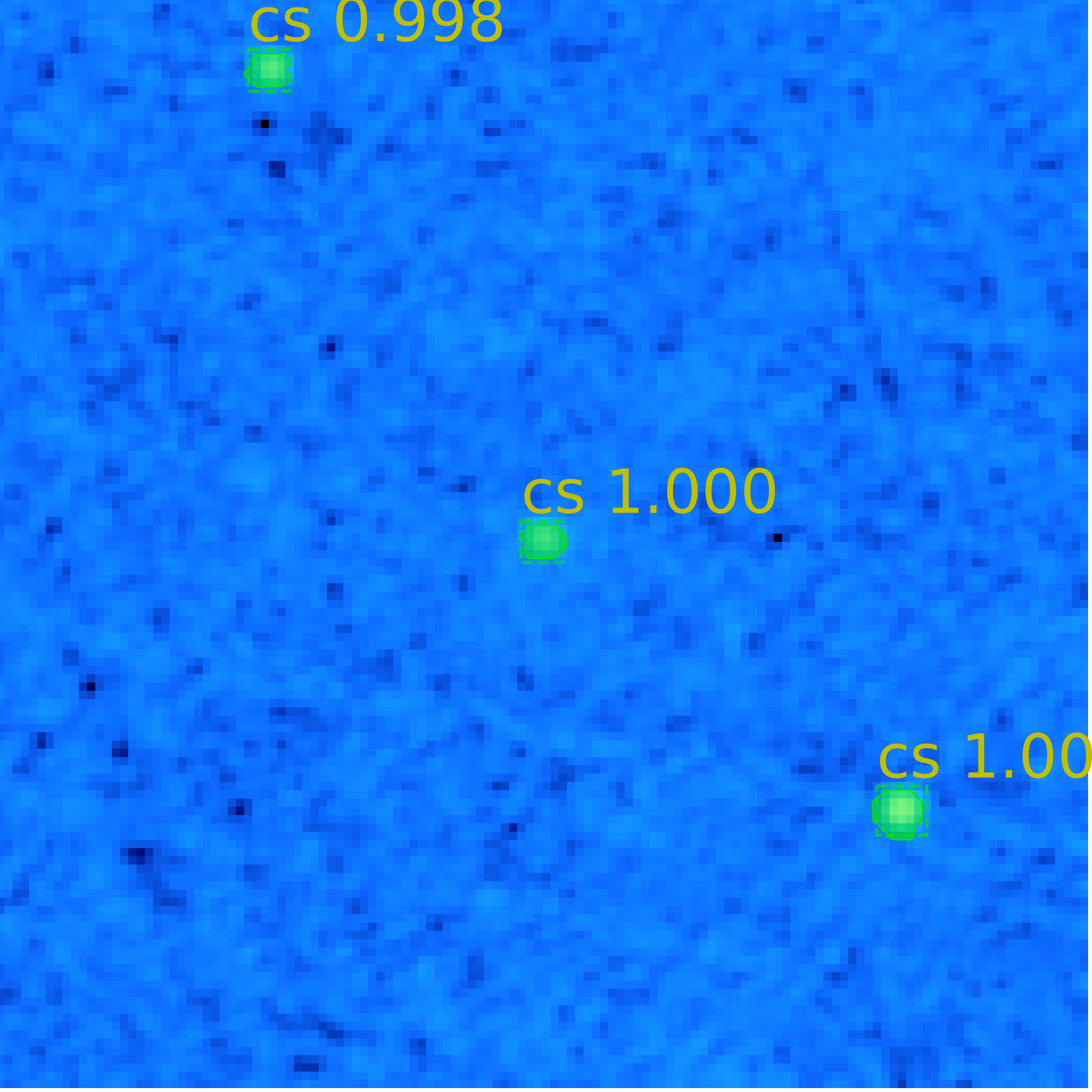}
\hspace{0.1mm}
\vspace{0.1mm}
\includegraphics[scale=0.15]{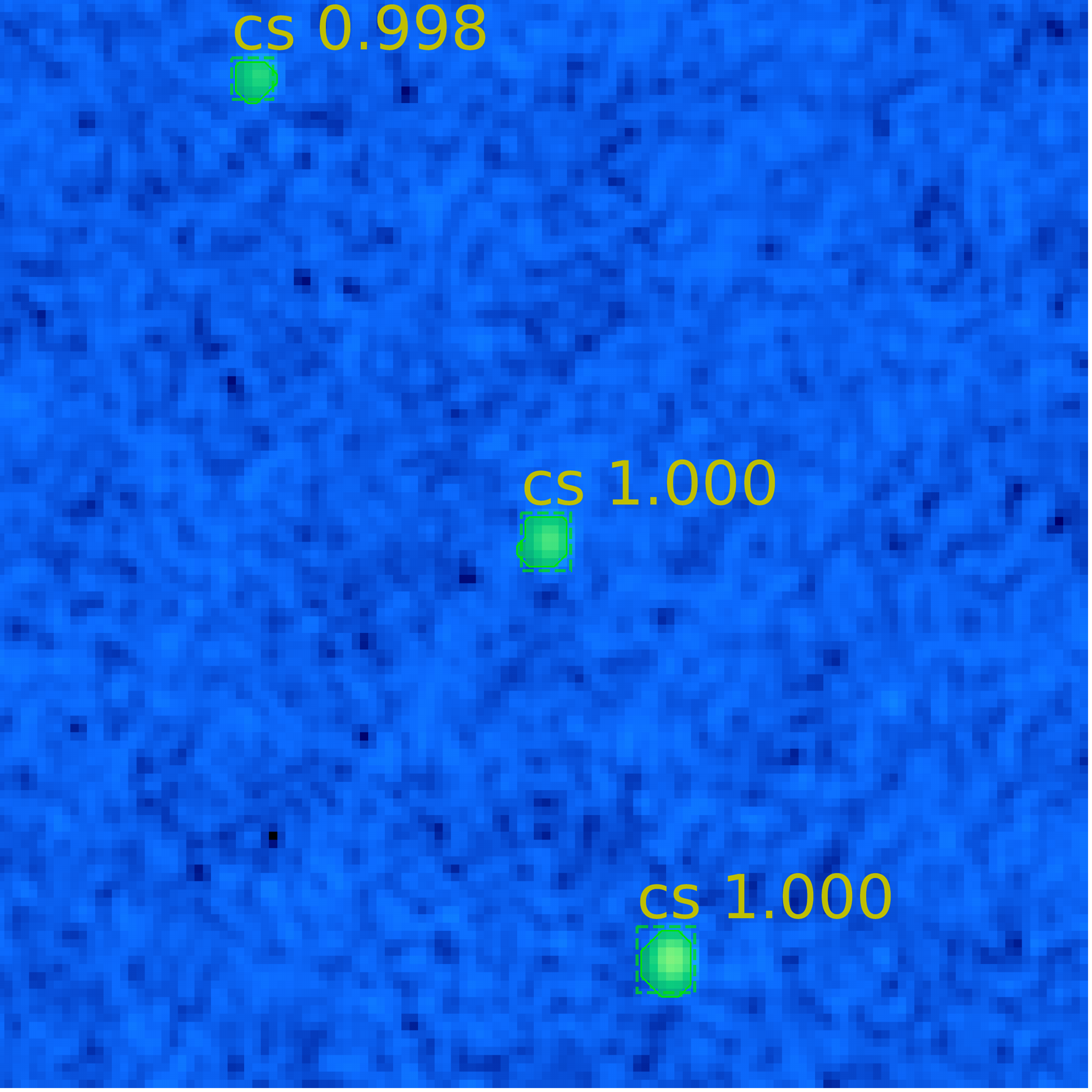}
\hspace{0.1mm}
\vspace{0.1mm}
\includegraphics[scale=0.15]{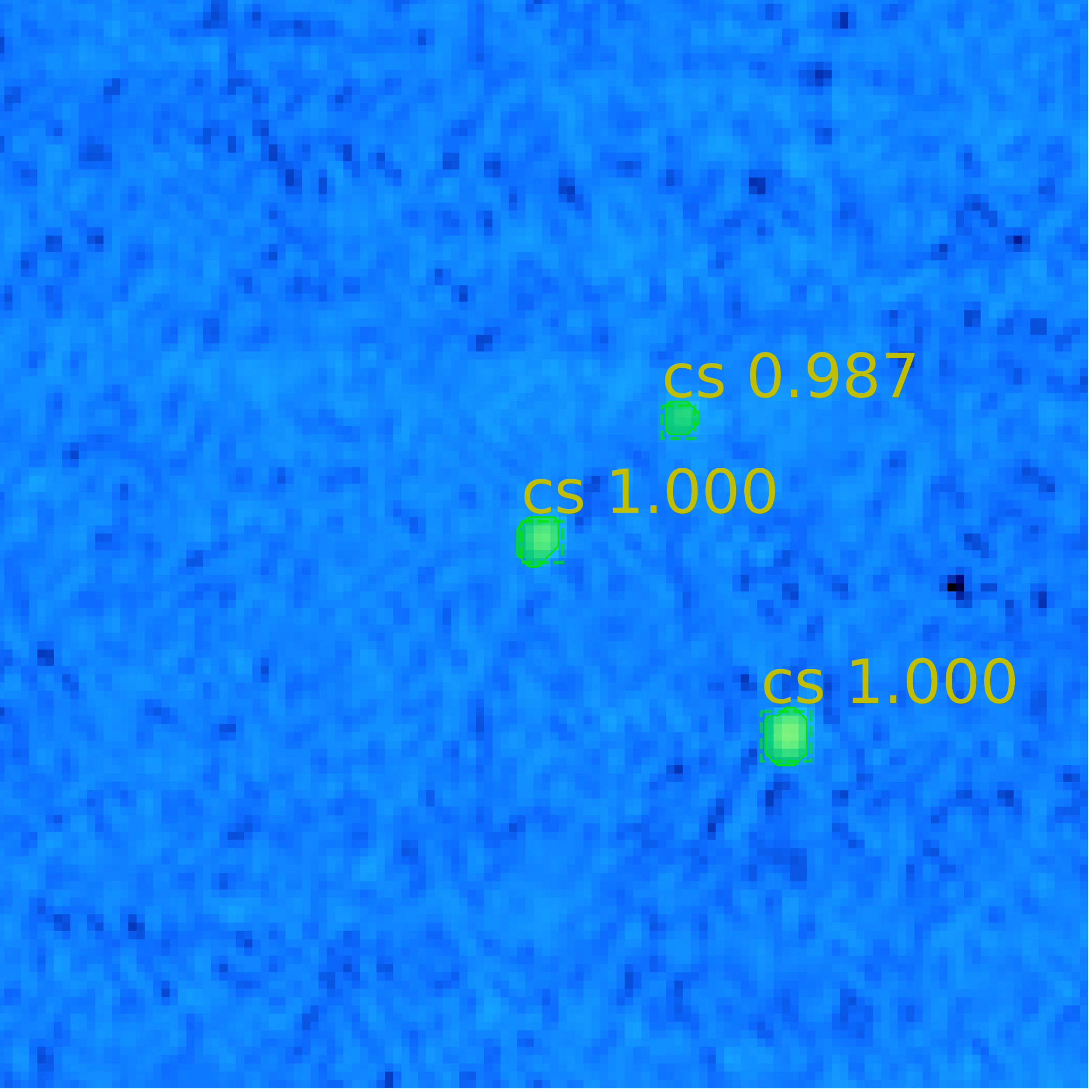}
\hspace{0.1mm}
\vspace{0.1mm}
\includegraphics[scale=0.15]{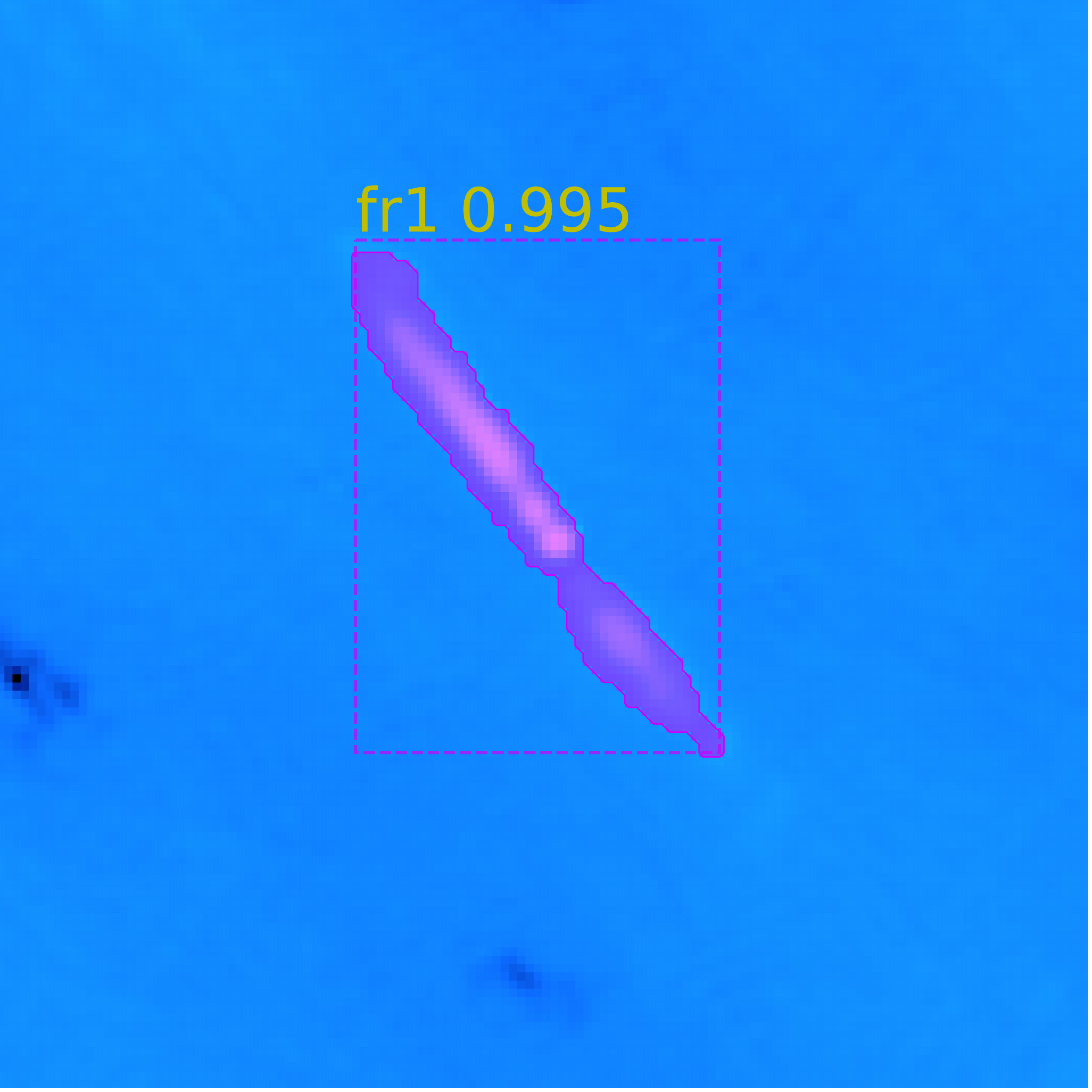}
\hspace{0.1mm}
\vspace{0.1mm}
\includegraphics[scale=0.15]{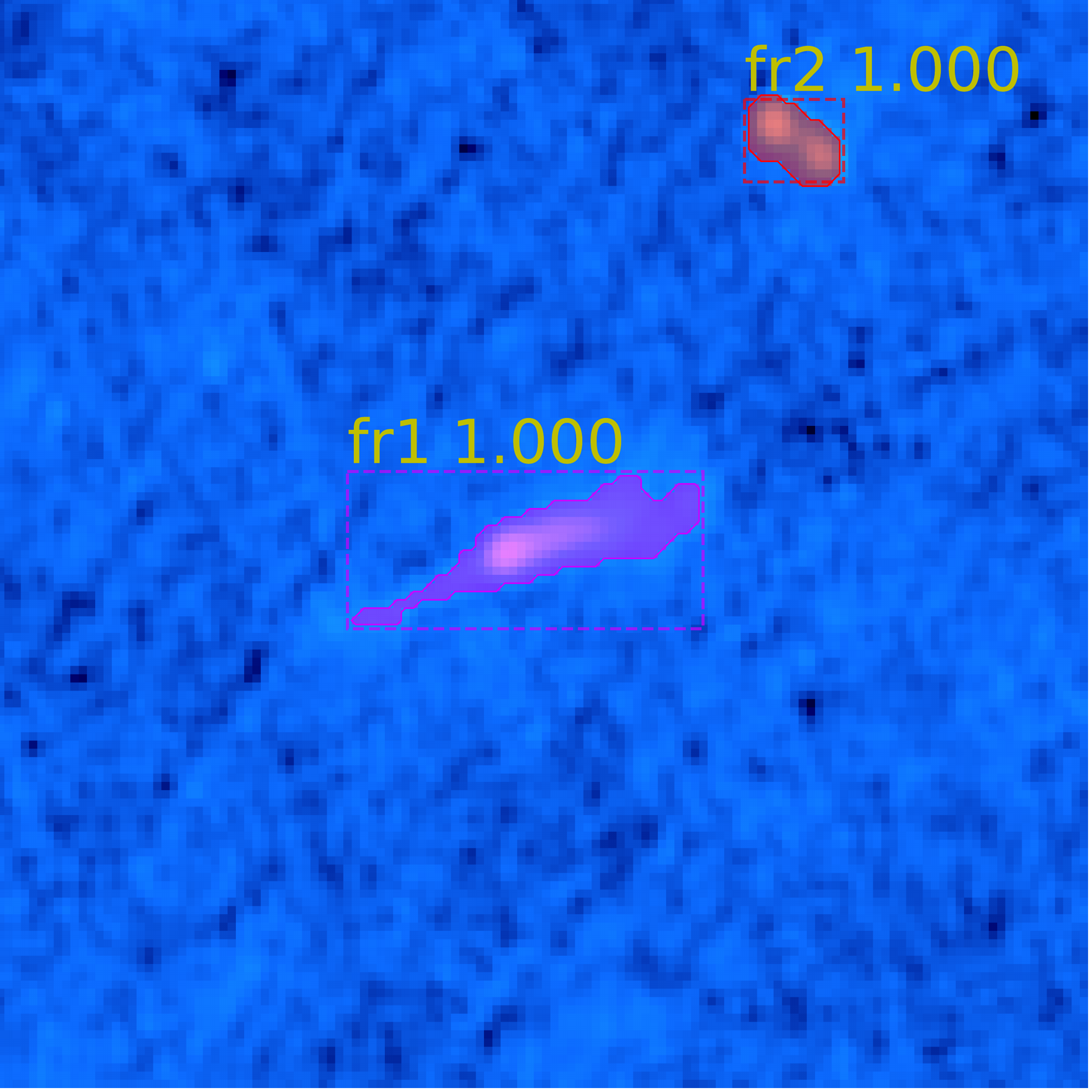}
\hspace{0.1mm}
\vspace{0.1mm}
\includegraphics[scale=0.15]{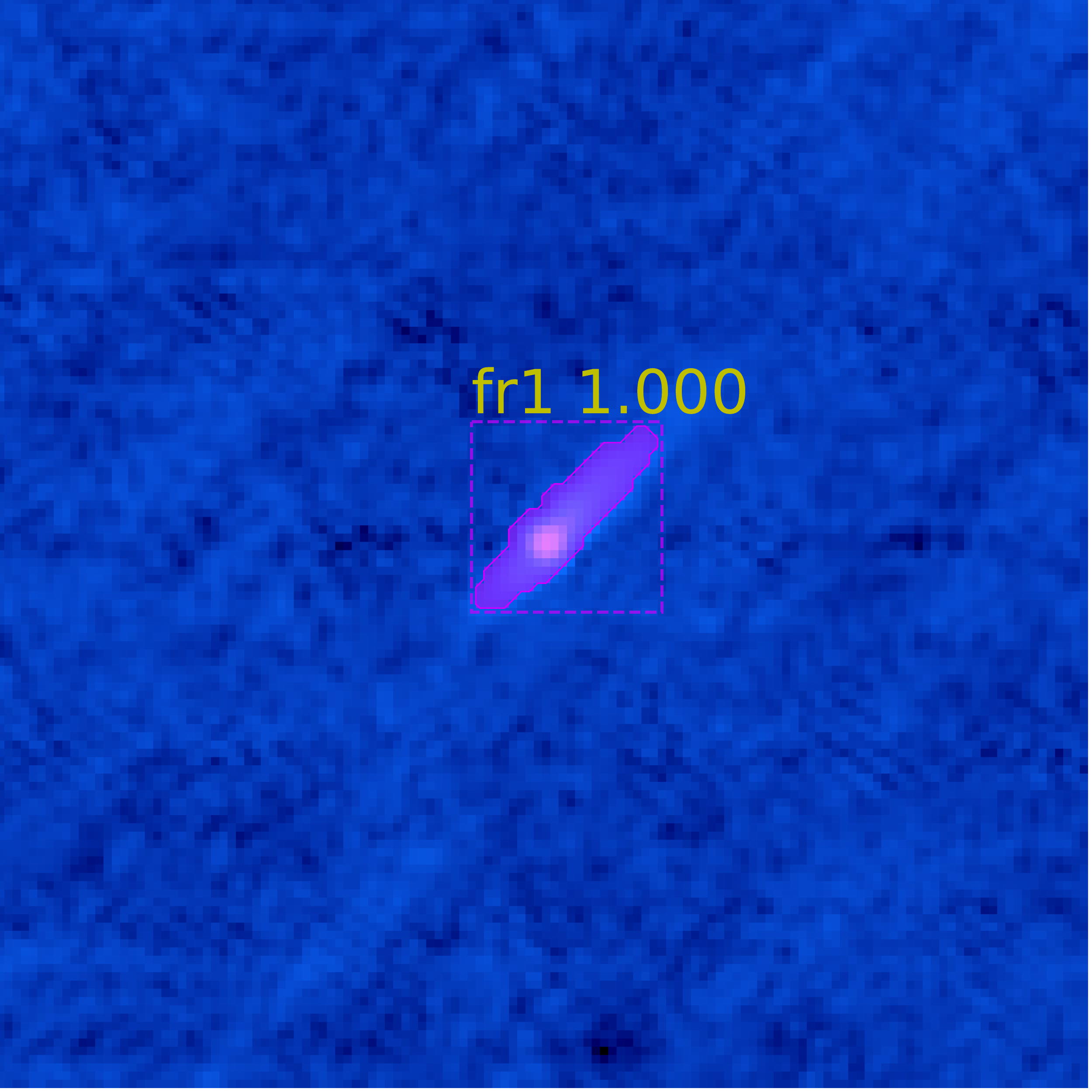}
\hspace{0.1mm}
\vspace{0.1mm}
\includegraphics[scale=0.15]{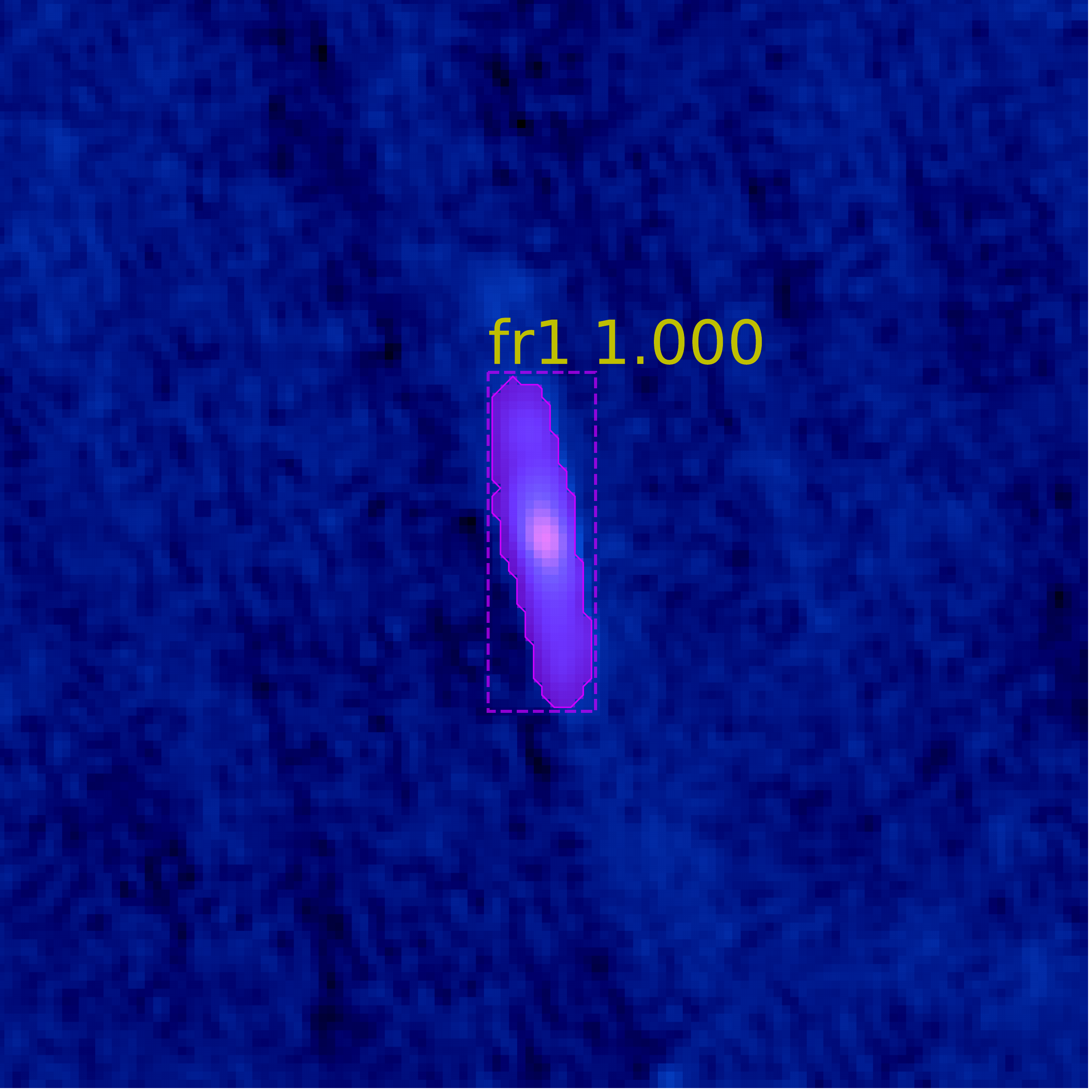}
\hspace{0.1mm}
\vspace{0.1mm}
\includegraphics[scale=0.15]{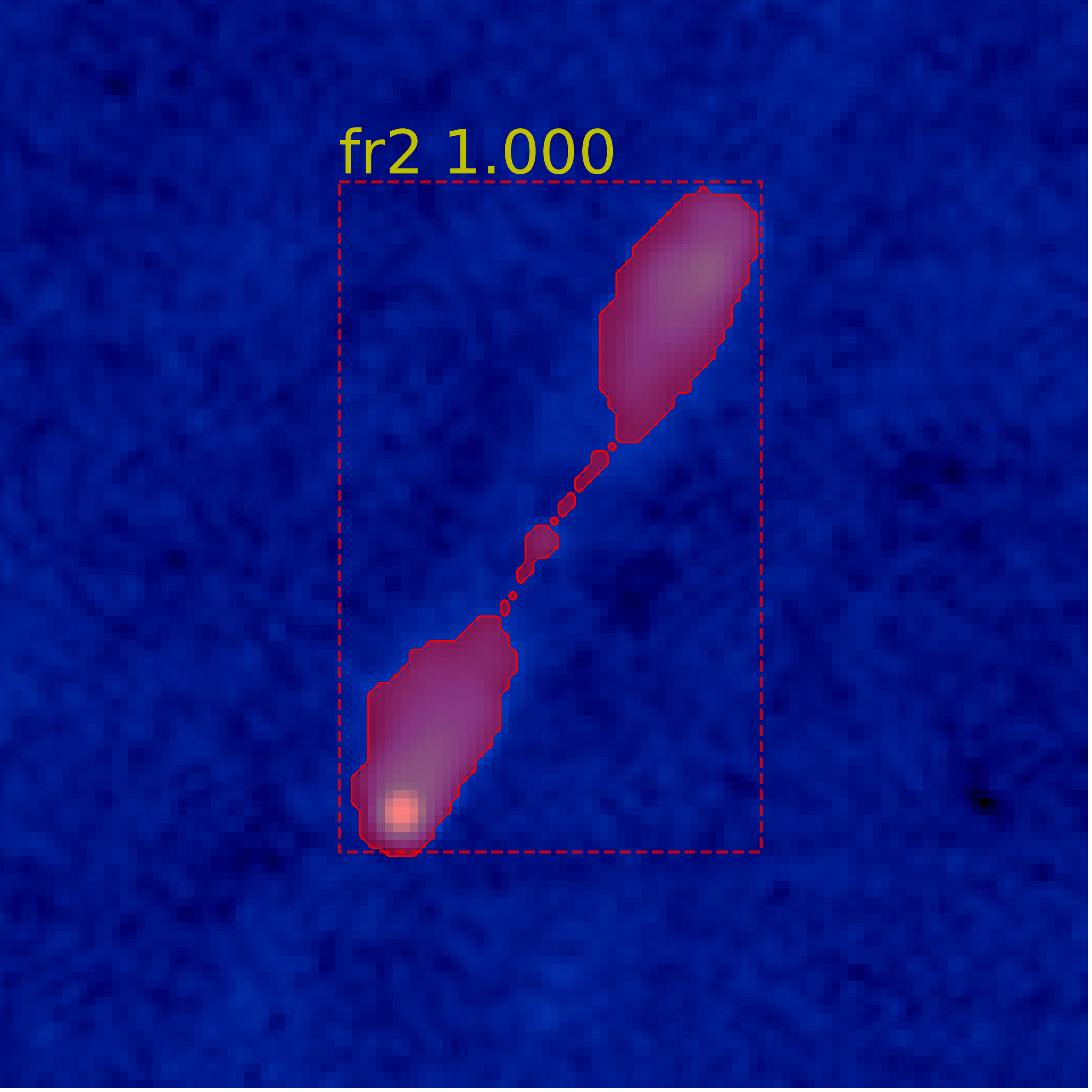}
\hspace{0.1mm}
\vspace{0.1mm}
\includegraphics[scale=0.15]{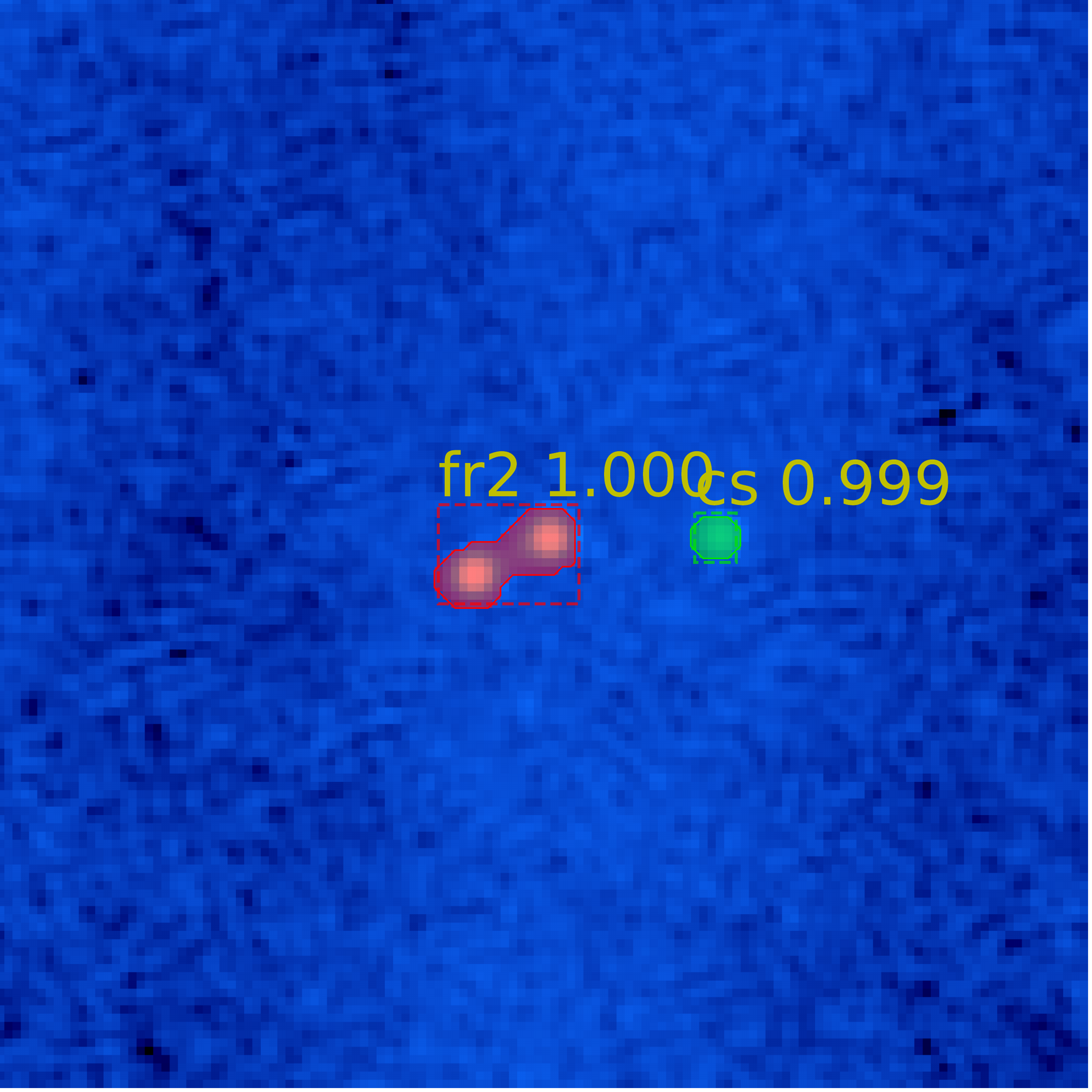}
\hspace{0.1mm}
\vspace{0.1mm}
\includegraphics[scale=0.15]{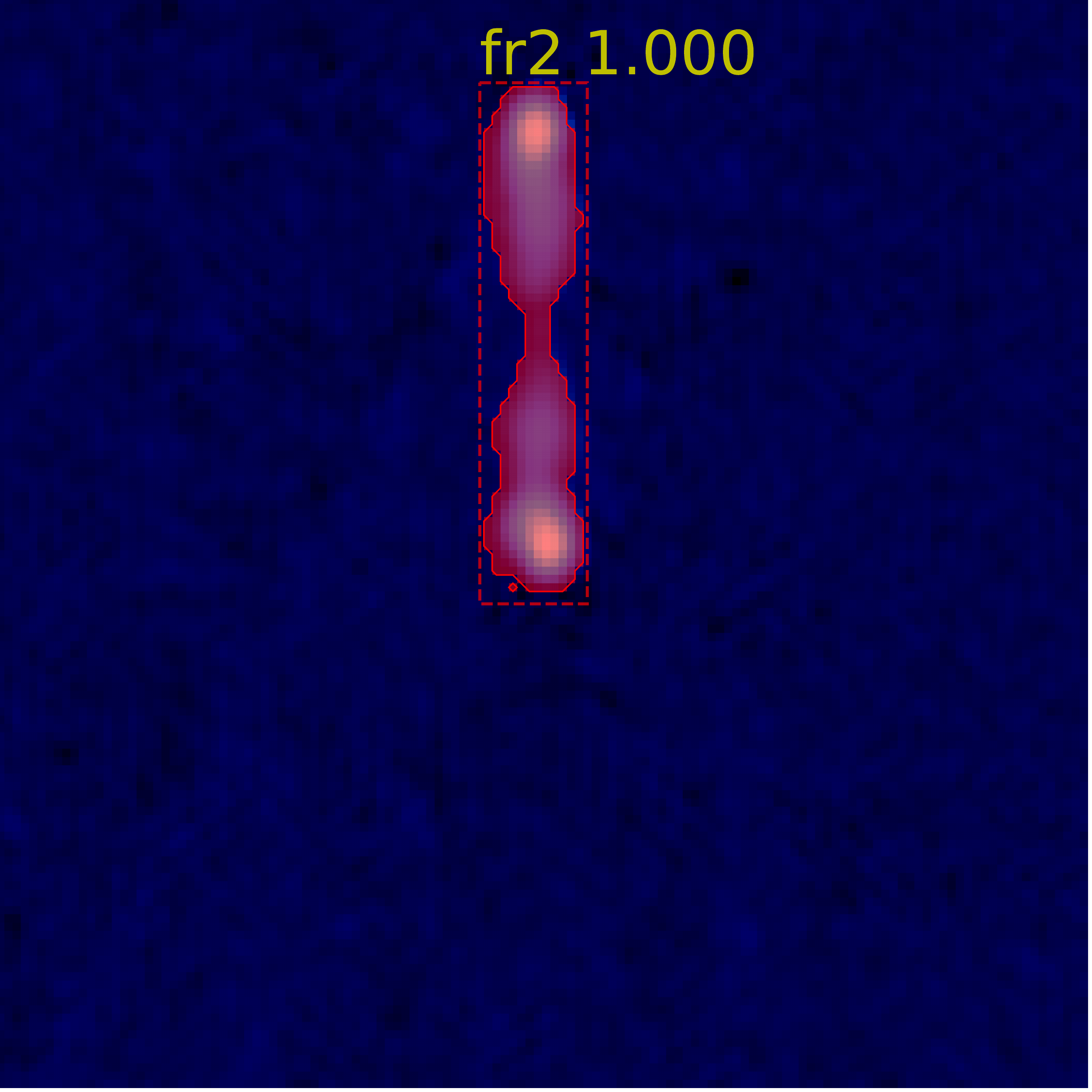}
\hspace{0.1mm}
\vspace{0.1mm}
\includegraphics[scale=0.15]{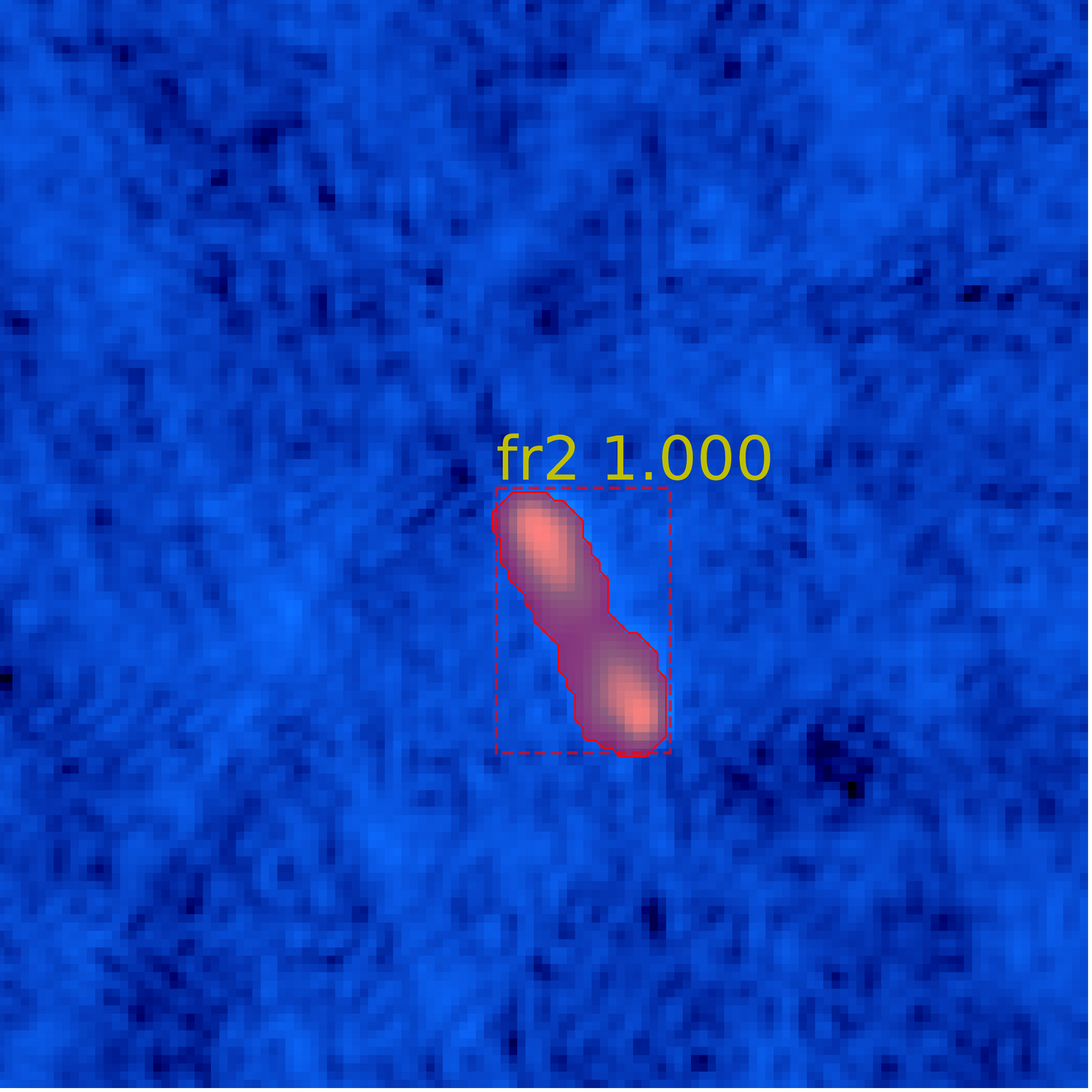}
\hspace{0.1mm}
\vspace{0.1mm}
\includegraphics[scale=0.15]{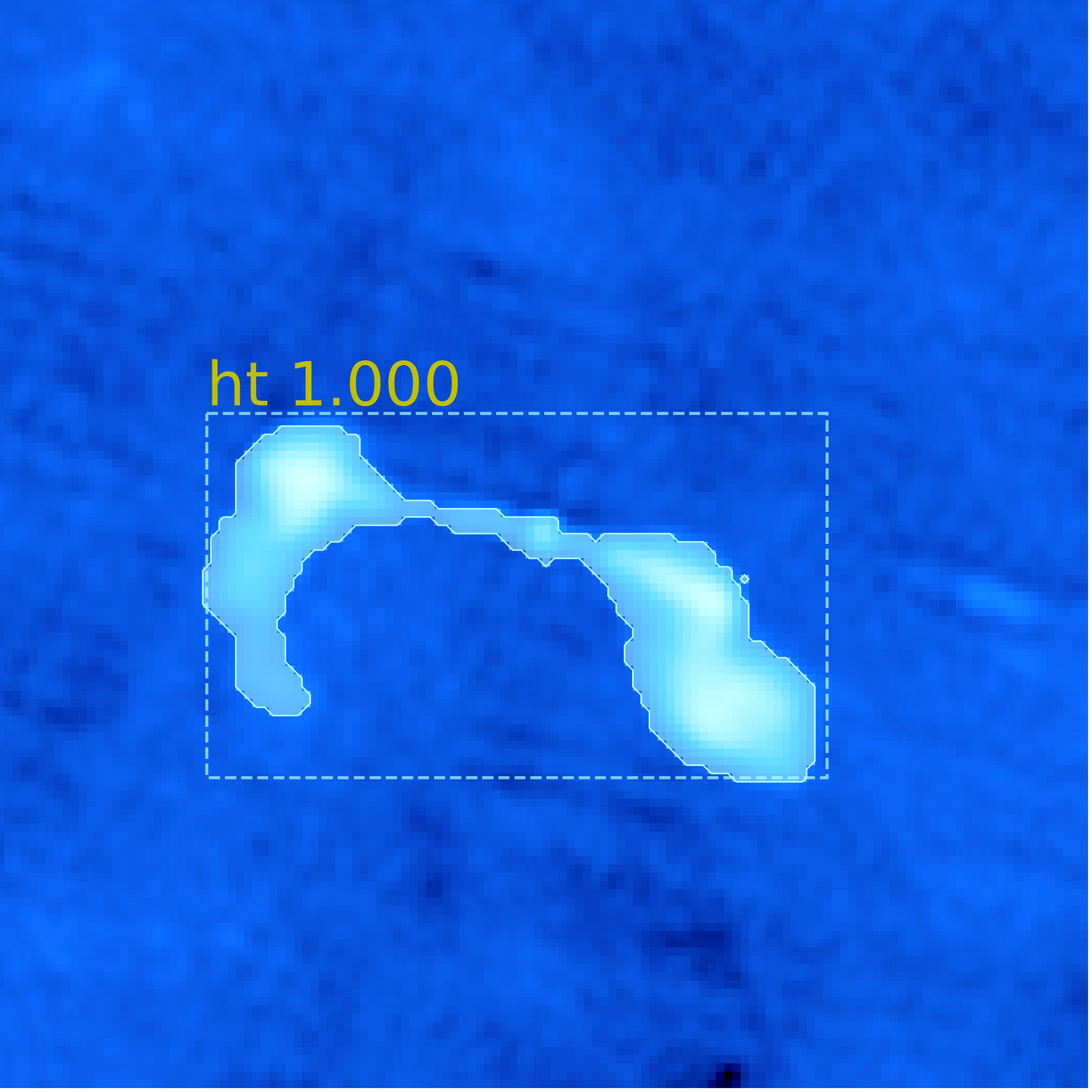}
\hspace{0.1mm}
\vspace{0.1mm}
\includegraphics[scale=0.15]{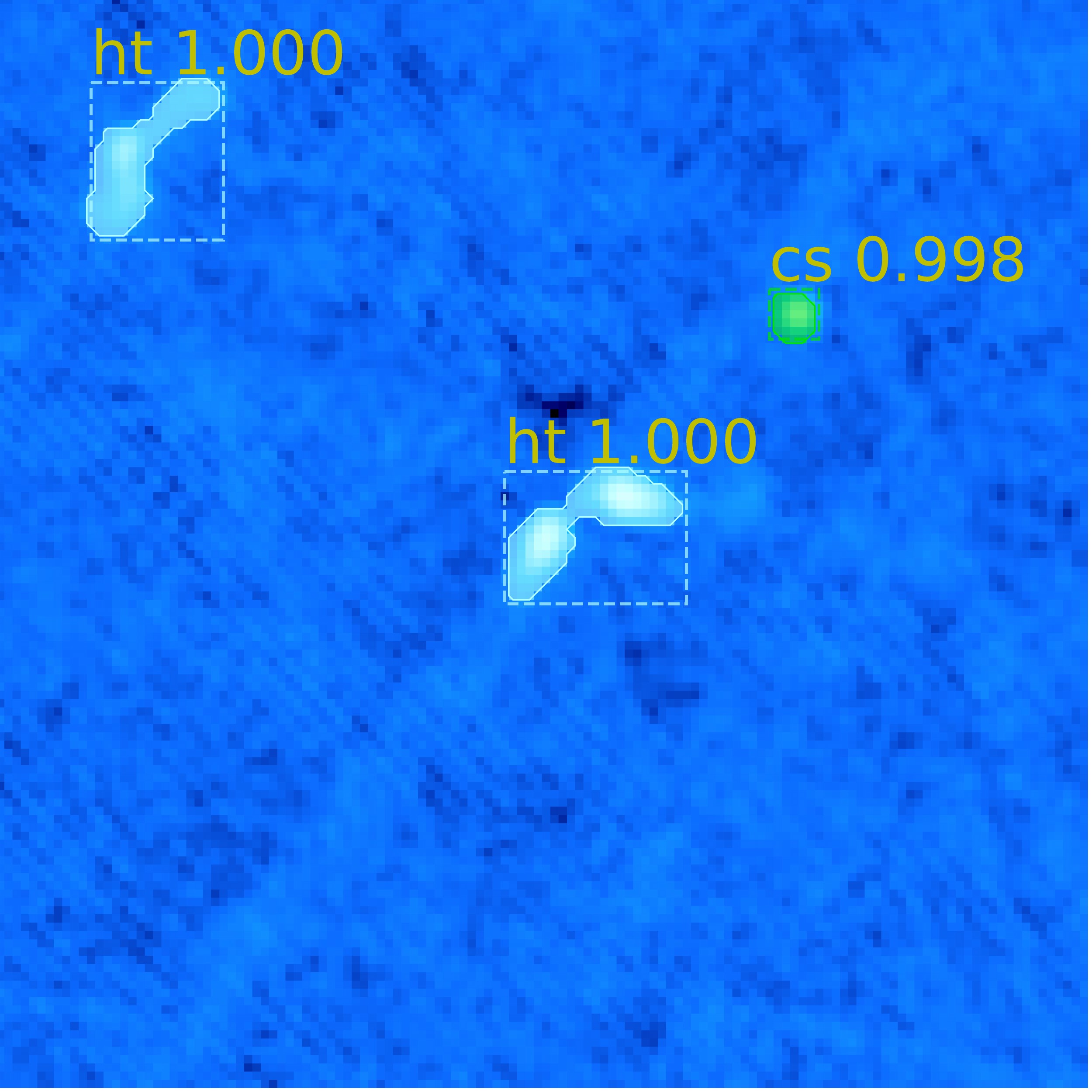}
\hspace{0.1mm}
\vspace{0.1mm}
\includegraphics[scale=0.15]{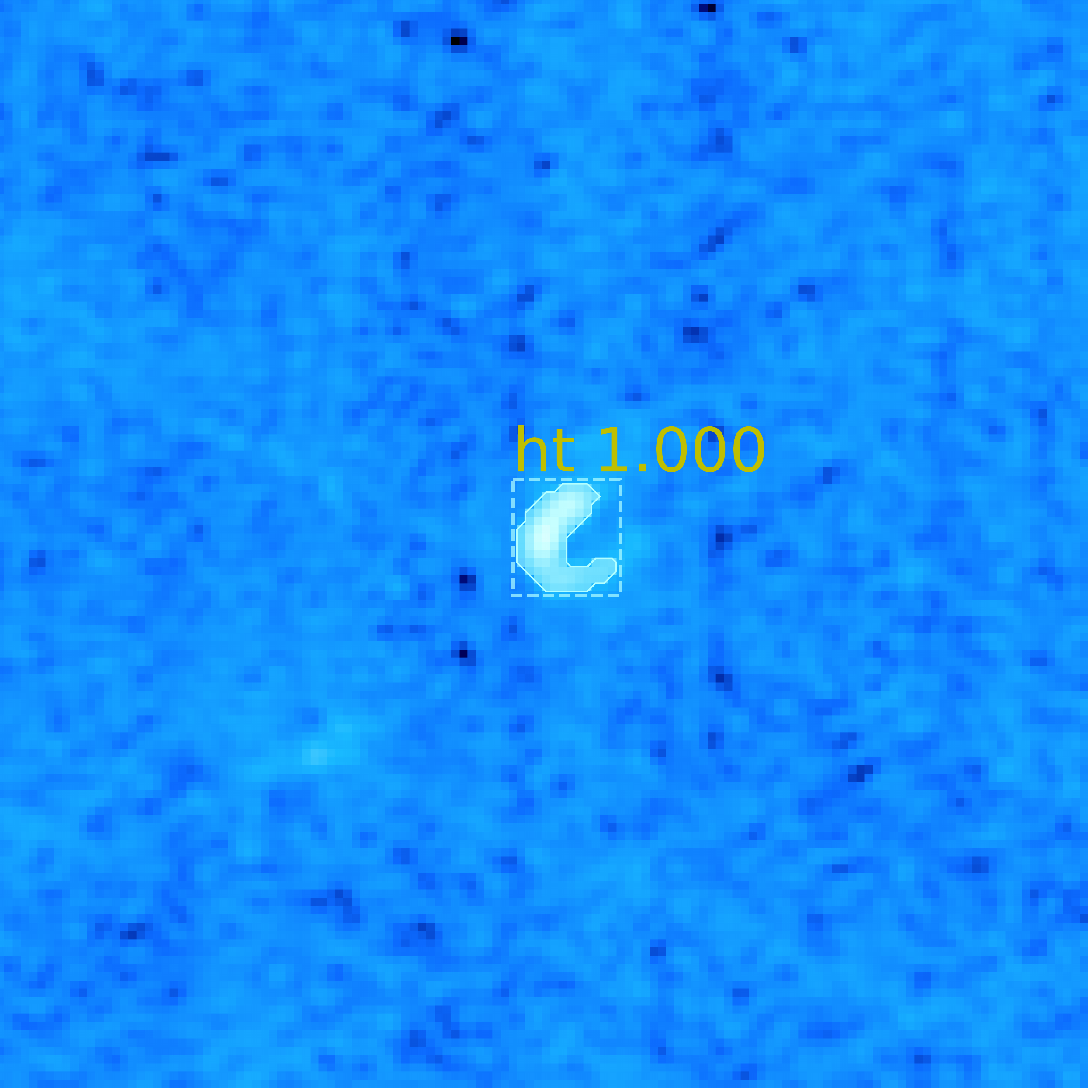}
\hspace{0.1mm}
\vspace{0.1mm}
\includegraphics[scale=0.15]{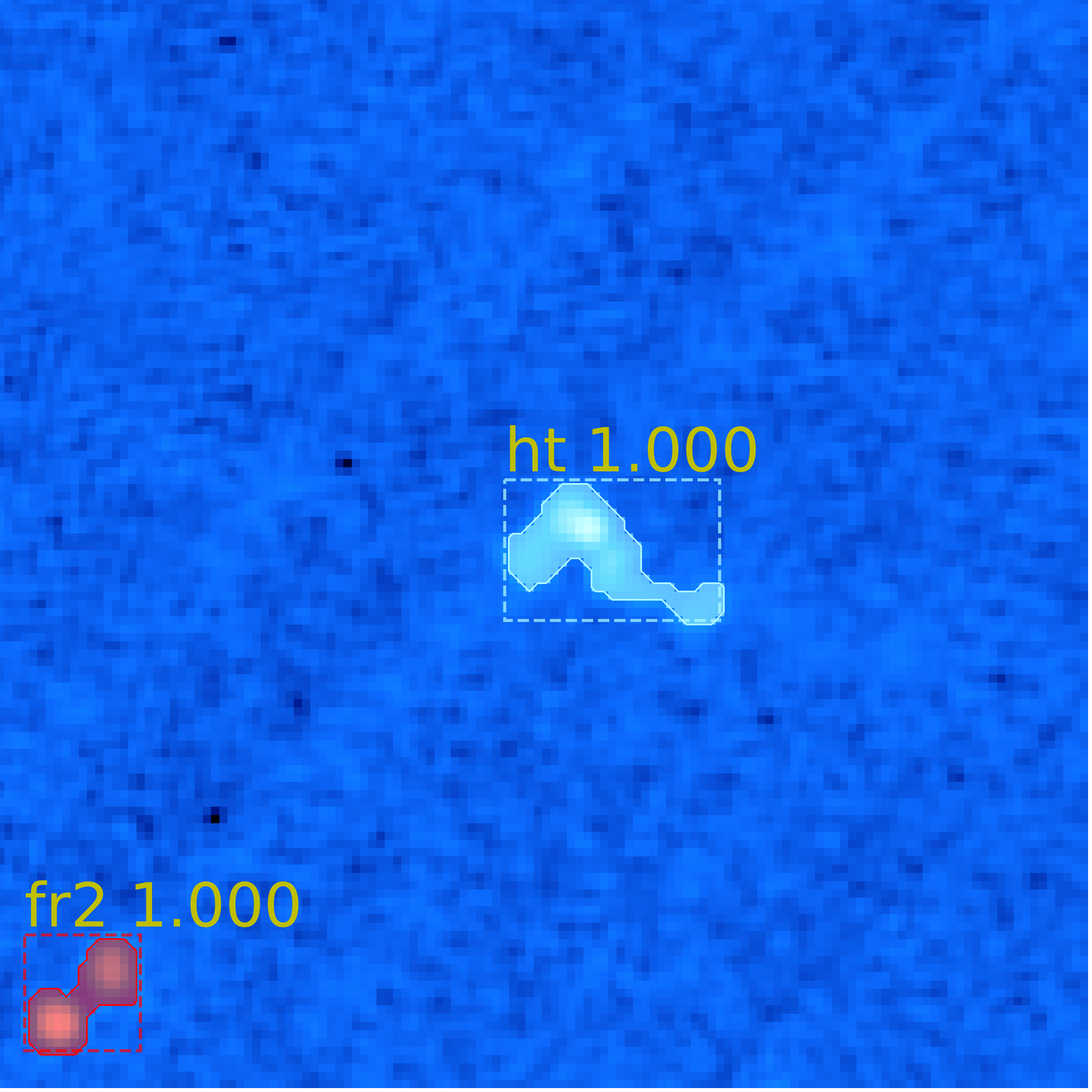}
\hspace{0.1mm}
\vspace{0.1mm}
\includegraphics[scale=0.15]{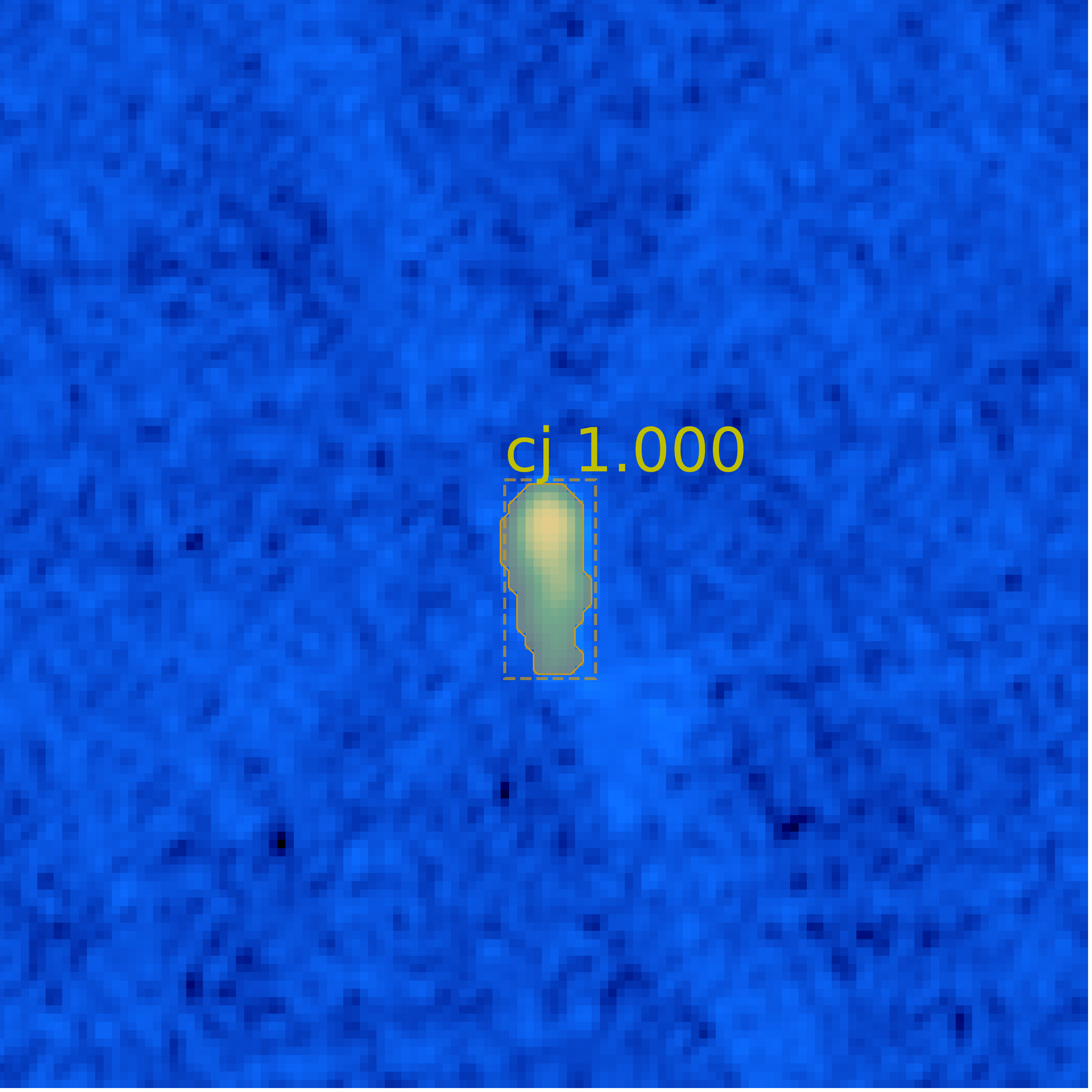}
\hspace{0.1mm}
\vspace{0.1mm}
\includegraphics[scale=0.15]{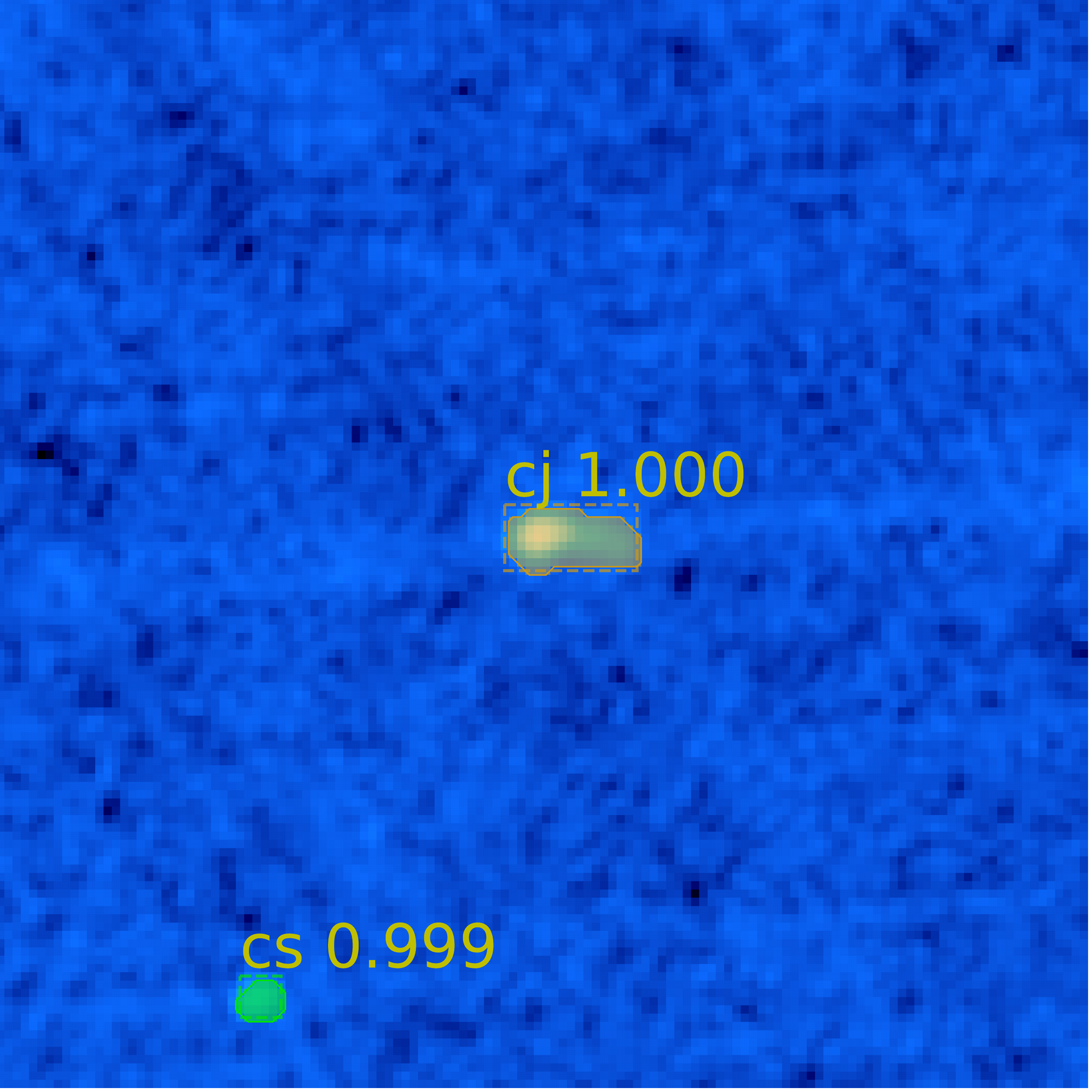}
\hspace{0.1mm}
\vspace{0.1mm}
\includegraphics[scale=0.15]{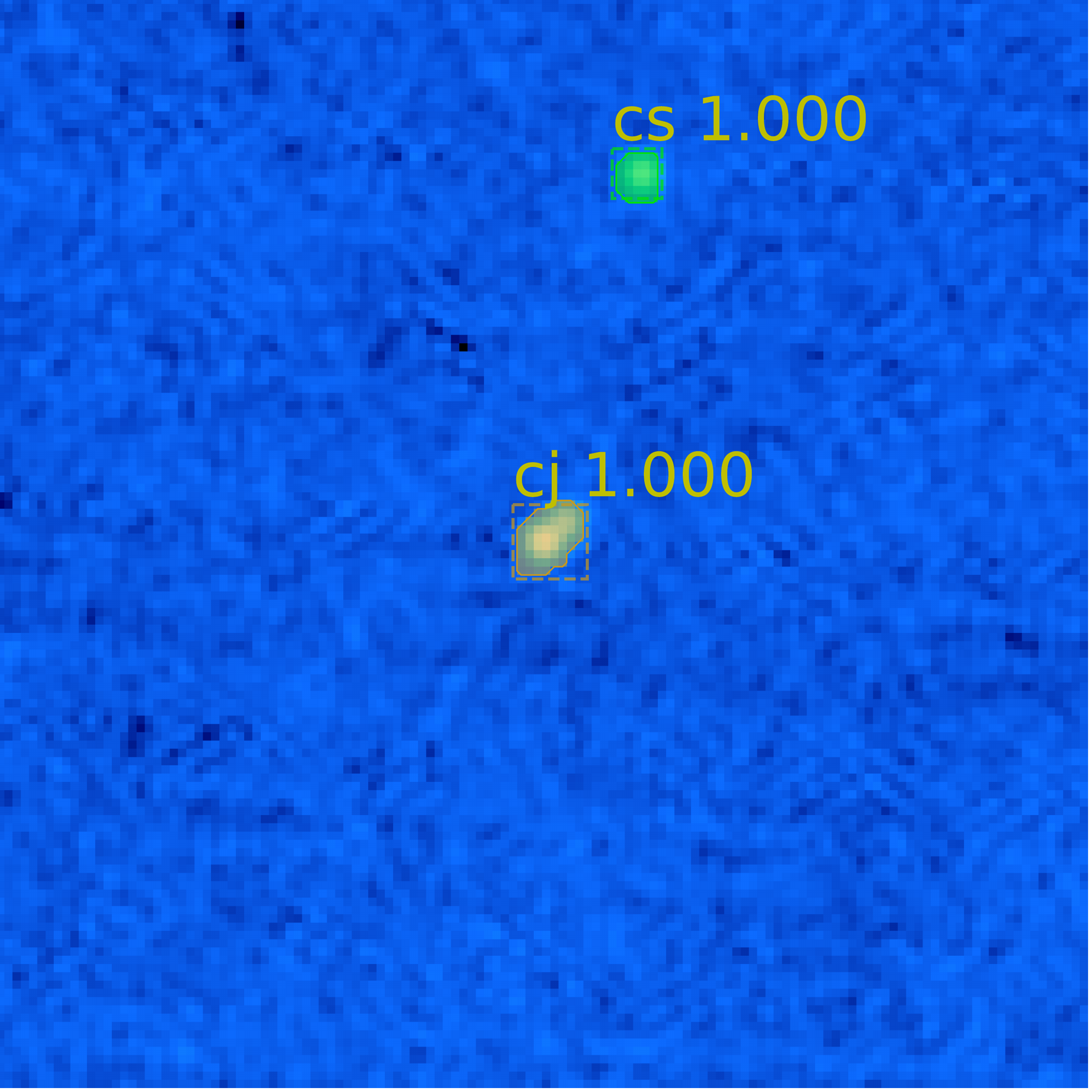}
\hspace{0.1mm}
\vspace{0.1mm}
\includegraphics[scale=0.15]{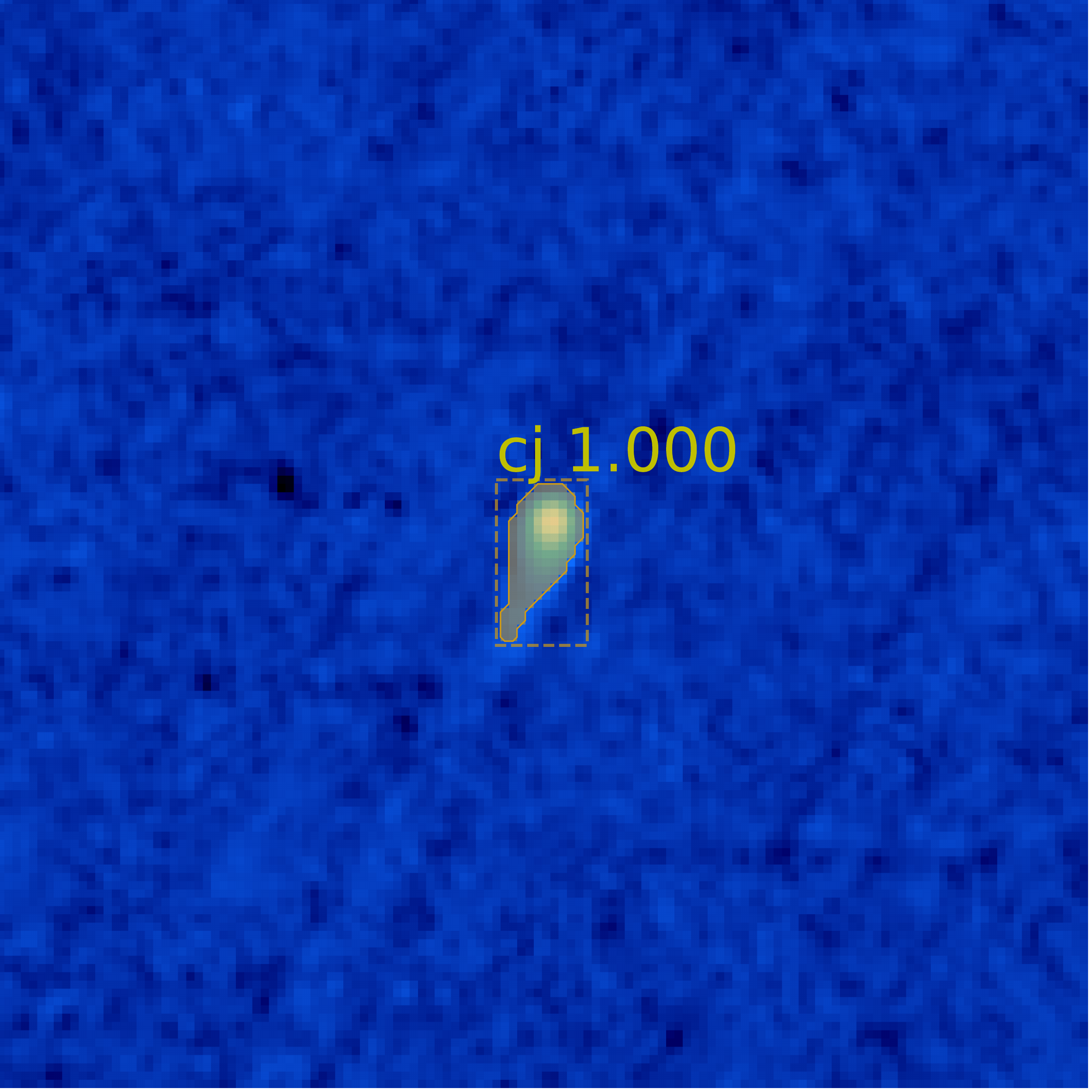}
\caption{Examples of segmentation and classification on the validation data set. The different colored rectangular boxes and masks represent the identified sources: green for CS (cs), magenta for FRI (fr1), red for FRII (fr2), light-cyan for HT (ht), and orange for CJ (cj). Each source is labeled with a class name and a score between 0 and 1 on the top left of the bounding box.}
\label{fig:valid-example}
\end{figure*}

Fig. \ref{fig:valid-example} shows examples of segmentation and classification on the validation dataset. The localization and size of the detected sources are labeled with five colored rectangular boxes and masks: green for CS, magenta for FRI, red for FRII, light-cyan for HT, and orange for CJ. Each source is also labeled with the associated class or label name with a probability score between 0 and 1 on the top-left of a rectangular box. In Fig. \ref{fig:valid-example}, all detected radio sources are segmented and classified with a higher score and refined masks, respectively. This suggests the high accuracy of the predicted masks by \HeTu-v2. 

\subsection{Inference experiments}\label{sec:inference}
\subsubsection{Building catalog} \label{sec:build_catalog}
To investigate the inference ability of trained \textsc{Hetu}-v2 model and the completeness, reliability, and accuracy of the catalog produced by this model, we used 946,366 processed images from the FIRST survey described in section \ref{sec:data}. Removing the repeated positions of components, the images contain a total of 946,366 components in the FIRST-14dec17 catalog. To speed up the prediction procedure, the images have been split into three parts ($i.e.$ 315,455, 315,455, and 315,456), and then run them simultaneously on three different GPU devices. The inferring experiments took about 17.5~h to finish.

\begin{figure*}[!ht]
\centering
\includegraphics[scale=1.0]{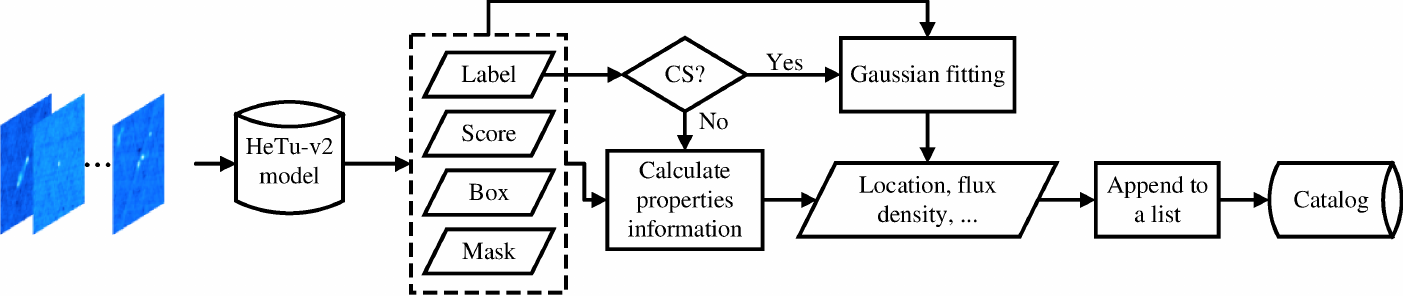}
\caption{Workflow of the \HeTu-v2 based source detector.}
\label{fig:wf-source-detector}
\end{figure*}

A batch prediction is performed on all images, and at the same time each predicted source is calculated 
for their properties information to form a catalog. Thus, the \HeTu-v2 is said to function as a source detector. 
The workflow of the \HeTu-v2 based source detector is shown in Fig. \ref{fig:wf-source-detector}. At first, the FIRST images are sequentially input to the \HeTu-v2 model and then output the prediction results of the detected sources, including predicted label, predicted score, predicted bounding box, 
and predicted mask. Next, depending on the prediction information, each predicted source properties such as location and flux density have been calculated. For each of CS sources, Gaussian fitting is performed using the IMFIT task of \miriad\ software \citep{sault1995retrospective}, which provides the fitting information: location ($i.e.$ right ascension and declination), local Root Mean Square (RMS) noise, peak flux density, integrated flux density, beam (major and minor axes length, and the position angle) in addition to the deconvolved beam information. For each of extended sources, the center position of the predicted box, local RMS noise, peak flux density at the maximum pixel value in the predicted box, and the total flux density of the detected source have been calculated. Here, the total flux density is given by Eq.~(\ref{eq:total_flux})\footnote{\url{https://github.com/mhardcastle/radioflux}}:
\begin{equation}\label{eq:total_flux}
{S_{{\rm{total}}}} = \left(\sum{{I_{{\rm{in}}}}[mask]}\right)/\left((2\pi  \cdot bmaj \cdot bmin)/G_{{\rm{factor}}}^2\right),
\end{equation}
where $mask$ is the predicted masks of the detected source in a type of Boolean. The function $\sum{{I_{{\rm{in}}}}[mask]}$ refers to the sum of the pixel values at the position where the $mask$ is $True$ in the input image. The $bmaj$ and $bmin$ represent the beam semi-major and semi-minor axis length of the input image in pixels scale. The $G_{\rm{factor}}$ is a factor of beam area and is defined as $G_{\rm{factor}}=2\sqrt {2({\rm In}(2))}$. At last, the information of each detected source is 
appended to a list and thereby \HeTu-v2 generates the initial catalog in a CSV file. The details of more catalog columns can be seen in Table~\ref{tab:catalog}. 

Since the image samples have overlapping sky regions and the sources detected on one image is detected again on other images, the generated initial catalog contains duplicate sources which need to be removed. Consequently, a post-processing program for de-duplication has been developed. Due to the extended source large size and complex structure, the small size (diameter $\leq$ 10 pixels) and the simple structure (Gaussian or point) of the CS sources, different methods are adopted for these two groups of sources. 

To begin with, the initial catalog is split into five sub-catalogs based on the label name. For CS sources, first, the initial CS sub-catalog performs intra-table cross-match on the location columns (centre\_ra and centre\_dec) using \stilts\ package \citep{2006ASPC..351..666T} within 5 arcsec ($i.e.$ the full width at half-maximum (FWHM) beam size in the FIRST survey) of the max error to generate the matched CSV file. Then, duplicate rows are removed from the matching results and keep the row of the first occurrence. Now the unmatched rows are appended to the remaining results of the matched CSV file to create a new CSV file as the final results. For each morphological class sub-catalog of extended sources, the bounding box of the $i$-th ($1 \le i \le {N_{\rm source}}$) row defining a location region, check whether the location from $(i+1)$-th to ${N_{\rm source}}$-th is within the region, and the sources within this region are regarded as duplicated with the $i$-th row and recorded in a TXT file. Here, ${N_{\rm source}}$ is the total number of rows or sources for the sub-catalog of extended sources. Following this method we can get all duplicate sources by looping through all rows. Finally, the duplicate sources in the last generated TXT file are removed from the sub-catalog. In the end, all processed sub-catalogs combined to obtain the final catalog.

\begin{table}[!ht]
\footnotesize
\centering
\begin{threeparttable}\caption{The columns information of output catalog by \HeTu-v2 based source detector. }\label{tab:catalog}
\doublerulesep 0.1pt \tabcolsep 4.8pt 
\begin{tabular}{p{0.2\columnwidth}p{0.7\columnwidth}}
\toprule
 Column name & Description \\ \hline
 source\_name & Predicted source name in JHHMMSSDDMMSS format. \\
image\_filename & Input image file name in PNG format. \\
 label & Predicted label name (cs: CS, fr1: FRI, fr2: FRII, ht: HT, cj: CJ). \\
 score &   Predicted score in the range of 0.0 to 1.0. \\
 box & Predicted bounding box points in the format xmin-ymin-xmax-ymax. \\
 mask & Predicted mask encode in run-length encoding (RLE) format. \\
local\_rms & Local noise level of input image in Jy$\cdot {\rm beam}^{-1}$. \\
peak\_flux &  Peak flux density of the fitted Gaussian model component in Jy$\cdot {\rm beam}^{-1}$ for CS sources, or flux density at the peak within the predicted box in Jy$\cdot {\rm beam}^{-1}$ for extended sources. \\
 err\_peak\_flux & Fitting error of the peak flux density in Jy$\cdot {\rm beam}^{-1}$, only for CS sources. \\
 int\_flux &  Integrated flux density of fitted Gaussian model component in Jy for CS sources, or total flux density of extended sources. \\
 err\_int\_flux &  Fitting error of the integrated flux density in Jy, only for CS sources. \\
 ra &  Right Ascension of the emission peak of the fitted Gaussian model component in degrees for CS sources, or Right Ascension of the emission peak within the predicted box in degrees for extended sources. \\
 dec &  Declination of the emission peak of the fitted Gaussian model component in degrees for CS sources, or Declination of the emission peak within the predicted box in degrees for extended sources. \\
 centre\_ra & Right Ascension at the center position of the predicted box in degrees. \\
 centre\_dec &  Declination at the center position of the predicted box in degrees. \\
 major &  Semi-major axis length of the fitted Gaussian model component in arcsec, only for CS sources. \\
 err\_major & Fitting error of the fitted semi-major axis length in arcsec, only for CS sources. \\
 minor & Semi-minor axis length of the fitted Gaussian model component in arcsec, only for CS sources. \\
 err\_minor & Fitting error of the fitted semi-minor axis length in arcsec, only for CS sources. \\
 pa & position angle of the fitted Gaussian model component in degrees, only for CS sources. \\
 err\_pa & Fitting error of the Gaussian model component's position angle in degrees, only for CS sources. \\
 deconv\_major & Deconvolved major axis in arcsec, only for CS sources.\\
 deconv\_minor& Deconvolved minor axis in arcsec, only for CS sources.\\
 deconv\_pa& Deconvolved position angle in degrees, only for CS sources.\\
\bottomrule
\end{tabular}
\end{threeparttable}
\end{table}

\subsubsection{Completeness of the catalog}
The final catalog (called FIRST-\HeTu) detected by \HeTu-v2 is shown in Table~\ref{tab:inference}. 
In Table~\ref{tab:inference}, \HeTu-v2 has detected a total of 835,435 sources, including 700,609 CS sources, 12,501 FRI sources, 47,582 FRII sources, 22,187 HT sources, and 52,557 CJ sources. We have compared FIRST-\HeTu\ with FIRST-14dec17 to investigate its completeness. 

The \happy\ program that was used to generate the FIRST-14dec17 catalog is a component-based source finder, which identifies a CS source with a single component while identifying an extended source with multiple discrete
components or sources. Diffuse emission between
close components cannot be recognized as Gaussian components,
so \happy\ cannot directly determine whether there is a connection between adjacent components, resulting in its lack of ability to automatically classify extended sources. Classification needs to be done by manual visual inspection in post-processing. Two methods have, therefore, been used to compare catalogs. For CS sources, we have used \stilts\ to perform cross-match on the fitted location columns between FIRST-\HeTu\ (ra and dec) and FIRST-14dec17 (RA and DEC) within a searching radius of 5 arcsec and obtained 689,877 CS sources detected by both. For extended sources, we have developed a script to automatically associate FIRST-14dec17 components with FIRST-\HeTu\ sources, which takes the FIRST-14dec17 components positioned within the predicted masks of the extended source as its associated components. 
There are 21,835 (FRI), 97,940 (FRII), 49,044 (HT), and 74,588 (CJ) FIRST-14dec17 components to be associated in different classes of extended sources.
Thus, a total of 933,284 FIRST-14dec17 components are associated with the sources detected by \HeTu-v2. The cross-matching rates achieve 98.6\%, showing excellent completeness of \HeTu-v2 detecting FIRST radio sources. The remaining 1.4\% of components that were not detected by \HeTu-v2 are limited by the current model's performance.

\begin{table}[!ht]
\footnotesize
\centering
\begin{threeparttable}\caption{Comparison the catalog results between FIRST-\HeTu\ and FIRST-14dec17}\label{tab:inference}
\doublerulesep 4pt \tabcolsep 12pt 
\begin{tabular}{p{0.1\columnwidth}p{0.1\columnwidth}p{0.5\columnwidth}}
\toprule
  Class & FIRST-\HeTu   & Components of the FIRST-14dec17 cross-matched or associated with the FIRST-\HeTu\ sources  \\\hline
  CS      & 700,609 & 689,877\\
  FRI    &  12,501  & 21,835 \\
  FRII    &  47,582  & 97,940 \\
  HT     & 22,186 & 49,044\\
  CJ   &52,557  & 74,588 \\ \hline
  Total & 835,435 & 933,284 \\
\bottomrule
\end{tabular}
\end{threeparttable}
\end{table}

Since \happy\ detection is related to the threshold value and strict acceptance criteria while \HeTu-v2 detection is based on the source morphology and unlimited on the flux density threshold and acceptance criteria \citep{1997ApJ...475..479W}, \HeTu-v2 can detect more CS sources than \happy. 
\HeTu-v2 has detected 10,732 CS sources that were not detected by \happy, with 3,311 of them having optical counterparts within a search radius of 5 arcsec from Sloan Digital Sky Survey Data Release 16 \citep[SDSS DR16;][]{2020ApJS..249....3A}. Among these 3,311 sources, approximately 75\% have a signal-to-noise ratio greater than or equal to 5, which is the threshold value used in \happy. 
 
The number of sources in the latest released catalogs of FRI, FRII, and HT sources are 122 \citep{capetti2017fricat}, 219 \citep{2017A&A...601A..81C}, and 717 \citep{2022ApJS..259...31S}, respectively. By the cross-matching script described in Section~\ref{sec:build_catalog}, we have found that the extended sources in the FIRST-\HeTu\ catalog completely cover the latest published results, and we have identified more new sources. Therefore, FIRST-\HeTu\ is currently the most complete catalog for the morphological classification and identification of extended sources in the FIRST survey.


On comparing the results obtained from \HeTu-v1 in Section~\ref{sec:data} with those from \HeTu-v2, we observe that the total number of CS sources detected by \HeTu-v1 ($i.e.$, 852,271) is higher than that detected by \HeTu-v2 ($i.e.$, 700,609). However, most of the CS sources detected by \HeTu-v1 but not by \HeTu-v2 are found to be false positives by visual inspection. This suggests that \HeTu-v2 detects fewer false positives in CS sources. Furthermore, the total number of detected extended sources by \HeTu-v2 is approximately 1.3 times that of \HeTu-v1, and all of these extended sources have associated components in the FIRST-14dec17 catalog, highlighting the significantly improved locating accuracy and detection performance of \HeTu-v2 on extended sources. These findings demonstrate that \HeTu-v2 has excellent accuracy on the FIRST survey and has established a high completeness catalog.

\subsubsection{Reliability and accuracy of the catalog}
 
\begin{figure}[!ht]
\centering
\includegraphics[scale=0.27]{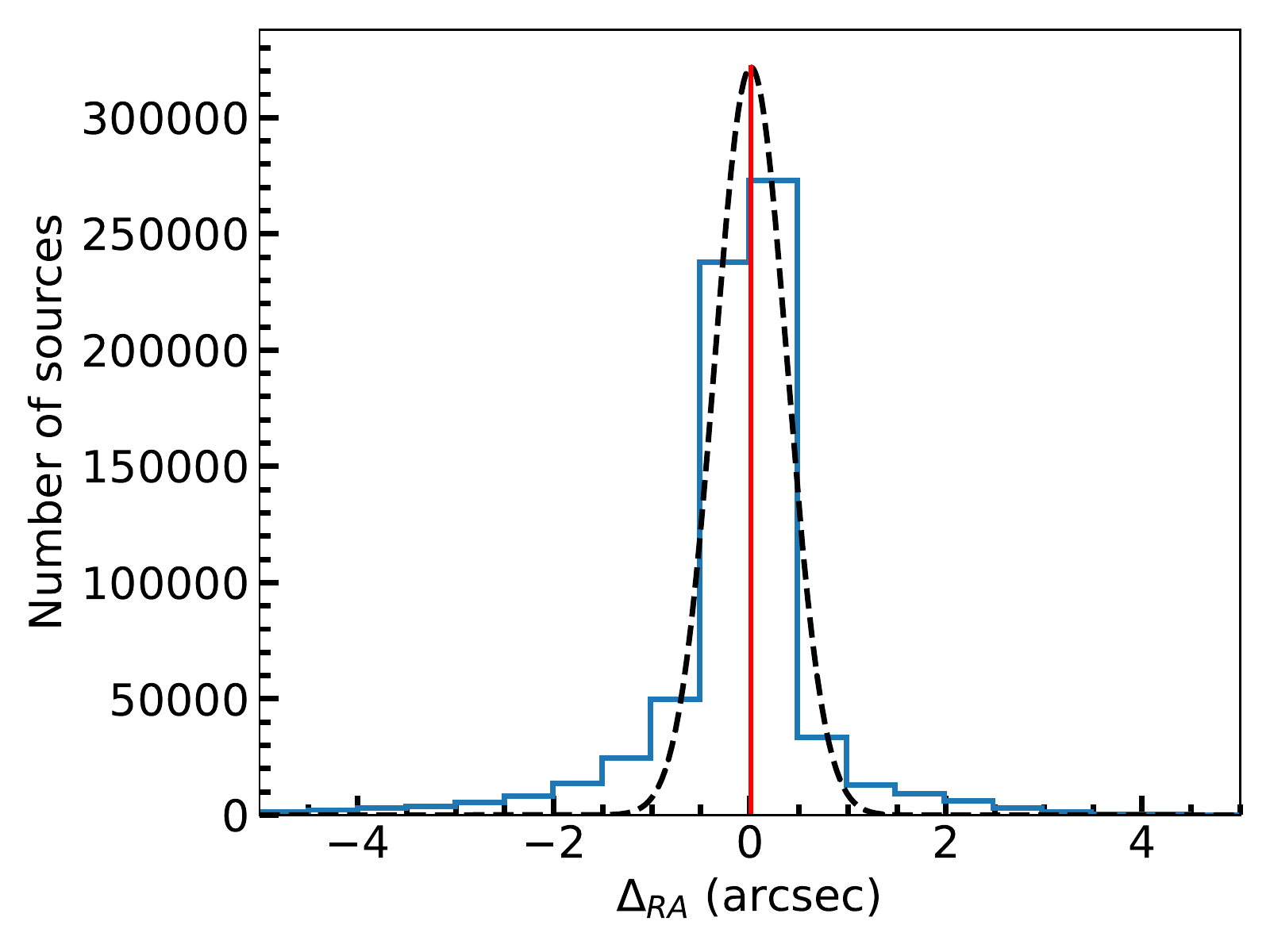}%
\includegraphics[scale=0.27]{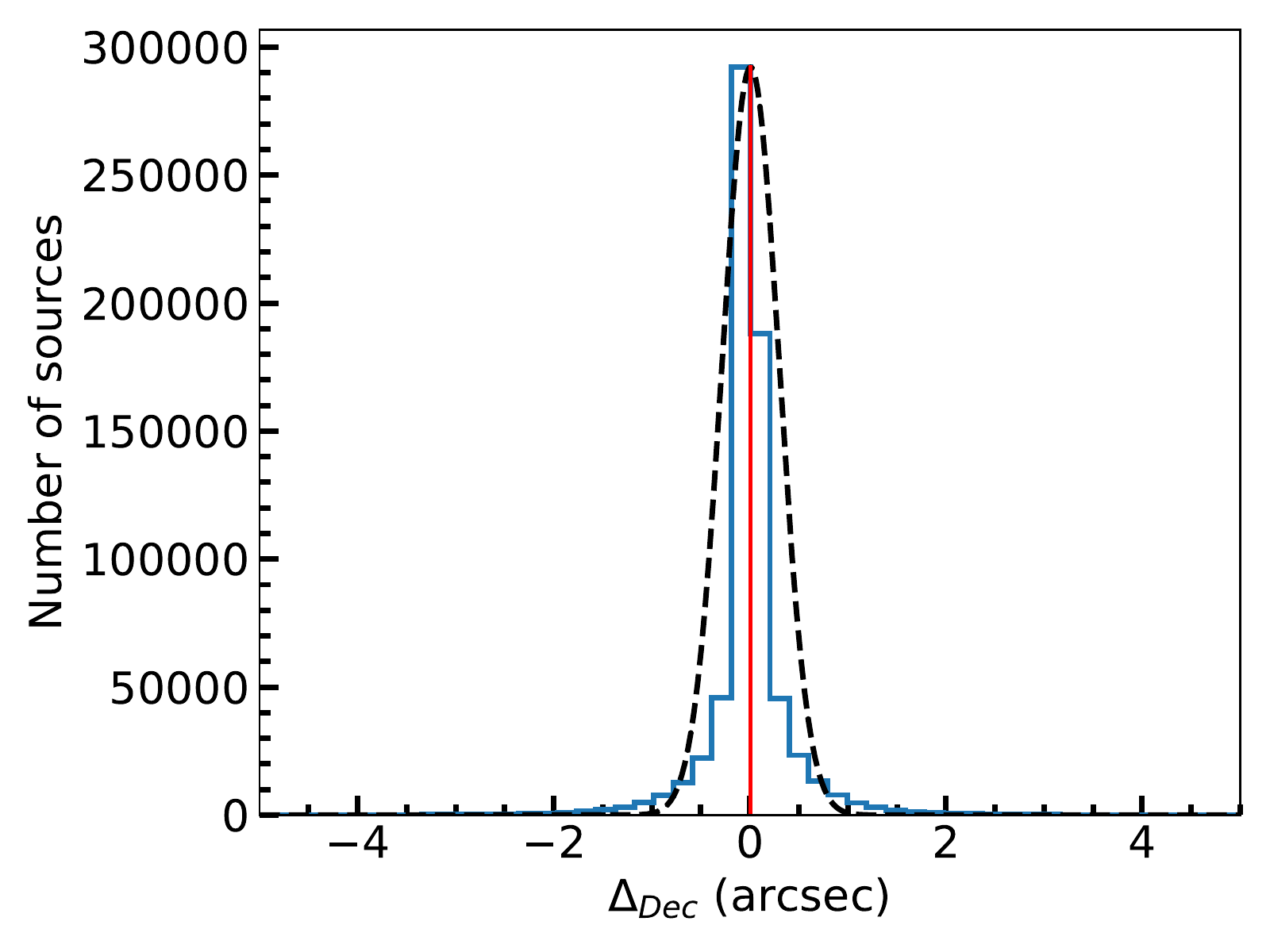}
\caption{Histograms of the difference in position coordinates ($i.e.$ RA and Dec) of CS sources detected by \HeTu-v2 compared to the matching FIRST-14dec17 sources, showing a mean offset of 0.0097 arcsec and 0.0073 arcsec on the RA axis and Dec axis, respectively. The black dotted curves show Gaussian fits to the histograms, and the red solid vertical lines indicate the mean offsets.}
\label{fig:ra_dec_density}
\end{figure}

For each CS source that is cross-matched with a FIRST-14dec17 component, we have measured the offset in position to check the precision of the positions of the CS sources detected by \HeTu-v2. The result is presented in Fig.~\ref{fig:ra_dec_density}, showing a mean offset of 0.0097 arcsec and 0.0072 arcsec on the Right Ascension (RA) axis and Declination (Dec) axis, respectively. The distributions in the left and right panels are concentrated around 0 arcsec.
This indicates that the positions of CS sources fitted by \HeTu-v2 are basically consistent with those of \happy. 

\begin{figure}[!ht]
\centering
\includegraphics[scale=0.5]{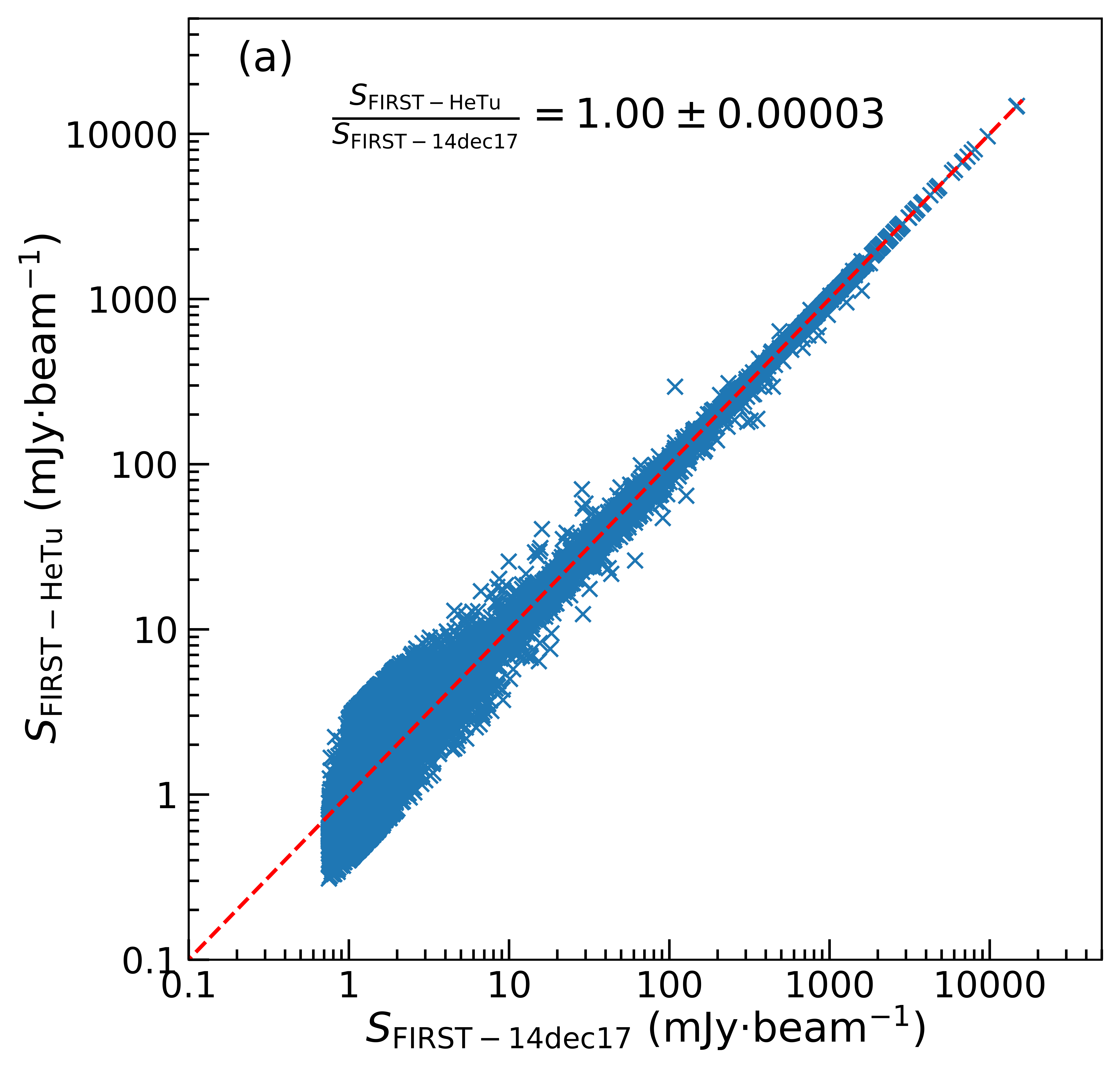}
\hspace{0.2mm}
\includegraphics[scale=0.5]{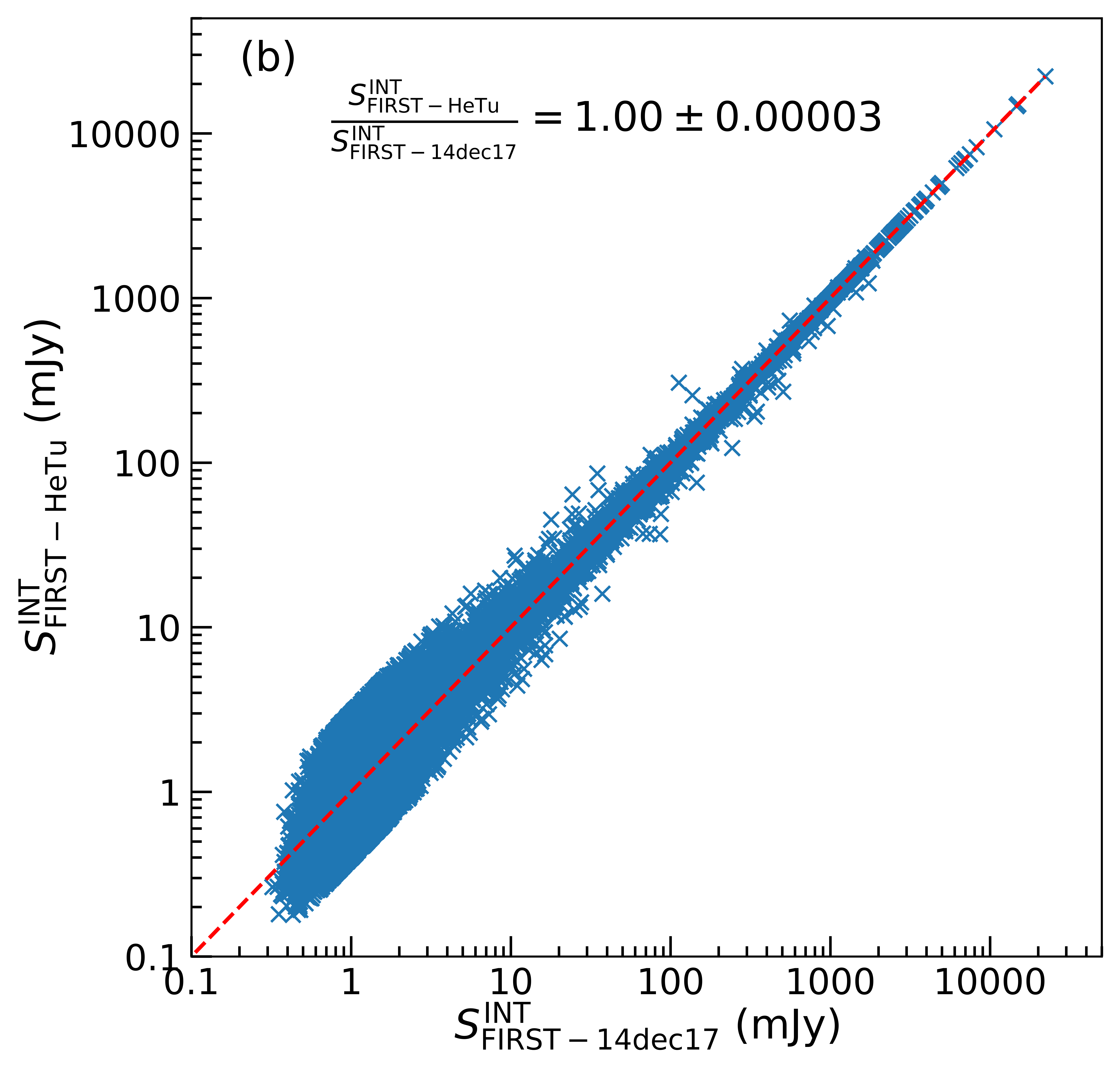}
\caption{Comparison of fitted source peak (a) and integrated (b) flux densities between FIRST-\HeTu\ and FIRST-14dec17 for all cross-matched CS sources. The red solid line is the equal flux line.}
\label{fig:flux_hetu_first}
\end{figure}

\begin{figure}[!ht]
\centering
\includegraphics[scale=0.27]{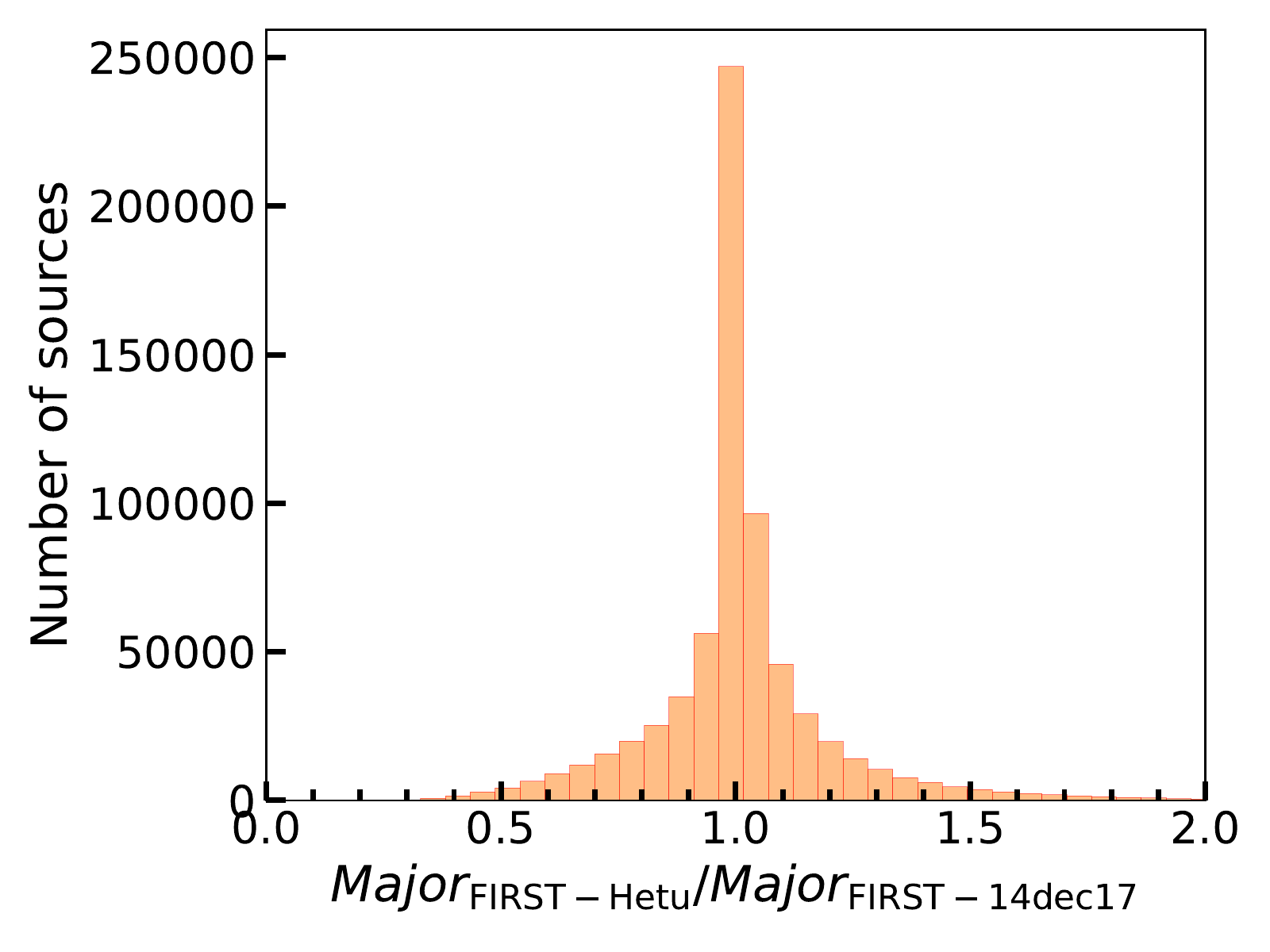}%
\includegraphics[scale=0.27]{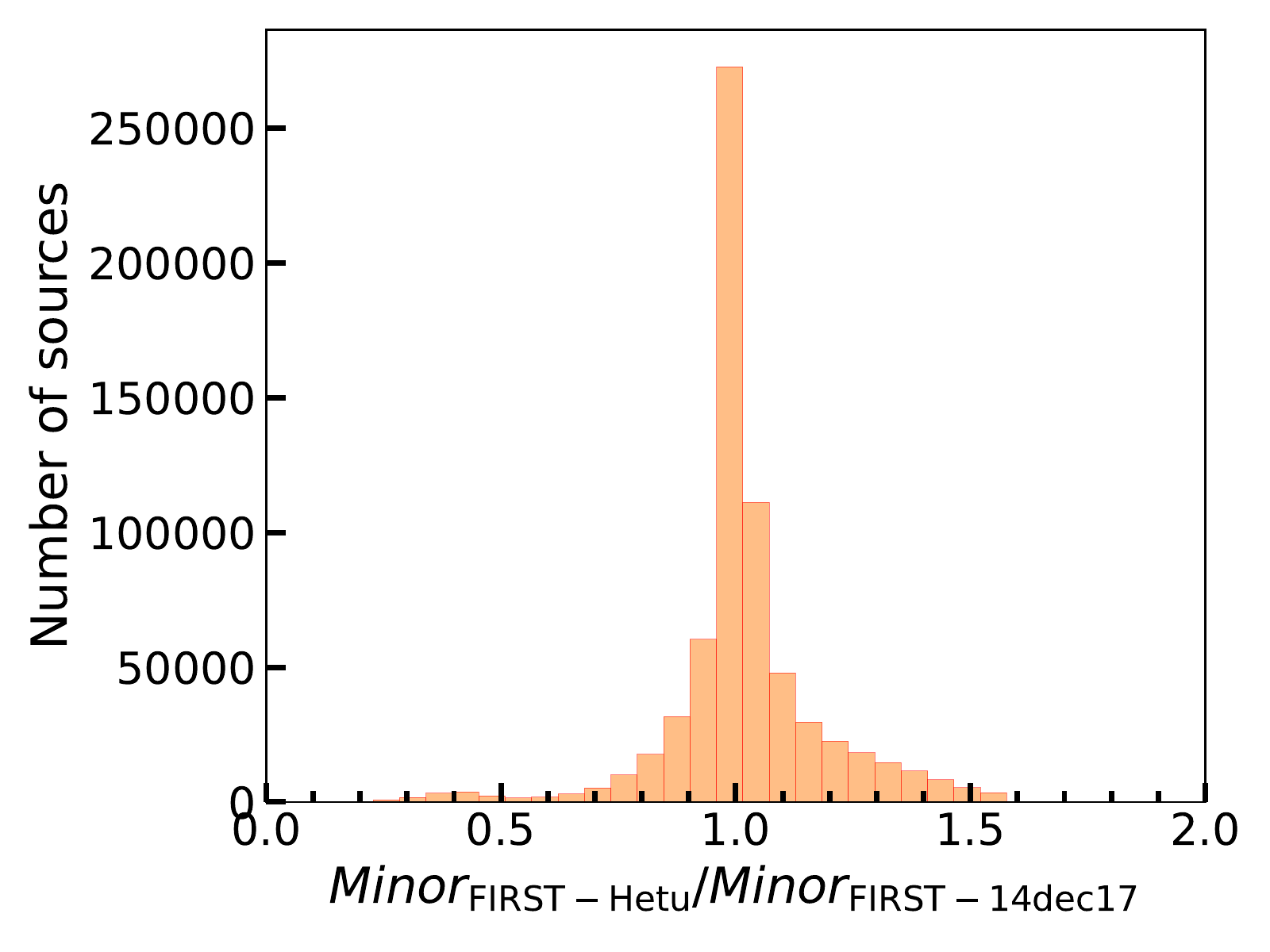}
\caption{Comparison of semi-major and semi-minor axis length of the fitted Gaussian model component between FIRST-\HeTu\ and FIRST-14dec17 for all cross-matched CS sources.}
\label{fig:beam_offsets}
\end{figure}

The flux density accuracy of the detected CS sources has been estimated by comparing the peak and integrated flux densities of cross-matched CS sources in the FIRST-\HeTu\ and FIRST-14dec17 catalogs. 
The results are shown in Fig.~\ref{fig:flux_hetu_first}. Defining $R = {S_{\rm FIRST-HeTu}}/{S_{\rm FIRST-14dec17}}$ as the ratio between the two peak flux densities, the median value of $R$ is 0.93 with a median absolute deviation of 0.08. The median and absolute deviation values of the ratio between two integrated flux (${S^{\rm INT}_{\rm FIRST-HeTu}}/{S^{\rm INT}_{\rm FIRST-14dec17}}$) densities are equivalent to those of $R$. In addition, in Fig.~\ref{fig:flux_hetu_first}, 
a linear fit to ${S_{\rm FIRST-HeTu}}$ vs. ${S_{\rm FIRST-14dec17}}$ gives the slope of 1.0 and intercept of 0.00003, the linear regression analysis of ${S^{\rm INT}_{\rm FIRST-HeTu}}$ vs. ${S^{\rm INT}_{\rm FIRST-14dec17}}$ yields the same slope and intercept as that of ${S_{\rm FIRST-HeTu}}$ vs. ${S_{\rm FIRST-14dec17}}$.
It indicates that the flux densities fitted by \HeTu-v2 and \happy\ are in basic agreement. 
The flux density shows larger fluctuations below 10 mJy, indicating that the flux density fitted by \HeTu-v2 is somewhat different from that by \happy\ in weak sources. Each fitted component by \happy\ lies in the contiguous set of pixels exceeding the threshold ($i.e.$ $5\sigma$), while the Gaussian fit by \HeTu-v2 is performed only in the pixels region within the predicted box. For weak point-like sources, the regions of the sources used for fitting by \HeTu-v2 are larger than those used in the \happy\ fitting. Furthermore, this also leads to differences in the shape and size of the Gaussian models that they fit. Fig. \ref{fig:beam_offsets} compares the semi-major and semi-minor axis lengths of the fitted Gaussian models between the two catalogs for all cross-matched CS sources, showing that the fitted Gaussian models by \HeTu-v2 are slightly larger than those by \happy.


\begin{figure}[!ht]
\centering
\includegraphics[scale=0.5]{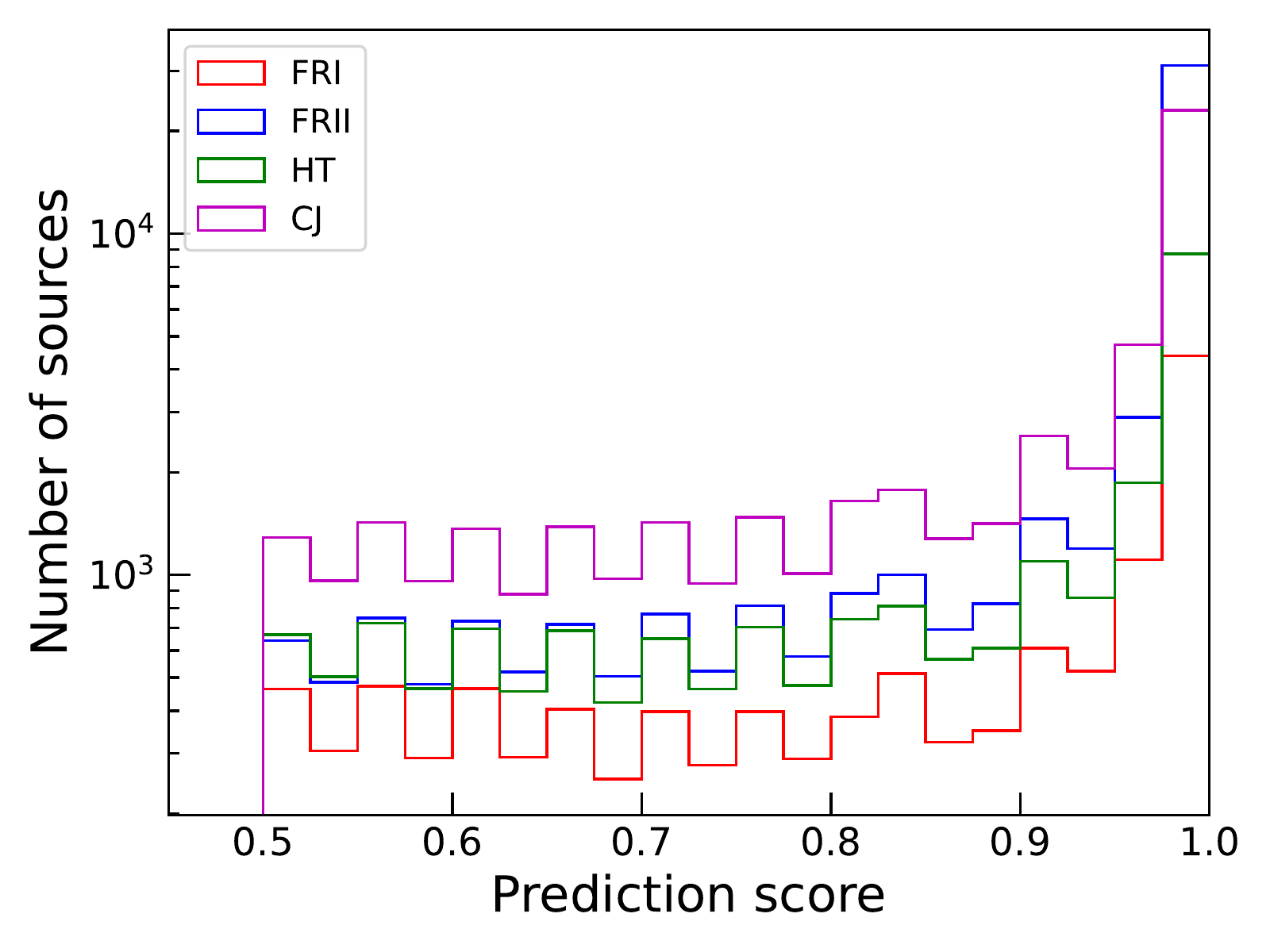}
\caption{Histograms of the prediction score of \HeTu-v2 detected extended sources.}
\label{fig:mask_score}
\end{figure}

By defining the source that cross-matched with FIRST-14dec17 as true positive, the detection precision for the CS source is 98.5\%. In addition, the analysis revealed that the ratio of integrated flux to peak flux for the cross-matched CS sources in FIRST-14dec17 is consistently less than 1.2, which suggests that these sources are indeed CS sources. For extended sources, we used the prediction score to estimate the detection accuracy; the higher the score, the higher the accuracy. Fig.~\ref{fig:mask_score} shows the prediction score histograms of the detected FRI, FRII, HT, and CJ sources. The histograms of four classes show protruded shoulders from 0.9 to 1.0 indicating that the majority of detected extended sources have high accuracy. The mean prediction scores of FRI, FRII, HT, and CJ sources have achieved 0.84, 0.93, 0.86, and 0.88, respectively. In summary, the FIRST-\HeTu\ is a reliable and accurate FIRST morphological catalog.

\subsection{Application prospects}\label{sec:practical}
\begin{figure}[!ht]
\centering
\includegraphics[scale=0.16]
{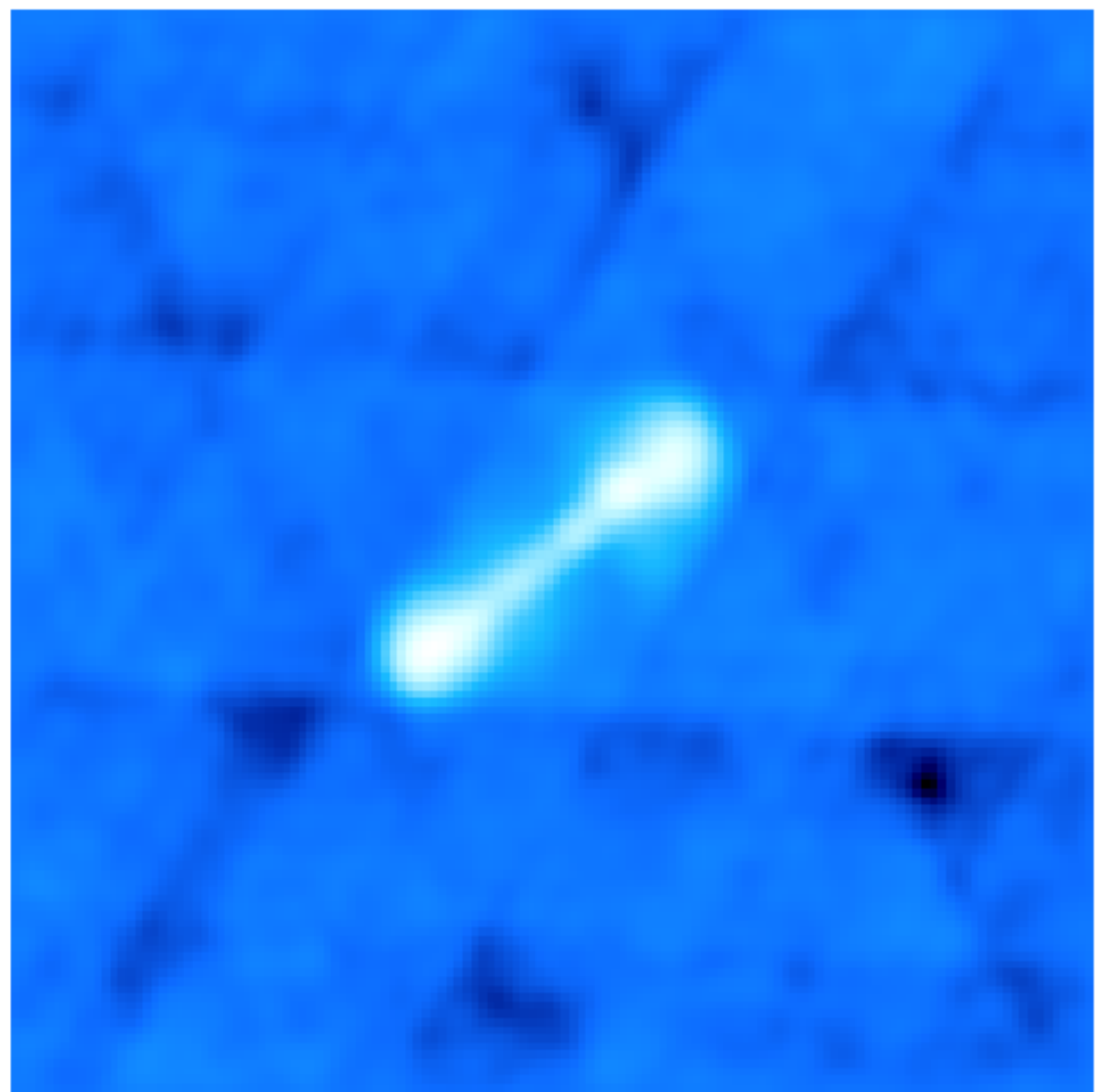}
\includegraphics[scale=0.16]{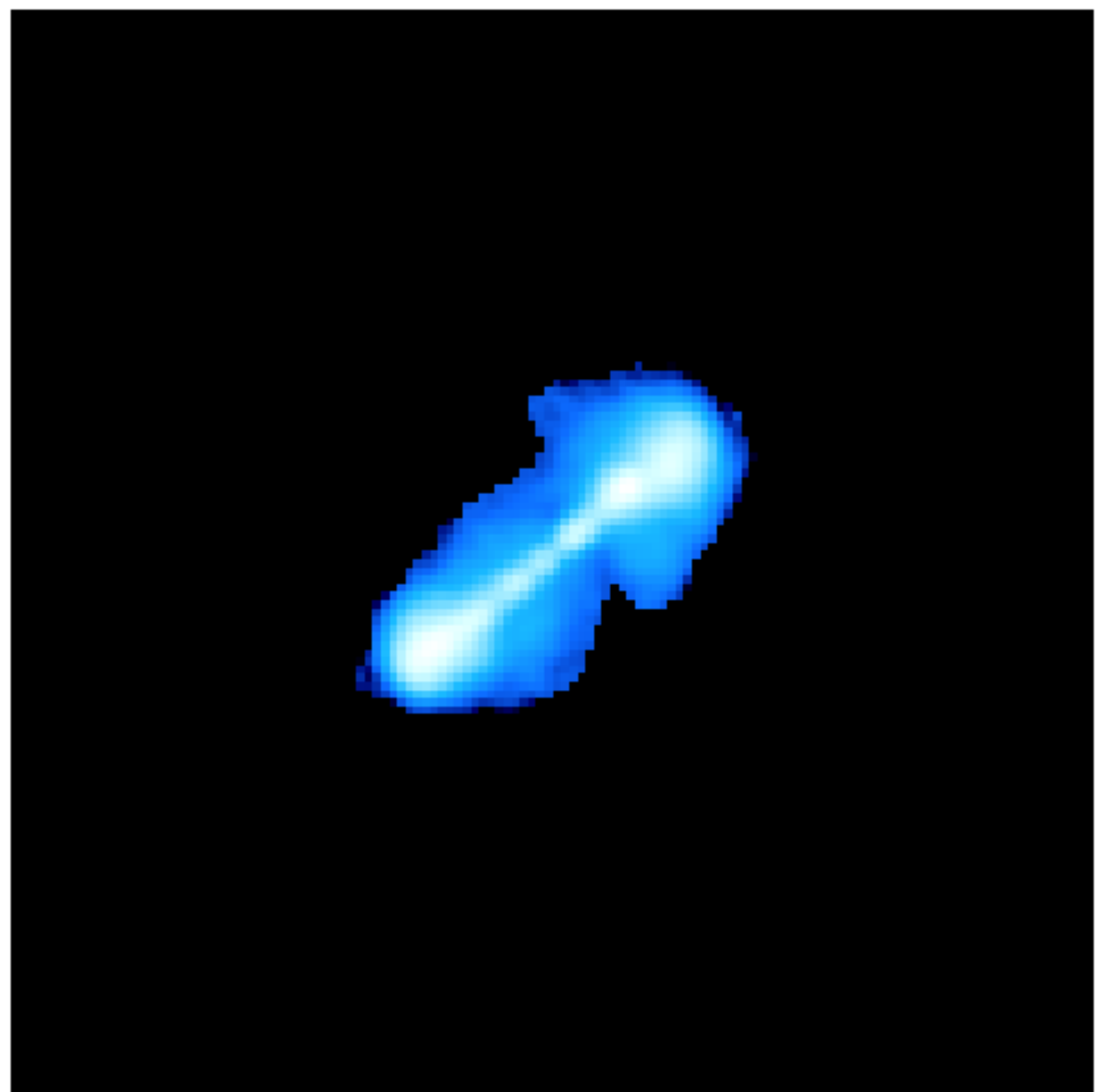}
\includegraphics[scale=0.16]{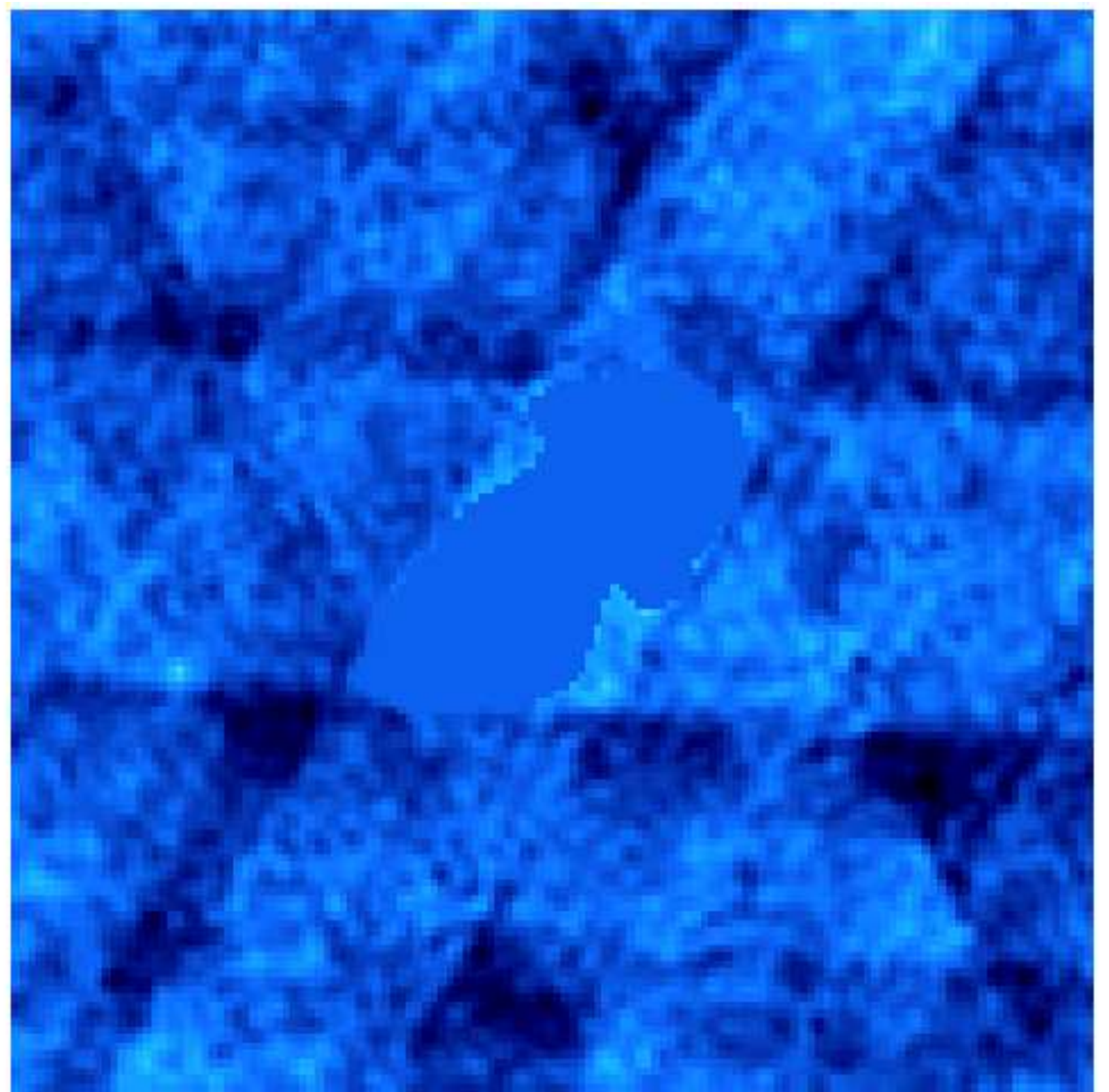}
\hspace{0.1mm}
\vspace{0.1mm}
\includegraphics[scale=0.16]{original.pdf}
\includegraphics[scale=0.16]{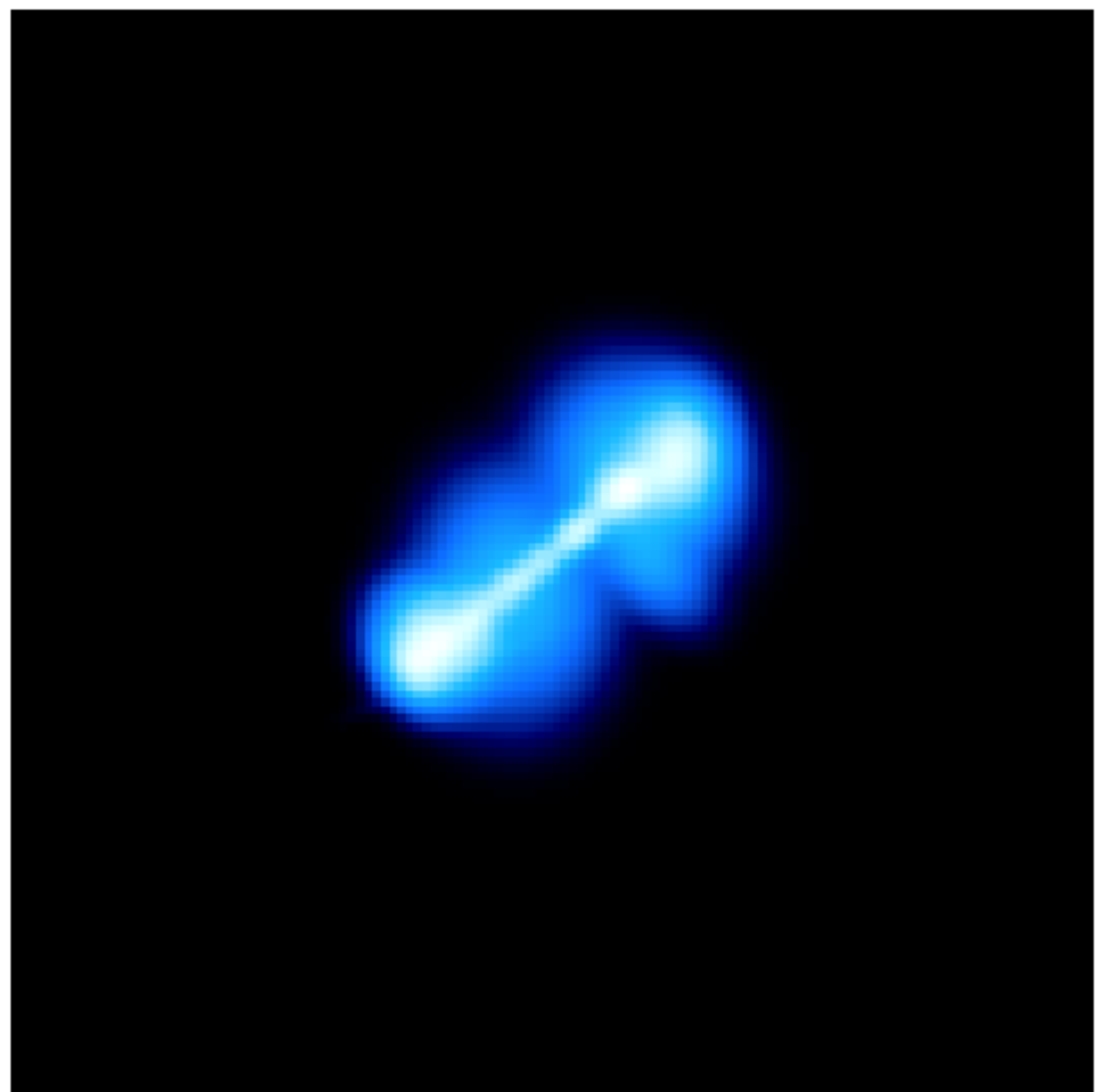}
\includegraphics[scale=0.16]{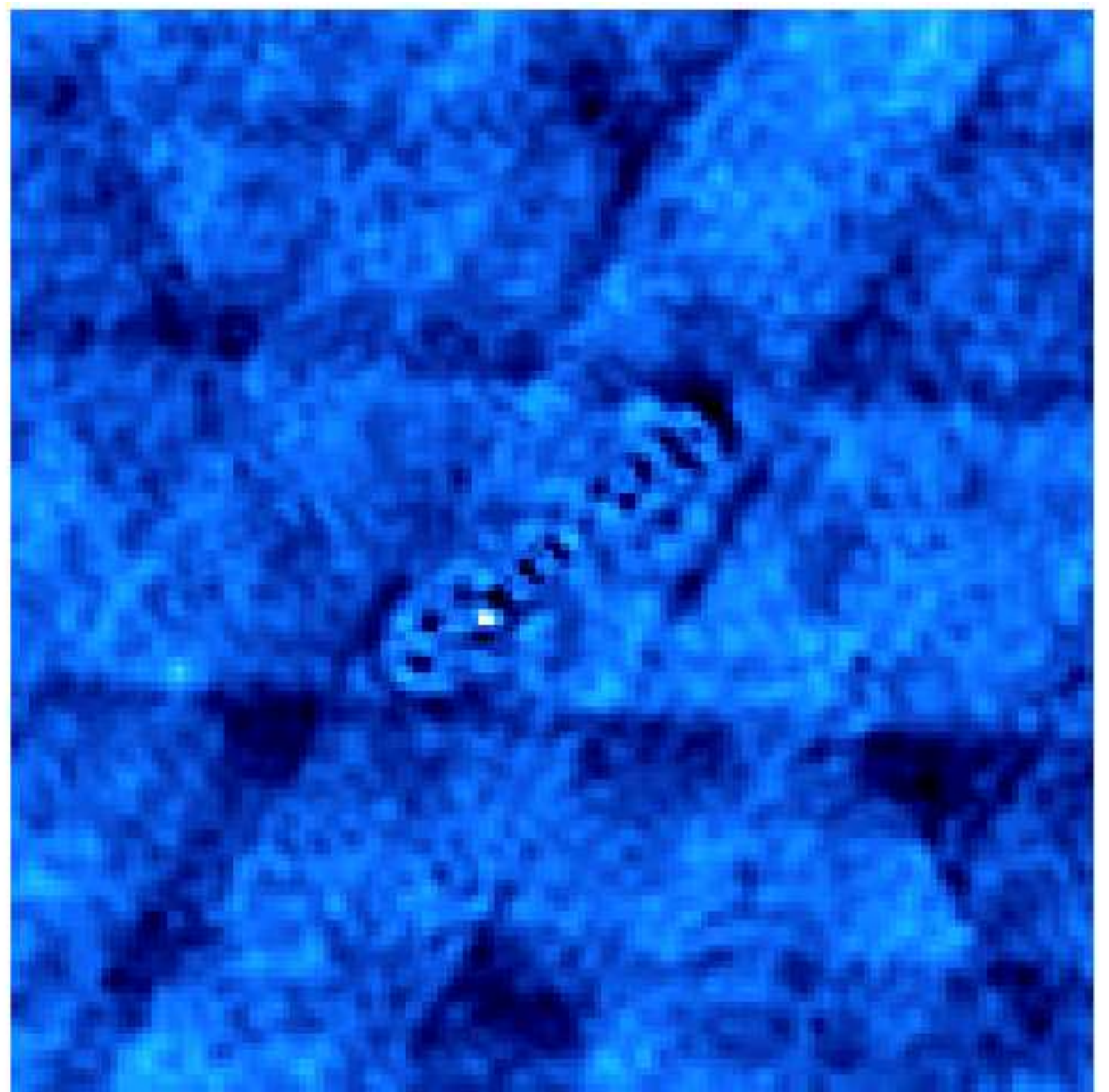}
\caption{Comparison of the model (middle column) and residual (right column) images from \HeTu-v2 (top row) and \pybdsf\ (bottom row) of source J085556.090+491113.15 (left column) from the FIRST survey. Here, \pybdsf\ having version v1.10.1 generates the model and residual images using the task `process\_image' with default parameters.}
\label{fig:skymodel}
\end{figure}

Building a catalog from radio images is the primary purpose and function of \HeTu-v2. 
The generated catalog by \HeTu-v2 has three new useful features compared to the traditional catalog: the morphological class for each source, the total flux density for the extended source, and the central location for each source. 
As a novel tool, \HeTu-v2 like traditional source-finding software, the created catalog results can be used to build sky models which can be used in the calibration steps of the standard data processing pipeline. For CS sources, the sky models are still created based on the Gaussian fitted information by \HeTu-v2. For extended sources, however, the prediction result by \HeTu-v2 allows building the sky model in pixel levels like the pixel segmentation-based source finder \profund\ \citep{hale2019radio}. Fig.~\ref{fig:skymodel} shows how \HeTu-v2 is able to trace the shape of the source and so model its radio emission. In the case shown here, \pybdsf\ does not adequately model the emission seen in the image. 
The residual images in Fig.~\ref{fig:skymodel} also show how \HeTu-v2 is tracing the shape well in comparison to the \pybdsf\ whose residual image appears to over-fit in areas, which can leave negative and excess positive residuals around the source.  

\HeTu-v2 is also able to perform automatic association of radio components like the recent works by \cite{2022arXiv220914226M} through a small amount of extended development. Based on the results of traditional source-finding software, this method can automatically and correctly combine the separated radio components that belong to a source, which can further help to find the corresponding optical host galaxy \citep{2015MNRAS.453.2326B} and infer the physical parameters such as the size and luminosity of the radio source. Traditionally, radio components association is used manually by visual inspection which is a time-consuming process even with extensive manpower for the huge amount of sources from the modern survey. In our method, the components of the source are associated by finding the components from the FIRST catalog within the predicted masks by \HeTu-v2. 

\begin{figure}[!ht]
\centering
\includegraphics[scale=0.15]{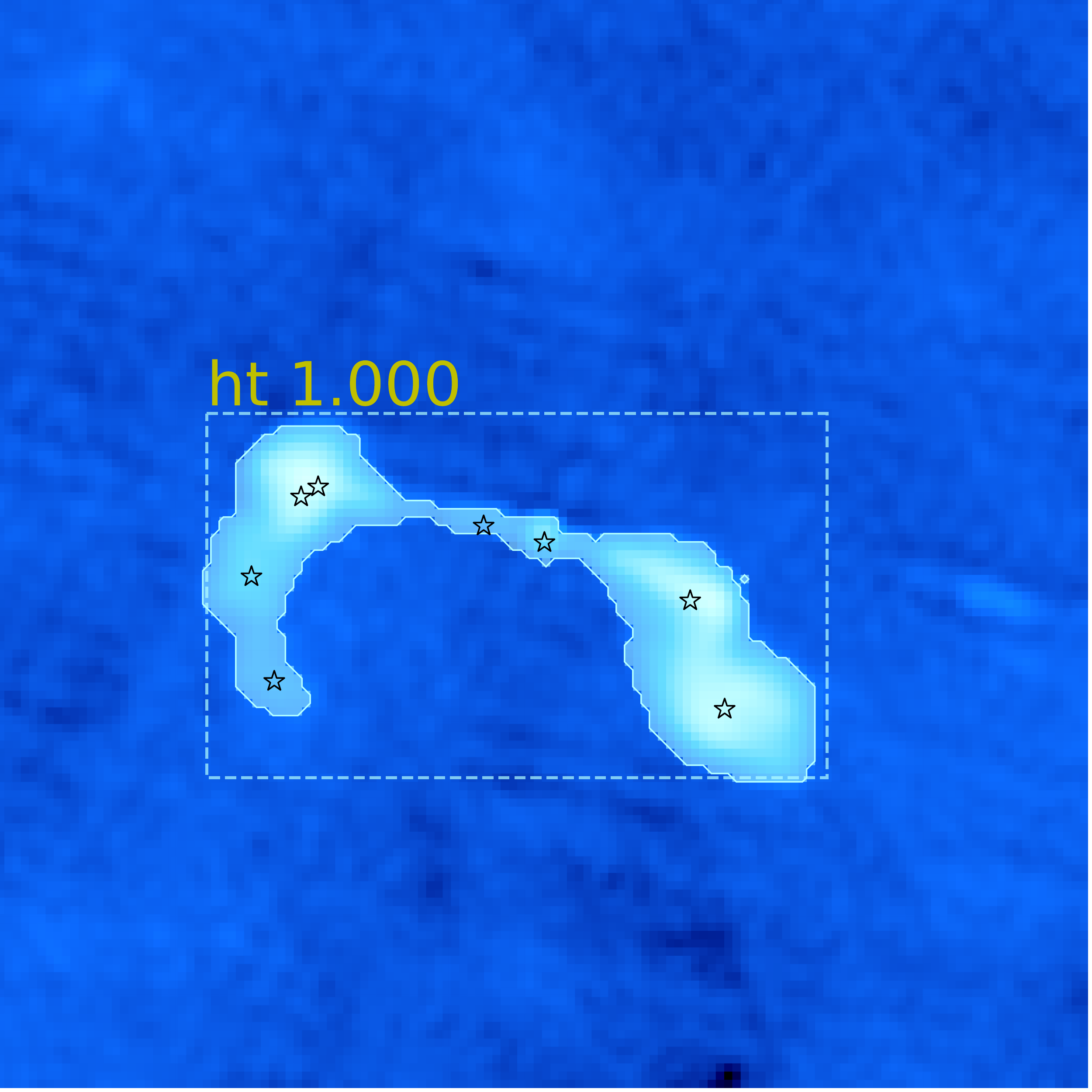}
\hspace{0.1mm}
\includegraphics[scale=0.15]{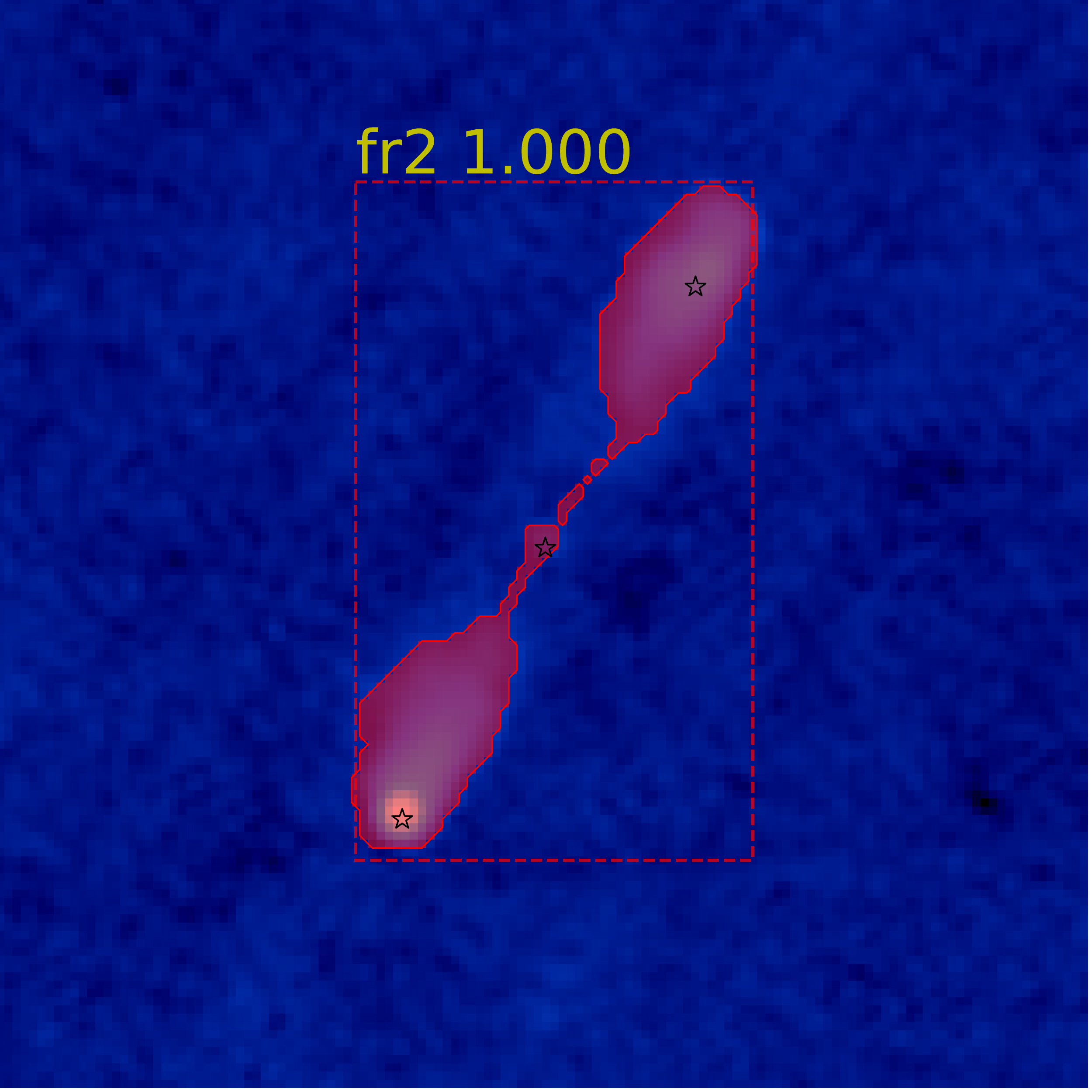}
\hspace{0.1mm}
\includegraphics[scale=0.15]{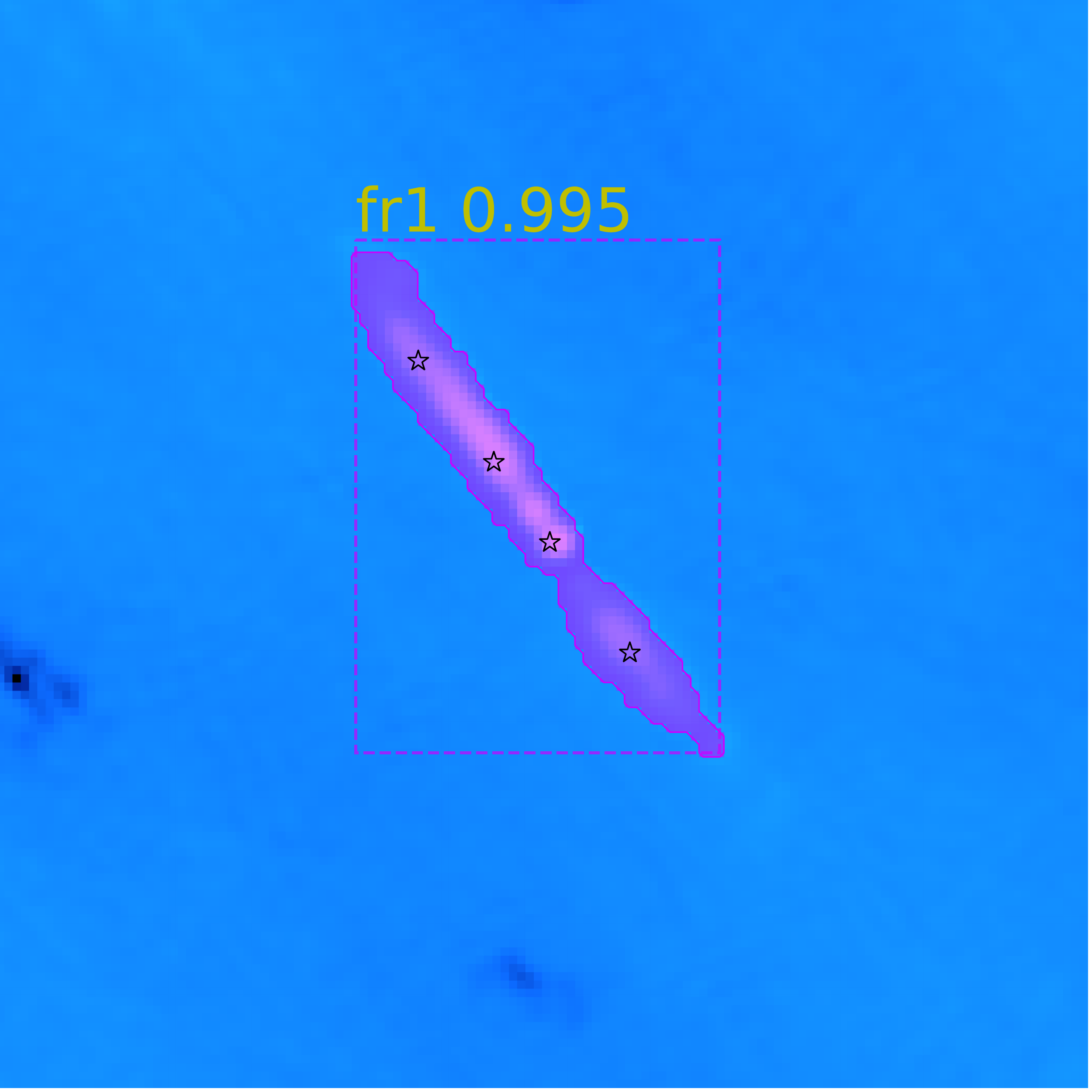}
\hspace{0.1mm}
\includegraphics[scale=0.15]{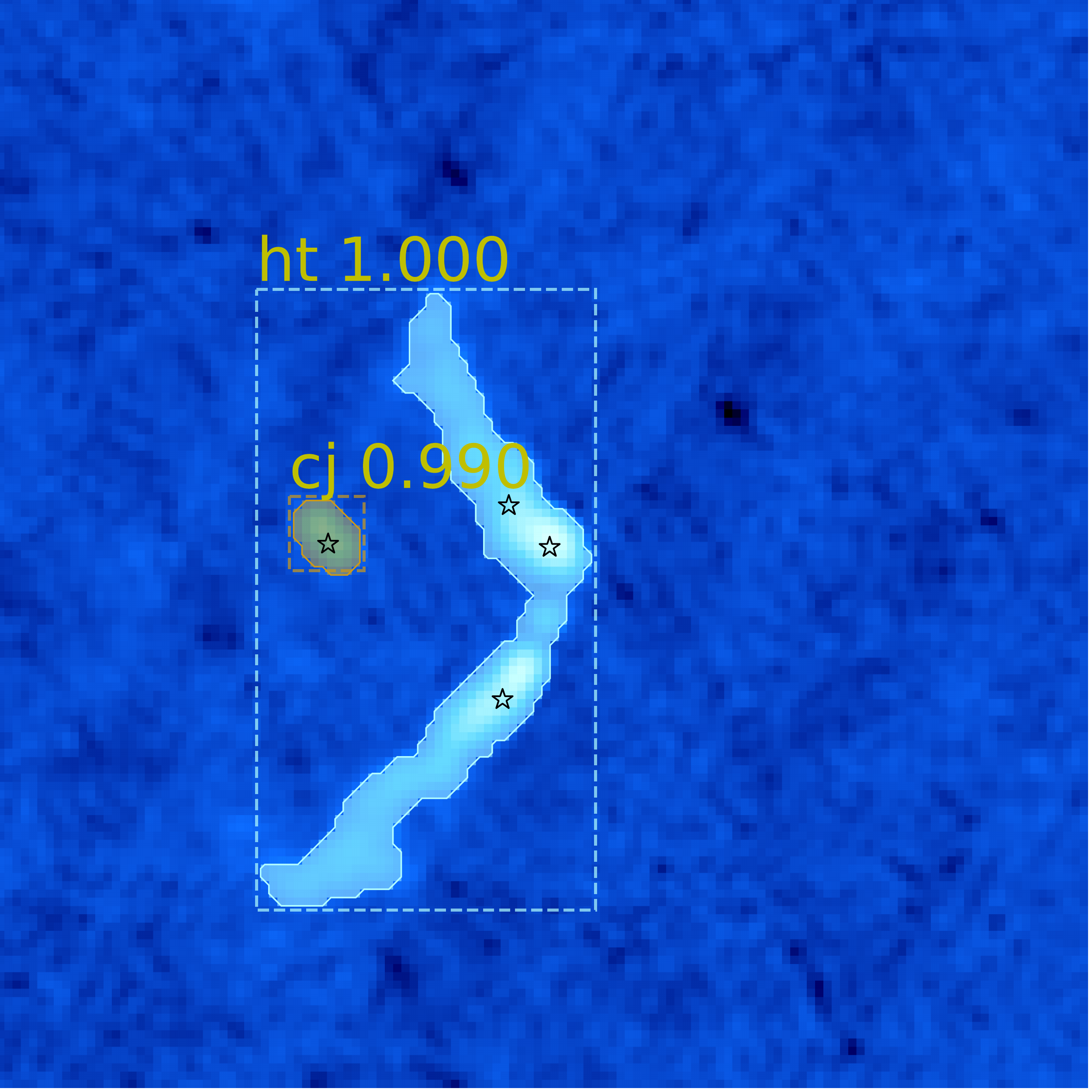}
\caption{Examples of automatic radio components association using \HeTu-v2. The different colored rectangular boxes and masks represent the identified sources: magenta for FRI (fr1), red for FRII (fr2), light-cyan for HT (ht), and orange for CJ (cj). Each source is labeled with a class name and a score between 0 and 1 on the top left of the bounding box. The black star indicates the position of the FIRST radio components.}
\label{fig:components}
\end{figure}

Fig.~\ref{fig:components} shows examples of automatic radio components association using \HeTu-v2. The components with black star markers that belong to a single radio source have been successfully associated. In the bottom-right panel, there is a CJ source inside the predicted bounding box of the HT source, and our method automatically ignores the CJ source as the associated components of the HT source. However, the method of \cite{2022arXiv220914226M} finds the components within the bounding box, which can treat the component of the CJ source as false positive for the HT source. This suggests that our method is more appropriate for automatically separating radio component associations.

\HeTu-v2 can also be used as a multi-class classifier when only focusing on determining whether the input image contains an interested celestial object and no need to pay attention to the location of the object and the number of celestial objects contained. \HeTu-v2 based classifier allows classifying radio images into 5 classes of radio galaxies: CS, FRI, FRII, HT, and CJ. Examples of radio galaxy classification can be seen in the first column of panels in Fig. \ref{fig:valid-example}. Such a classifier can automatically provide a large number of morphologically classified samples to scientists, which greatly helps to investigate and address key cosmological questions, such as the formation and evolution of SMBH.

\section{Summary and conclusions}\label{sec:conlu}
Radio source detection is generally done by component-based Gaussian fitting or pixels segmentation in source-finding software. These methods do not work well on extended sources and the classification of components needs to be done manually at a later stage. Such a laborious approach is not 
applicable to modern large-scale radio surveys.
The novel method based on the deep learning object detection algorithm using Mask R-CNN can automatically segment and classify radio sources. However, the obtained segmentation masks are still coarse at the boundaries of sources. In this paper, we developed a source detector called \HeTu-v2 for automated detection and classification of radio sources in high-quality segmentation. 
Based on the combination of Mask R-CNN and Transformer block network, \HeTu-v2 has achieved excellent performance on FIRST radio source detection with mean $AP_{\rm @50:5:95}$ of 77.8\% which improved by 15.6 points and 11.3 points compared to \HeTu-v1 and Mask R-CNN alone. \HeTu-v2 is able to distinguish the five most common different types of radio source morphologies: CS, FRI, FRII, HT, and CJ. \HeTu-v2 has established a new reliable morphological catalog for the FIRST survey with 98.6\% of completeness and up to 98.5\% accuracy compared to the latest release catalog of FIRST. \HeTu-v2 also works well on other astronomical tasks of building sky models, radio components association, and radio galaxies classification.

For the FIRST survey, the input images for \HeTu-v2 were generated by image cutouts centered on the FIRST-14dec17 catalog. This method is highly suitable for the FIRST survey, where the original images are not available, and guarantees the completeness of detection results to a great extent. However, it may not be as efficient when applied to other radio survey data due to the total area of all cutouts is much larger than the original total area of the survey. To apply the \HeTu-v2 model to other modern radio surveys, the following 
image cutout method can be adopted. Modern radio continuum surveys often generate image data with very large sizes. For instance, in the EMU survey, each image has a pixel size of approximately $16,000\times14,000$. Radio sources appear small and clustered in the image ($\sim$ 25,000 components per image), making it difficult to accurately detect and recognize them directly from the large image using \HeTu-v2. To address this issue, a sliding cropping method is typically employed to extract multiple small images from the larger image, which are then subjected to detection. To start, the large image is cropped into smaller fixed-sized images ($M \times M$ pixels) sequentially from the top-left corner to the bottom-right corner. To avoid some sources being segmented and cut off between two small images, an overlap ($N_{\rm overlap}$ pixels) region is set between them. The value of $M$ can vary between 132 and 250, whereas $N_{\rm overlap}$ corresponds to the pixel size of the largest radio source within the survey area. Finally, the results obtained from each small image are merged, and any duplicate sources detected within overlapping regions are removed to obtain the final result for the entire large image area.



\section*{CRediT authorship contribution statement}
{\bf B. Lao}: Conceptualization, Methodology, Software, Writing - original draft, Writing - review \& editing. {\bf S. Jaiswal}: Data Processing \& analysis, Writing - review \& editing. {\bf Z. Zhao}: Network Design, Writing - original draft. {\bf L. Lin}: Data Labels. {\bf J. Wang}: Workflow Design. {\bf X. Sun}: Data analysis, Writing - review \& editing. {\bf S.-L. Qin}: Writing - review \& editing.

\section*{Declaration of competing interest}
The authors declare that they have no known competing financial interests or personal relationships that could have appeared
to influence the work reported in this paper.

\section*{Data availability}
The data used in this study are available
at the FIRST survey archive (\url{https://third.ucllnl.org}). The software code is publicly available at \url{https://github.com/lao19881213/RGC-Mask-Transfiner}. 

\section*{Acknowledgements}
This work was supported by National SKA Program of China (2022SKA0120101). The preliminary experiments used resources of China SKA Regional Centre prototype \citep{An2019,An2022} funded by National SKA Program of China (2022SKA0130103) and the National Key R\&D Programme of China (2018YFA0404603). We thank Tao An and Ailin Wang for the helpful discussions.
\appendix



\bibliographystyle{elsarticle-harv} 
\bibliography{example}





\end{document}